\documentclass[prb,preprint]{revtex4-1} 
\usepackage{amsmath, amssymb, graphics}
\usepackage{dsfont}
\usepackage{rotating}
\usepackage{listings}

\usepackage{titlesec}

\usepackage{amsmath, amssymb, graphics}

\usepackage{url}

\usepackage{amsfonts} 
\usepackage{graphicx} %

\newcommand{\mathsym}[1]{{}}

\DeclareRobustCommand{\bfrac}[2]{%
  \mathchoice{\frac{\raisebox{-0.4ex}{$#1$}}{\raisebox{0.4ex}{$#2$}}}%
             {\frac{\raisebox{-0.4ex}{$\scriptstyle#1$}}{\raisebox{0.1ex}{$\scriptstyle#2$}}}%
             {\frac{#1}{#2}}%
             {\frac{#1}{#2}}%
}

\begin{document}

\title{Very special special functions: Chebyshev polynomials of a discrete variable and their physical applications}

\author{David J. Siminovitch}
\email{siminovitch@uleth.ca}
\affiliation{Department of Physics and Astronomy, The University of Lethbridge, 
Lethbridge,  Alberta T1K 3M4 }

\date{\today}

\begin{abstract}
Over nearly six decades, the Chebyshev polynomials of a discrete real variable have found applications in  
spin physics, spin tomography, in the development of  operator expansions, and in defining tensor operator equivalents. The properties of these polynomials are discussed in detail, and then examples are provided to illustrate the diversity of their applications in magnetic resonance. These examples include the use of the Chebyshev polynomial operators as an orthonormal basis to expand rotation operators, projection operators, and the 
Stratonovich-Weyl operator. The duality of the Chebyshev polynomials of a discrete real variable as Clebsch-Gordan coupling coefficients is noted and exploited, and it is shown that the Chebyshev polynomial operators can be recoupled as a rank-zero composite tensor defined by the product of a spin and spatial tensor. These application examples demonstrate that the Chebyshev polynomials of a discrete real variable are a unique nexus for spin physics,  special functions, angular momentum (re)coupling, and irreducible representations of the rotation group. 
\end{abstract}

\maketitle

\newpage
\baselineskip=10mm
\flushleft

\section{Introduction}

\subsection{The Majorana Formula}

Abragam introduces quantum mechanics in his nuclear magnetic resonance  text ``The Principles of Nuclear Magnetism" \cite{abragamtext} with two fundamental equations,  
the Schr{\"o}dinger equation,  and  the spin transition probability formula first derived by Majorana \cite{emajorana}. Majorana's  formula 
\cite{emajorana} provides the answer to the simplest and most fundamental question one can ask of a spin system: what is the probability for finding a spin in the state with magnetic projection number 
$m^{\prime}$ at time $t$ knowing that it was previously in the state $m$ at time $t=0$? More precisely, consider a spin in a uniform, static magnetic field {\bf H}$_0 =H_0 \, {\bf \hat{k}}$, whose direction can be taken to be the axis of quantization. Under the application of a perpendicular alternating radiofrequency field ${\bf H}_1 = H_1 \cos \psi t \, {\bf \hat{i}}$ in the laboratory frame which causes transitions,  Majorana's formula 
\cite{emajorana}, as quoted by  Abragam 
\cite{abragamtext}, gives the probability in a spin-$j$ system of a transition from a state of magnetic quantum 
number $m$ at time $t=0$ to one of magnetic quantum number $m^{\prime}$ at time $t$:
\begin{eqnarray}
\mbox{ P}^{(j)}_{mm^{\prime}}(t) & = & (j-m)! (j+m)! (j-m^{\prime})! (j+m^{\prime})! \left( \cos  \tfrac{1}{2} \alpha  \right)^{\!4j} \times \nonumber \\
& & 
\;\;\;\;\left[ \sum_{\lambda=0}^{2j} \displaystyle\frac{(-1)^r   \left( \tan \frac{1}{2} \alpha \right)^{2\lambda -m+m^{\prime} }}
{\lambda!(\lambda-m+m^{\prime})!(j-m^{\prime}-\lambda)!(j+m-\lambda)!}     \right]^{\!2} \label{abramajorana} \\
\mbox{where} \;\;\;  \sin  \tfrac{1}{2} \alpha & = & \sin \Theta \, \sin \bfrac{\psi}{2} \label{probangles} \\
\psi&  =  & \left|\gamma {\bf H}_e\right| t \\
{\bf H}_e & = & \left[ H_0 + \frac{\omega}{\gamma} \right] {\bf \hat{k}} + H_1 \,  {\bf \hat{i}}
\end{eqnarray}
In a frame rotating with angular frequency $\omega$, the direction of the effective field {\bf H}$_e$ is defined by the polar angle $\Theta$,  and  $\left|\gamma {\bf H}_e\right| $  is the Larmor precession frequency of the spin magnetic moment around  the effective field {\bf H}$_e$.   Recognizing that Majorana's elegant derivation \cite{emajorana} of the spin transition probability relied on the theory of the irreducible representations of the group of rotations, Abragam 
\cite{abragamtext}  simply stated the formula of 
Eq.(\ref{abramajorana}), and did not discuss its derivation. We follow suit, since fortunately, there are a number of excellent textbook discussions 
\cite{ramsey, corio,gilmore,biedenharn, thompson, Esposito} of the Majorana formula of Eq.(1) which fill in the details of the derivation, and also provide some history.   

\subsection{The Meckler formula}  An alternative derivation of the Majorana formula \cite{emajorana} was given by Bloch and Rabi \cite{rabi2}.  Despite an improvement in the  symmetry of the formula in the indices $m$ and $m^{\prime}$, their version \cite{rabi2} was much the same as that of the original  \cite{emajorana}. 
Then in 1958, Meckler \cite{meckler:majorana} published a remarkably simple version of the Majorana formula \cite{emajorana}, relying on a novel projection operator method. 
 Meckler \cite{meckler:majorana,meckler:angular}  took a very unorthodox approach to calculating the transition probability, using projection operators expanded in a Chebyshev polynomial operator basis $f_L^{(j)}( {\bf \hat{n}} \cdot {\bf J}) $ (see Table I) never previously used in magnetic resonance or in any other physical application. The final result of Meckler's calculation \cite{meckler:majorana} took 
 the  following very concise form
\begin{eqnarray}
\mbox{ P}^{(j)}_{mm^{\prime}}(t) & = &  \left| \langle {\bf \hat{b}},m^{\prime} |\,{\bf \hat{a}},m \rangle \right|^{2} \label{modtrans}   \\
 & = & \sum_{\lambda=0}^{2j} f_{\lambda}^{(j)} (m) \; f_{\lambda} ^{(j)}(m^{\prime})\;  P_{\lambda}({\bf \hat{a}} \cdot {\bf \hat{b}}) \label{firstmeck} \\
\mbox{where} \;\;\;  {\bf \hat{a}} \cdot {\bf \hat{b}} & = & \cos \beta(t) 
\end{eqnarray}
Meckler's  formula 
 \cite{meckler:majorana,meckler:angular} provides the answer to a slightly different version of the query answered by the 
Majorana formula  \cite{emajorana} : given a spin initially quantized along a unit vector 
${\bf \hat{a}}$ with component $m$, what is the probability that it is quantized along a unit vector ${\bf \hat{b}}$ with component 
$m^{\prime}$ at a later time $t$?
Meckler's  formula 
 \cite{meckler:majorana,meckler:angular}  makes use of just two special functions, the well-known Legendre polynomials $ P_{\lambda}(\cos \beta (t))$, and the lesser-known Chebyshev polynomials 
$ f_{\lambda} ^{(j)}(m) =\langle jm| \; f_{\lambda}^{(j)} ( J_z)  \;  |jm \rangle   $ of a discrete variable \cite{bateman,nikiforov2,Nikiforov, vilenkin:specfuncbook}, the diagonal matrix elements of the Chebyshev polynomial  operators
 $f_{\lambda} ^{(j)}(J_z) $. These latter operators and their matrix elements $ f_{\lambda} ^{(j)} (m) $  are the subject of this article. To avoid any confusion from the outset, the Chebyshev polynomials 
 $ f_{\lambda}^{(j)}(m) $ of a discrete variable \cite{bateman,nikiforov2,Nikiforov, vilenkin:specfuncbook} should not be confused with the more commonly known and used Chebyshev polynomials of the first  ($T_n(x)$) and second ($U_n(x)$)  kind \cite{arken, olver, tem:bk}.

Why should  the Meckler version  \cite{meckler:majorana,meckler:angular} of the Majorana formula \cite{emajorana} be of interest? In answering this question,  we should note the following: 

(1) The conciseness of the Meckler version  \cite{meckler:majorana,meckler:angular}
 of the Majorana formula  \cite{emajorana} is evident upon a comparison of Eqs.(\ref{abramajorana}) and (\ref{firstmeck}). The Meckler formula 
 \cite{meckler:majorana,meckler:angular}   takes the form of Fourier-Legendre series, which as we shall see in Section {\bf 4.3.1}, can in certain cases be summed to yield simple, closed-form expressions for ``spin-flip" transition probabilities such as $\mbox{P}^{(j)}_{j,-j} (t) $ or $\mbox{P}^{(j)}_{j-1,-(j-1)} (t) $. 

(2) The Meckler version 
  \cite{meckler:majorana,meckler:angular}  of the Majorana formula \cite{emajorana}  does what Majorana's original version \cite{emajorana} does not:  Meckler's  version 
 \cite{meckler:majorana,meckler:angular}   isolates the dependence on the magnetic projection numbers of the initial and final states in a 
Chebyshev polynomial product term $\left[ f_{\lambda}^{(j)} (m) \; f_{\lambda} ^{(j)}(m^{\prime}) \right]$, leaving the time-dependence  isolated in a Legendre polynomial term $P_{\lambda}(\cos \beta (t))$. In this respect, it is reminiscent of the Clebsch-Gordan coefficient expansion of the rotation matrices ${\cal D}_{m m^{\prime}}^{(j)}(\psi, {\bf \hat{n}}) $ which isolates the dependence on the indices $m$ and $m^{\prime}$ from the time-dependent terms $\psi(t)$ and ${\bf \hat{n}}(t)$  \cite{varshal1:ang, sim:beyond}. 

(3) In his reduction of the Majorana problem  \cite{emajorana}  to the calculation of the joint probability of quantization along two different axes, Meckler \cite{meckler:majorana,meckler:angular} was the first to exploit Chebyshev polynomials of a discrete variable  
\cite{bateman,nikiforov2,Nikiforov, vilenkin:specfuncbook}  in a physics application.   As this article describes, other physics applications of these special functions would follow over the next six decades, including  applications in   
spin physics, spin tomography, and in the development of  operator expansions and equivalents. During this six decade period, it would appear that the 
Chebyshev 
orthonormal basis operators $f_{\lambda}^{(j)} ( {\bf \hat{n}} \cdot {\bf J}) $, first introduced by Meckler \cite{meckler:angular} in 1959, were 
independently rediscovered twice thereafter, by Corio \cite{corio:siam} in 1975, and then by Filippov and Man'ko 
\cite{filippov4} in 2010.

\subsection{Very special special  functions}
Special functions \cite{askey} such as Legendre polynomials, Bessel functions, Chebyshev polynomials and Hermite polynomials are essential tools in mathematical physics \cite{tem:bk,arken,butkov,riley}. As Michiel Hazewinkel  has noted \cite{vilenkin:specfuncbook}, ``Special functions are, well, special." Are some special functions more special than others? In this article we describe some physical applications of some very special, special functions, 
the Chebyshev polynomials of a discrete real variable  \cite{bateman,nikiforov2,Nikiforov, vilenkin:specfuncbook}. These special functions are a member of the family of classical orthogonal polynomials of a discrete variable known as the Hahn polynomials: \cite{bateman,nikiforov2,Nikiforov,vilenkin:specfuncbook} 

Why are  the Hahn polynomials, and the Chebyshev polynomials in particular,  so very special? At least from the point of view of physical applications, it is hard to find any other special function  that has been more obscure. Contributing to this obscurity is the fact that Hahn polynomials have never been discussed in any of the texts commonly  used in undergraduate mathematical physics courses.   Is this obscurity deserved, and might we expect this obscurity to change? Whereas no mention of the Hahn polynomials can be found in  the classic handbook on mathematical functions by Abramowitz and Stegun \cite{abramo}, the Bateman project carried out by Erd\'elyi and colleagues 
\cite{bateman} and  the recently published successor to  Abramowitz and Stegun \cite{abramo}, the NIST Handbook by Olver and colleagues \cite{olver}, both  contain excellent summaries of the properties of classical orthogonal polynomials of a discrete 
variable \cite{bateman,nikiforov2,Nikiforov, vilenkin:specfuncbook}.  Just in the last decade,  in a substantial body of work,  Filippov and Man'ko and coworkers  \cite{ filippov4, filippov2:thesis, fillipov1:qubit, filippov3, manko:spintomo} have shown the promise of Chebyshev polynomials of a discrete variable  for spin tomographic applications,  and this article can be used as an introduction to these applications. 

From the point of view of physics applications, the Chebyshev polynomials of a discrete real variable \cite{bateman,nikiforov2,Nikiforov,vilenkin:specfuncbook} possess some striking properties, and to make that point, 
the sceptical and curious reader might wish to answer the following questions:

1. In a spin-$j$ system, the spin transition probability $\mbox{ P}^{(j)}_{mm^{\prime}}(t) $  can be written as a Fourier-Legendre series whose expansion  coefficents can be expressed in terms of one (and only one) special function. What is that special function?

2. Projection operators, including the coherent state projector, can be written in terms of one (and only one) special function. What is that special function?

3. Ignoring a phase-factor, the Clebsch-Gordan coupling coefficients $ C_{jmj-m}^{\lambda 0}$ of angular momentum theory 
\cite{brinksatch:ang,varshal1:ang} are identical to what special function?

4. In a spin-$j$ system, what special function operator $g^{(j)}\!({\bf \hat{n}} \cdot {\bf J})$  provides a unique orthonormal Hermitian expansion basis for the angle-axis  $(\psi, {\bf \hat{n}})$ parametrized rotation operator $\hat{{\cal D}}^{(j)} \!(\psi, {\bf \hat{n}}) = e^{i \psi({\bf \hat{n}} \cdot {\bf J}) }$ ?

5. The trace of the rotation operator $\hat{{\cal D}}^{(j)} \!(\psi, {\bf \hat{n}}) = e^{i \psi({\bf \hat{n}} \cdot {\bf J}) }$ defines the character 
$ \chi^{(j)}(\psi) = \mbox{Tr} \! \left[ \hat{{\cal D}}^{(j)} \!(\psi, {\bf \hat{n}})   \right]$ \cite{varshal1:ang} of irreducible representations of the rotation group.  The trace of the product of this  rotation operator and a special function operator is proportional to the generalized characters $\chi_{\lambda}^{(j)}(\psi) $ \cite{varshal1:ang} of the rotation group. What is that special function operator?

6. What special function $g^{(j)}(J_z)$ of the operator variable $J_z$ for a spin-$j$ system is identical to the projection-zero spin polarization operators
\cite{varshal1:ang}  $\hat{T}_{\lambda 0}^{(j)} $?

7. What special function operator $g^{(j)}( {\bf \hat{n}} \cdot {\bf J}$) for a spin-$j$ system can be recoupled as a rank-zero irreducible composite tensor defined by the direct product of two rank-$\lambda$ tensors, one the spin tensor $ {\bf T}_{\lambda}({\bf J}) $,  and the other, the spatial Racah tensor 
${\bf C}_{\lambda}({\bf \hat{n}})$?

8. What special function operator $g^{(j)}( {\bf \hat{n}} \cdot{\bf J}$) for a spin-$j$ system has matrix elements expanded in spherical harmonics, with Clebsch-Gordan coefficients as the expansion coefficients?

9. The density operator ${\hat \rho}$ for a spin-$j$ system can be tomographically reconstructed 
\cite{filippov4, filippov2:thesis, fillipov1:qubit, filippov3, manko:spintomo, klimovchumakov, ariano} using one (and only one) special function operator 
$h^{(j)}\! ( {\bf \hat{n}} \cdot {\bf J}$). What is that special function operator $h^{(j)}\! ( {\bf \hat{n}}\cdot {\bf J}$)?

10. For a spin-$j$ system, the spin polarization operators  \cite{varshal1:ang}  $\hat{T}^{(j)}_{\lambda \mu} $ may be viewed as an integral transformation of a unique special function polynomial operator $g^{(j)}( {\bf \hat{n}} \cdot {\bf J}$)  from the continuous variables $(\theta, \phi) $ (which define 
${\bf \hat{n}} \equiv {\bf \hat{n}}(\theta, \phi) $) to the discrete variables $\lambda, \mu$ which define the spin polarization operators \cite{varshal1:ang} 
$\hat{T}^{(j)}_{\lambda \mu} $. What is that special function operator $g^{(j)}( {\bf \hat{n}} \cdot {\bf J}$)?

The surprising answer to all these questions is the same, the special functions known as the Chebyshev polynomials of a discrete real 
variable  \cite{bateman,nikiforov2,Nikiforov,vilenkin:specfuncbook}.

\subsection{Article Organization} This article is organized as follows: we begin in Section {\bf 2} with a brief summary of the properties of the Chebyshev polynomials  $ f^{(j)}  _L(m)   $ of a discrete real variable $m\;\;(m=-j, -j+1, \ldots, j-1, j)$,  and of the 
Chebyshev 
orthonormal basis operators $f_L^{(j)}( {\bf \hat{n}} \cdot {\bf J}) $. Note that  the eigenvalues of $ ({\bf \hat{n}} \cdot {\bf J})$ are discrete real variables, and  that because of ``space quantization" \cite{frenchtaylor}, the projection $({\bf \hat{n}} \cdot {\bf J}) =J_n$ can therefore only take on $(2j+1)$ possible values. Sections {\bf 3} through {\bf 6} discuss several examples of physical applications of Chebyshev polynomials, beginning in  Section {\bf 3} with an 
introduction to projection operators and their use in  the calculation of transition probabilities. Section {\bf 4}  is devoted to a discussion of Meckler's formula \cite{meckler:majorana,meckler:angular}  for the calculation of spin transition probabilites. Section {\bf 5} illustrates how the Chebyshev polynomial operators provide a Hermitian orthonormal basis for expanding the rotation and Stratonovich-Weyl operators
\cite{varilly2:moyal, klimov:distr, heissweigert, klimovchumakov}, and for tomographic reconstruction of the density 
operator  \cite{filippov4, filippov2:thesis, fillipov1:qubit, filippov3, manko:spintomo, klimovchumakov,ariano}.   Section {\bf 6} shows that  the Chebyshev polynomial operators can be recoupled as the product of spin and spatial tensors. Section {\bf 7} discusses how the Chebyshev polynomial operators 
$f_{\lambda} ^{(j)}(J_z)$ can be used to develop operator equivalents for any irreducible tensor operator. Concluding remarks in Section {\bf 8} highlight some of the unique features of Chebyshev polynomials and their applications in magnetic resonance.

\section{Chebyshev Polynomial Properties}
This section is devoted to a brief summary of the most important properties of the Chebyshev polynomials of a discrete
 variable  \cite{bateman,nikiforov2,Nikiforov,vilenkin:specfuncbook}. More detailed discussions can be found in specialized monographs \cite{vilenkin:specfuncbook,Nikiforov,olver,bateman,nikiforov2} and in  the original literature \cite{meckler:majorana,meckler:angular,corio:siam,filippov2:thesis,corio:ortho,werb:tensor,NormRay}. We begin in Section {\bf 2.1} with a discussion of the properties of 
the Chebyshev polynomial scalars $f_{\lambda}^{(j)}(m)=\langle jm| \; f_{\lambda}^{(j)} ( J_z)  \;  |jm \rangle  $, followed in Section {\bf 2.2} with a discussion of the properties of the Chebyshev polynomial
operators $f_{\lambda}^{(j)}( {\bf \hat{n}} \cdot {\bf J}) $. As  matters of notation are concerned, throughout this article we make exclusive use of the notation adopted by Filippov and Man'ko and co-workers \cite{ filippov4, filippov2:thesis, fillipov1:qubit, filippov3, manko:spintomo} for the 
Chebyshev polynomial scalars $ f^{(j)}  _{\lambda}(m)   $  and the Chebyshev polynomial operators   $f_{\lambda}^{(j)}( {\bf \hat{n}} \cdot {\bf J}) $. Table II compares this notation with that originally used by 
Meckler  \cite{meckler:majorana,meckler:angular}, who was the first to introduce Chebyshev polynomials of a discrete variable in physics applications, and that used by Corio \cite{corio:siam,corio:ortho}. 

Aside from the Racah polynomials 
\cite{Nikiforov}, which are equivalent to the Racah angular momentum coupling coefficients \cite{varshal1:ang}  or the 
Wigner 6$j$-symbols \cite{varshal1:ang}, the  Chebyshev polynomials $ f_{\lambda}^{(j)} (m) $ are distinguished as the only special functions which are equivalent to an angular momentum coupling coefficient. The Chebyshev polynomials $ f_{\lambda}^{(j)} (m) $ are, to within a phase-factor, equivalent \cite{meckler:majorana, meckler:angular,NormRay} to 
Clebsch-Gordan coupling coefficients  \cite{varshal1:ang} $C_{jmj-m}^{\lambda \, 0}$
\begin{equation}
f_{\lambda}^{(j)} (m) = (-1)^{j-m} \;  C_{jmj-m}^{\lambda \, 0} \label{chebyclebsch}
\end{equation}
This equivalence is neither obvious nor anticipated.
The first statement of this striking and surprising equivalence between a special function, the Chebyshev polynomials of a discrete variable 
$ f_{\lambda}^{(j)} (m) $  \cite{bateman,nikiforov2,Nikiforov,vilenkin:specfuncbook},  and an angular momentum coupling coefficient, the Clebsch-Gordan coefficient $ C_{jmj-m}^{\lambda \, 0}$,  appears to have made by Meckler \cite{meckler:majorana} in 1958, which he followed a year later 
with a proof \cite{meckler:angular}. Quite independently, in 1958, Gelfand et al. \cite{gelfanda, gelfand} noticed an analogy between Clebsch-Gordan coefficients and the Jacobi polynomial special functions. In retrospect, this analogy is not a surprise given the relationships established since between the Hahn and Jacobi polynomial special functions \cite{Koornwinder}, and the fact the Chebyshev polynomials of a discrete real variable are a special case of the Hahn polynomials \cite{bateman,nikiforov2, Nikiforov, vilenkin:specfuncbook}. As a result of the observations made by Meckler \cite{meckler:majorana, meckler:angular} and  Gelfand et al. \cite{gelfanda, gelfand}, and later work on the connections between Racah polynomials and Wigner 6$j$-symbols  by Askey and Wilson
 \cite{AskWil,Wil},  in effect, Clebsch-Gordan coefficients and Wigner 6$j$-symbols could be recognized in the theory of special functions as discrete analogs of Jacobi polynomials. Meckler's recognition \cite{meckler:majorana} and elegant proof of the Chebyshev polynomial duality \cite{meckler:angular}  was essentially forgotten until Normand and Raynal \cite{NormRay} rediscovered and proved the same equivalence 25 years later, unaware of Meckler's pioneering work \cite{meckler:majorana, meckler:angular}.

\subsection{Chebyshev Polynomial Scalars: $f_{\lambda}^{(j)}(m)$}
\subsubsection{Chebyshev Polynomials defined as Special Function Solutions of a Difference Equation}

The differential equation of the form 
\begin{equation}
\sigma(x) y^{\prime \prime} + \tau(x) y^{\prime} + \lambda y =0 \label{hypo}
\end{equation}
where $\sigma(x)$ is a polynomial of degree 2, $\tau(x)$ is a polynomial of degree 1, and $\lambda$ is a constant is called a hypergeometric type differential equation, whose solutions are called hypergeometric functions. If $y_m(x)$ and $y_n(x)$ are eigensolutions of this equation, with eigenvalues $\lambda_m$ and $\lambda_n$, respectively, then orthogonality of these solutions on the interval $(a,b)$  with respect to a weight function $w(x)$ can be defined as
\cite{leites}
\begin{equation}
\int_a^b y_m(x)\, y_n(x) \,w(x) \, dx =0 \;\;\;\;\;\;\;\; (n \neq m)
\end{equation}

The role of the differentiation operator $d/dx$ in the case of classical orthogonal polynomials is played by $\Delta$ (the forward-difference operator)   and by  $\nabla$ (the backward-difference operator)  in the case of the classical orthogonal polynomials of a discrete variable. These operators are defined as 
\cite{olver}
\begin{eqnarray}
\Delta \! \left[ f(x)\right] & =  & f(x+1)-f(x) \\
\nabla \!  \left[  f(x)\right]  & = & f(x) -f(x-1)
\end{eqnarray}
The difference equation which approximates Eq.(\ref{hypo}) on the uniform lattice is \cite{leites}
\begin{eqnarray}
\Delta [\sigma(x) w(x) \nabla  y] + \lambda w(x) y & = & 0 \label{diffeq} \\
\mbox{where}   \;\;\;\;\; \Delta [\sigma(x) w(x)] & = & \tau(x) w(x) 
\end{eqnarray}
Orthogonality on the uniform lattice is defined as \cite{leites}
\begin{equation}
\sum_{x_i=a}^{b-1} y_m(x_i)\, y_n(x_i) \,w(x_i) =0 \label{ortholattice}
\end{equation}
Hahn polynomials $h_n^{(\alpha,\beta)}(x,N)$, along with Meixner, Krawtchouk, and Charlier polynomials, belong to the classical orthogonal polynomials of a discrete variable \cite{bateman,nikiforov2,Nikiforov,vilenkin:specfuncbook,olver}, or more aptly, to polynomials orthogonal on a discrete set of points \cite{nikiforov2}. A figure illustrating the relationships between these polynomials in the Askey scheme \cite{AskeyWilson} can be found in Olver et al. \cite{olver}. 
The Hahn polynomials $h_n^{(\alpha,\beta)}(x,N)$ are polynomial solutions of the difference equation (\ref{diffeq}) defined by 
\begin{eqnarray}
\alpha,\beta &  >  & -1  \nonumber  \\
\sigma(x) & = & x \, (N+\alpha -x)  \nonumber   \\
w(x) & = & \displaystyle\frac{\Gamma(N+\alpha-x) \;  \Gamma(\beta +1+x)}{\Gamma(x+1) \; \Gamma(N-x)}  \nonumber   \\
(a,b) & = & (0,N)  \nonumber   \\
\tau(x) & = & (\beta +1)(N-1) -(\alpha+ \beta +2) \, x
\end{eqnarray}
A special case of the Hahn polynomials are the Chebyshev polynomials of a discrete variable 
$t_n(x,N) \equiv h_n^{(0,0)}(x,N)$ \cite{bateman,nikiforov2,Nikiforov,vilenkin:specfuncbook,olver}, defined as 
\begin{eqnarray}
\alpha \;\; = \;\; \beta &  = & 0 \nonumber \\
\sigma(x) & = & x(N -x) \nonumber  \\
w(x) & = & 1 \nonumber   \\
(a,b) & = & (0,N) \nonumber  \\
\tau(x) & = & N-1-2x
\end{eqnarray}
In this article, we shall discuss a normalized version of the Chebyshev polynomials  $t_n(x,N)$ defined in terms of $L, j$ and $m \in  [-j,j]$ as
\begin{eqnarray}
 f_L^{(j)} (m) & =  & F(L,j) \;t_L(j+m,2j+1) \label{normdef} \\
\mbox{where}\;\;\;\;  n & = & L \nonumber   \\
x & = & j+m \equiv \langle  jm|\,  j \mathds{1}  + J_z \,  |jm \rangle\;\;\;\;\;\;\;\;(m \in [-j,j], \mbox{so}\;\; x \in [0,N-1]) \nonumber  \\
N & = & 2j+1
\end{eqnarray}
The normalization function $F(L,j)$ and the Bateman project definition  \cite{bateman}  of the Chebyshev polynomials $t_L(j+m,2j+1)$ which we use  in 
Eq.(\ref{normdef})  to define the normalized version of the Chebyshev polynomials $ f_L^{(j)} (m)$ which are the subject of this article,    are defined in Table III.

\subsubsection{Parity Properties}
In common with the Legendre polynomials $P_L(x)$, the parity of the Chebyshev polynomials $ f_L^{(j)}(m)$ is determined by their degree $L$:
\begin{equation}
f_L^{(j)}(-m) = (-1)^L \, f_L^{(j)}(m) \label{parity}
\end{equation}
Symmetry properties of the Clebsch-Gordan coefficients  \cite{varshal1:ang} lead to a simple proof of this parity relation:
\begin{eqnarray}
f_L^{(j)}(-m) & = & (-1)^{-m-j} \,\boxed{ C^{L0}_{j-mjm}} \label{parity1}\\
& = & (-1)^{-m-j} \,  (-1)^{2j} \, \boxed{ (-1)^{-L}}\, C^{L0}_{jmj-m}  \label{parity2} \\
& = & (-1)^{L}\,\boxed{(-1)^{j-m}} \, C^{L0}_{jmj-m}  \label{parity3} \\
& = & (-1)^{L}\, (-1)^{m-j} \, C^{L0}_{jmj-m} \\
& = &  (-1)^L \, f_L^{(j)}(m) 
\end{eqnarray}
The Clebsch-Gordan coefficient in the ``boxed" term of Eq.(\ref{parity1}) has been rewritten in Eq.(\ref{parity2}) using the following symmetry 
property \cite{varshal1:ang}
\begin{equation}
C^{c\gamma}_{a\alpha b\beta} = (-1)^{a+b-c} \; C^{c\gamma}_{b\beta a\alpha } 
\end{equation}
Exploiting the fact that both $L$ and $j-m$ are integers, the ``boxed" terms of Eqs.(\ref{parity2}) and (\ref{parity3}) have been rewritten as
\begin{eqnarray}
 (-1)^{-L} & = & (-1)^{L} \\
(-1)^{j-m} & = & (-1)^{m-j} 
\end{eqnarray}

\subsubsection{Generating the Chebyshev Polynomials $ f_L^{(j)}(m)$}
In order to generate the Chebyshev polynomials, there are four possible approaches, which we now summarize. 
\paragraph{Bateman's definition} Eq.(\ref{normdef}) of Section {\bf 2.1.1} defines the Chebyshev polynomials $ f_L^{(j)}(m)$ in terms of  
Bateman's Chebyshev polynomials   \cite{bateman}  $t_L(j+m,2j+1) $. These polynomials  can be evaluated using the forward-difference operator definition  given in Table III. 

\paragraph{Clebsch-Gordan coefficients} The equivalence of Eq.(\ref{chebyclebsch}) offers the opportunity to prove properties of the Chebyshev polynomials $ f_{\lambda}^{(j)} (m) $ using well-known properties of Clebsch-Gordan coefficients  \cite{varshal1:ang}, an opportunity we will frequently take advantage of. On the other hand, it also provides  a very direct method of generating the Chebyshev polynomials $ f_{\lambda}^{(j)} (m) $  using representations of the Clebsch-Gordan coefficients 
$C^{c\gamma}_{a\alpha b \beta}$  in the form of algebraic sums  \cite{varshal1:ang}.  As an example, the following 
Clebsch-Gordan coefficient representation due to Wigner \cite{varshal1:ang, wignerrep}
\begin{eqnarray}
C^{c\gamma}_{a\alpha b \beta} & = & \delta_{\gamma, \alpha + \beta} \; \Delta(abc) 
\left[ \frac{(c+\gamma)! (c-\gamma)! (2c+1)}{(a+\alpha)!(a-\alpha)!(b+\beta)!(b-\beta)!} \right]^{\!1/2} \nonumber \\
& & \times \sum_z \frac{(-1)^{b+\beta+z} (c+b+ \alpha -z)!(a-\alpha +z)!}{z!(c-a+b-z)!(c+\gamma-z)!(a-b-\gamma +z)!} \label{wigrep1}\\
\mbox{where} \;\;\;\;  \Delta(abc) & = & \left[\frac{(a+b-c)! (a-b+c)! (-a+b+c)!}{(a+b+c+1)!}   \right]^{\!1/2} \label{wigrep2}
\end{eqnarray}
can be used to generate the Chebyshev polynomial $ f_{2}^{(1)} (m) $. In this case, the fixed parameters in Eq.(\ref{wigrep1}) take the values
\begin{eqnarray}
 a & = & b = 1 \\
c & = & 2 \\
\alpha & = & -\beta = m \\
\gamma & = & 0 \\
\Delta(abc) &  =  & 1/\sqrt{30}
\end{eqnarray}
and since the summation index $z$  in Eq.(\ref{wigrep1}) assumes integer values for which all the factorial arguments are non-negative, $z$ can only assume the values 0, 1, and 2.  Using the Clebsch-Gordan coefficient representation of Eq.(\ref{wigrep1}) for $C^{20}_{1m1-m}$, we easily find the following algebraic sum for $f_{2}^{(1)} (m) $
\begin{eqnarray}
f_{2}^{(1)} (m) & = & \frac{(-1)^{2(1-m)}}{2\sqrt{6}} \left[ \frac{(m+3)!(1-m)! -4(m+2)!(2-m)! +(m+1)!(3-m)!}{(m+1)!(1-m)!} \right] \;\;\; \;\;\; \\
& = & \frac{1}{\sqrt{6}} \left[ 3m^2-2 \right]
\end{eqnarray}
in agreement with the polynomial form given in Table I. 

\paragraph{Recursion relation} The right column of Table III states the Chebyshev polynomial recursion relation  which can be used to generate these polynomials. It also compares the Chebyshev polynomial recursion relation \cite{meckler:majorana, meckler:angular, corio:siam,filippov2:thesis, corio:ortho} with the equivalent  Clebsch-Gordan coefficient 
 recursion relation \cite{varshal1:ang}.  

\paragraph{Legendre polynomial operators} As described in Appendix A, the Chebyshev polynomials $  f_{\lambda} ^{(j)} (m) $ can be calculated from the diagonal matrix elements of the Legendre 
polynomial operators $ \overline{P}_{\lambda}({\bf \hat{n}} \cdot {\bf J}) $ \cite{zemach} or  $P_{\lambda}({\bf J})$ \cite{schwinger:majorana} according to the following relations:
\begin{eqnarray}
f_{\lambda} ^{(j)} (m)  & = &\sqrt{ \frac{2\lambda+1}{2j+1}  } \left[  \left[ {\bf J}^2 \right]^{\!l}  \right]^{\!-1/2} \! 
 \langle jm| \, \overline{P}_{\lambda}({\bf \hat{n}} \cdot {\bf J}) \, |jm \rangle \\
 f_{\lambda} ^{(j)} (m)  & = &\sqrt{ \frac{2\lambda+1}{2j+1}  } \; \langle jm|  \, P_{\lambda}({\bf J})\, |jm \rangle 
\end{eqnarray}

\subsection{Chebyshev Polynomial Operators: $f_{\lambda}^{(j)}( {\bf \hat{n}} \cdot {\bf J}) $}

In 1975, Corio  \cite{corio:siam} presented a method for expanding an arbitrary function of a component of angular momentum 
$g({\bf \hat{n}} \cdot {\bf J)}$ in terms of the orthonormal Chebyshev polynomial operators  $f_{\lambda}^{(j)}({\bf \hat{n}} \cdot {\bf J})$. By replacing 
$m$ with $({\bf \hat{n}} \cdot {\bf J})$, Corio \cite{corio:siam} remarked that the recursion relation for the Chebyshev polynomial scalars $f_{\lambda}^{(j)}(m)$ (see Table III), together with an  initial operator $f_{0}^{(j)}({\bf \hat{n}} \cdot {\bf J})$, could be used to compile a table of the 
Chebyshev polynomial operators $f_{\lambda}^{(j)}({\bf \hat{n}} \cdot {\bf J})$. 
I have generated all the elements of Table I using this procedure, starting with 
\begin{equation}
    f_0^{(j)}({\bf \hat{n}} \cdot {\bf J}) = \displaystyle\frac{ \mathds{1} }{\sqrt{2j+1} } 
\end{equation}
All of the operator elements $f_{\lambda}^{(j)}({\bf \hat{n}} \cdot {\bf J})$ in Table I agree with the equivalent operator elements 
 $f_{\lambda}^{(j)}( {\bf \hat{n}} \cdot {\bf J}) $  determined by Filippov and Man'ko \cite{filippov4, filippov2:thesis}. 

Table IV compares the traces, matrix elements, and Hermitian conjugates of the spin polarization operators $\hat{T}_{\lambda \mu}^{(j)} $ with those  of the spin polarization operator expansion \cite{filippov2:thesis} of  the Chebyshev polynomial operators 
$f_{\lambda}^{(j)}({\bf \hat{n}} \cdot {\bf J})= \sum_{\mu=-\lambda}^{\lambda} C_{\lambda \mu}^{\star}({\bf \hat{n}}) \; \,
\hat{T}^{(j)}_{\lambda \mu}$. In the next three sections, we briefly discuss the traces, matrix elements and Hermitian conjugates of the Chebyshev polynomial operators $f_{\lambda}^{(j)}({\bf \hat{n}} \cdot {\bf J})$. Table IV also compares a spin polarization operator $\hat{T}_{\lambda \mu}^{(j)} $ expansion of the density operator $\hat \rho$ with a Chebyshev polynomial operator $f_{\lambda}^{(j)}({\bf \hat{n}} \cdot {\bf J})$ expansion. The latter expansion, an example of  tomographic reconstruction of the density operator $\hat \rho$ 
\cite{filippov4, filippov2:thesis, fillipov1:qubit, filippov3, manko:spintomo, klimovchumakov, ariano}, will be discussed in Section {\bf 5.3}. 

\subsubsection{Traces}
In this section, we state and prove the following traces for the Chebyshev polynomial operators $f^{(j)}_{{\lambda}^{\prime}}({\bf \hat{n}} \cdot {\bf J})$:
 \begin{eqnarray}
\mbox{Tr} \! \left[ f^{(j)}_{\lambda}({\bf \hat{n}} \cdot {\bf J}) \; f^{(j)}_{{\lambda}^{\prime}}({\bf \hat{n}} \cdot {\bf J}) \right]  & = &    
\sum_{\mu=-\lambda}^{\lambda} C_{\lambda \mu}^{\star}({\bf \hat{n}}) \; C_{\lambda \mu}({\bf \hat{n}}) = 1 \label{traceone}\\
\mbox{Tr} \!  \left[ f^{(j)}_{\lambda}({\bf \hat{n}} \cdot {\bf J}) \; f^{(j)} _{{\lambda}^{\prime}}({\bf \hat{n}}^{\prime} \! \cdot {\bf J}) \right] 
& = & \sum_{\mu=-\lambda}^{\lambda} C_{\lambda \mu}^{\star}({\bf \hat{n}}) \; C_{\lambda \mu}({\bf \hat{n}}^{\prime}) 
=  P_{\lambda}({\bf \hat{n}} \cdot {\bf \hat{n}}^{\prime}) \label{tracetwo}
\end{eqnarray}
The trace result in Eq.(\ref{traceone}), the statement of orthonormality \cite{corio:ortho} for the 
Chebyshev polynomial operators $f^{(j)}_{{\lambda}}({\bf \hat{n}} \cdot {\bf J})$,  can be verified as follows
\begin{eqnarray}
 &  & \mbox{Tr} \! \left[ f^{(j)}_{\lambda}({\bf \hat{n}} \cdot {\bf J}) \; f^{(j)}_{{\lambda}^{\prime}}({\bf \hat{n}} \cdot {\bf J}) \right] \nonumber \\
& = &  \mbox{Tr}  \!  \left[\; \sum_{\mu=-\lambda}^{\lambda} C_{\lambda \mu}^{\star}({\bf \hat{n}}) \;  \hat{T}_{\lambda \mu}^{(j)}  \;
 \boxed{ \sum_{\mu^{\prime}=-\lambda^{\prime}}^{\lambda^{\prime}} C_{\lambda^{\prime} \mu^{\prime}}^{\star} ({\bf \hat{n}})\;
 \hat{T}_{\lambda^{\prime} \mu^{\prime}}^{(j)} }\; \right] 
 \label{trac1} \\
& = &  \mbox{Tr}  \!  \left[\; \sum_{\mu=-\lambda}^{\lambda} C_{\lambda \mu}^{\star}({\bf \hat{n}}) \;  \hat{T}_{\lambda \mu}^{(j)}  \;
\boxed{ \sum_{\mu^{\prime}=-\lambda^{\prime}}^{\lambda^{\prime}} C_{\lambda^{\prime} \mu^{\prime}} ({\bf \hat{n}})\;
 \left[\hat{T}_{\lambda^{\prime} \mu^{\prime}}^{(j)} \right]^{\dagger} } \;\right] 
 \label{trac2} \\
& = & \sum_{\mu=-\lambda}^{\lambda} \, \sum_{\mu^{\prime}=-\lambda^{\prime}}^{\lambda^{\prime}} 
C_{\lambda \mu}^{\star}({\bf \hat{n}}) \; C_{\lambda^{\prime} \mu^{\prime}} ({\bf \hat{n}})\; 
  \mbox{Tr} \left[ \, \boxed{ \hat{T}_{\lambda \mu}^{(j)}  \left[\hat{T}_{\lambda^{\prime} \mu^{\prime}}^{(j)} \right]^{\dagger} }  \;  \right]    \label{trac3}    \\
& = & \sum_{\mu=-\lambda}^{\lambda} C_{\lambda \mu}^{\star}({\bf \hat{n}}) \; C_{\lambda \mu}({\bf \hat{n}}) =
\boxed{\sum_{\mu=-\lambda}^{\lambda} | C_{\lambda \mu}({\bf \hat{n}}) |^2 =1} \label{trac4}
\end{eqnarray}
The ``boxed" term of Eq.(\ref{trac1}) has been replaced with the ``boxed" term of Eq.(\ref{trac2})  using the following
 properties  \cite{varshal1:ang,brinksatch:ang} of the spin polarization operators $\hat{T}_{\lambda \mu}^{(j)} $ and the Racah spherical harmonics 
$C_{\lambda \mu}({\bf \hat{n}})$:
\begin{eqnarray}
 \left[\hat{T}_{\lambda \mu}^{(j)} \right]^{\! \dagger} & = & (-1)^{\mu} \; \hat{T}_{\lambda -\mu}^{(j)} \\
C_{\lambda -\mu}({\bf \hat{n}}) & = & (-1)^{\mu} \; C_{\lambda \mu}^{\star}({\bf \hat{n}}) 
\end{eqnarray}
The double summation of Eq.(\ref{trac3}) has been reduced to a single summation in Eq.(\ref{trac4}) using the following normalization identity for the spin polarization operators \cite{varshal1:ang}
\begin{equation}
 \mbox{Tr} \! \left[\; \left[\hat{T}_{\lambda^{\prime} \mu^{\prime}}^{(j)} \right]^{\dagger}  \hat{T}_{\lambda \mu}^{(j)} \right] = 
\delta_{\lambda \lambda^{\prime}}\; \delta_{\mu \mu^{\prime}}
\end{equation}
The final simplification, the ``boxed" term of Eq.(\ref{trac4}), is just the sum rule \cite{brinksatch:ang,varshal1:ang} for the Racah spherical harmonics. 

In verifying the trace result of Eq.(\ref{tracetwo}), nothing would change in the calculation of
 $ \mbox{Tr}\!  \left[ f^{(j)}_{\lambda}({\bf \hat{n}} \cdot {\bf J}) \; f^{(j)}_{{\lambda}^{\prime}}({\bf \hat{n}}^{\prime} \cdot {\bf J})\right]$ except that the final simplication made above in Eq.(\ref{trac4})  would now require  use of the spherical harmonics addition theorem \cite{brinksatch:ang,varshal1:ang} since now
${\bf \hat{n}}  \neq {\bf \hat{n}}^{\prime}$:
\begin{eqnarray}
 \mbox{Tr} \! \left[ f^{(j)}_{\lambda}({\bf \hat{n}} \cdot {\bf J}) \; f^{(j)} _{{\lambda}^{\prime}}({\bf \hat{n}}^{\prime} \cdot {\bf J}) \right] 
& = & \sum_{\mu=-\lambda}^{\lambda} C_{\lambda \mu}^{\star}({\bf \hat{n}}) \; C_{\lambda \mu}({\bf \hat{n}}^{\prime}) \\
& = & P_{\lambda}({\bf \hat{n}} \cdot {\bf \hat{n}}^{\prime}) \label{traceLegendre}
\end{eqnarray}
Meckler \cite{meckler:angular} provided the first proof of this most important trace relation for the Chebyshev polynomial operators $f^{(j)}_{\lambda}({\bf \hat{n}} \cdot {\bf J}) $. As we shall see in Section {\bf 4.1.1}, Meckler then  took advantage of this relation in his projection operator approach
 \cite{meckler:majorana, meckler:angular} to calculate the spin transition probablity $\mbox{P}^{(j)}_{mm^{\prime}}(t)$.  In Sections {\bf 5.2} and 
{\bf 5.3},  we show how this trace relation can be  used to evaluate traces that define delta functions which involve integrations of operators and  spin tomograms   
on the sphere ${\bf S}^2$.

\subsubsection{Matrix Elements and Orthogonality Relations}
\paragraph{Operator Expansions} In order to calculate matrix elements of the Chebyshev polynomial operators $f_{\lambda}^{(j)} ( {\bf \hat{n}} \cdot {\bf J})$ and $f_{\lambda}^{(j)} (J_z) $, we exploit relations between these operators and the spin 
polarization operators \cite{varshal1:ang} $\hat{T}_{\lambda \mu}^{(j)}$. As we discuss in Section {\bf 6.1.2}, the Chebyshev polynomial operators $f_{\lambda}^{(j)} ( {\bf \hat{n}} \cdot {\bf J})$ and $f_{\lambda}^{(j)} (J_z) $   can be expressed in terms of the spin 
polarization operators \cite{varshal1:ang} $\hat{T}_{\lambda \mu}^{(j)}$ and the Racah spherical harmonics $C_{\lambda \mu}(\theta, \phi)$  as follows \cite{filippov2:thesis}
\begin{eqnarray}
 f_{\lambda}^{(j)}({\bf \hat{n}} \cdot {\bf J} )  & =   &
   \sum_{\mu=-\lambda}^{\lambda} C_{\lambda \mu}^{\star}(\theta, \phi) \; \hat{T}_{\lambda \mu}^{(j)} \equiv 
   \sum_{\mu=-\lambda}^{\lambda} C_{\lambda \mu}^{\star}( {\bf \hat{n}}) \; \hat{T}_{\lambda \mu}^{(j)}  \label{jn}\\
 f_{\lambda}^{(j)}( {\bf \hat{z}} \cdot {\bf J}) \equiv  f_{\lambda}^{(j)}(J_z) & = & 
 \sum_{\mu=-\lambda}^{\lambda} \boxed{ C_{\lambda \mu}^{\star}(0,0)} \; \hat{T}_{\lambda \mu}^{(j)} \label{jz1} \\
& = & \sum_{\mu=-\lambda}^{\lambda}  \; \boxed{\delta_{\mu 0}} \; \hat{T}_{\lambda \mu}^{(j)}  \label{jz2} \\
& = &  \hat{T}_{\lambda 0}^{(j)} \label{jz3}
\end{eqnarray}
In Eq.(\ref{jn}), ${\bf \hat{n}} \equiv (\theta, \phi)$ denotes a quantization axis defined by polar angles $(\theta, \phi)$ with respect to the 
 ${\bf \hat{z}} $-axis. The spherical harmonic in the ``boxed" term of Eq.(\ref{jz1}) has been replaced by the Kronecker delta function in the ``boxed" term of Eq.(\ref{jz2} ) using the properties of the spherical harmonics \cite{varshal1:ang}. 

Eq.(\ref{jz3}) shows that the Chebyshev polynomial operators $f_{\lambda}^{(j)} \!(J_z) $ and the projection-zero spin polarization operators \cite{varshal1:ang} $\hat{T}_{\lambda 0}^{(j)}$ are equivalent, a fundamental result which we shall take advantage of in subsequent sections. Although it has been proved in many ways \cite{meckler:angular,filippov2:thesis,corio:ortho,NormRay,werb:tensor,fillipov1:qubit}, it was Meckler \cite{meckler:angular} who first recognized that $f_{\lambda}^{(j)} (J_z) $ was proportional to the projection-zero spin polarization operators \cite{varshal1:ang} $\hat{T}_{\lambda 0}^{(j)}$.

Inverting Eq.(\ref{jn}) to express the spin polarization operators $\hat{T}^{(j)}_{\lambda \mu}$  in terms of the  Chebyshev polynomial operators 
$f_{\lambda}^{(j)} \! \left( {\bf \hat{n}} \cdot {\bf J} \right)$ is easily achieved using  the orthogonality relations \cite{brinksatch:ang} for the Racah spherical harmonics 
\begin{equation}
(2L+1)  \int_0^{\pi} d\theta \sin \theta \int_0^{2\pi} d\phi  \;  C_{LM}^{\star}(\theta, \phi) \; C_{L^{\prime} M^{\prime}}(\theta, \phi)  = 
\delta_{LL^{\prime}} \; \delta_{MM^{\prime}} \, 4 \pi \label{racor}
\end{equation}
with the result that
\begin{eqnarray}
 \hat{T}^{(j)}_{\lambda \mu} &  =  &
 \frac{2\lambda +1 }{4\pi} \int_{{\bf S}^2}  C_{\lambda \mu}({\bf \hat{n}}) \; f_{\lambda}^{(j)} \! \left( {\bf \hat{n}} \cdot {\bf J} \right) \, d{\bf \hat{n}}  \label{spintensorcheby} \\
\mbox{where} \;\;\; d{\bf \hat{n}} & \equiv & d\Omega=\sin \theta \, d\theta \, d\phi 
\end{eqnarray}
This relation may be viewed as an integral transformation of the Chebyshev polynomial operators 
$f_{\lambda}^{(j)} \! \left( {\bf \hat{n}} \cdot {\bf J} \right)$ from the continuous variables $(\theta, \phi) $ (which define 
${\bf \hat{n}} \equiv {\bf \hat{n}}(\theta, \phi) $) to the discrete variables $\lambda, \mu$ which define the spin polarization operators 
$\hat{T}^{(j)}_{\lambda \mu} $. 

Inverting Eq.(\ref{jn}) to express the Racah spherical harmonics $C_{\lambda \mu}({\bf \hat{n}})$  in terms of the  Chebyshev polynomial operators 
$f_{\lambda}^{(j)} \! \left( {\bf \hat{n}} \cdot {\bf J} \right)$ is easily achieved using the trace relation for the spin polarization operators 
$\hat{T}^{(j)}_{\lambda \mu}$ given in Table IV, with the result that 
\begin{eqnarray}
C_{\lambda \mu}({\bf \hat{n}}) & = & \mbox{Tr} \! \left[ \hat{T}^{(j)}_{\lambda \mu} \; f_{\lambda}^{(j)} \! \left( {\bf \hat{n}} \cdot {\bf J} \right)  \right] 
\label{raccheb1} \\
& = &  \mbox{Tr} \! \left[ \; \boxed{ 
\frac{2\lambda +1 }{4\pi} \int_{{\bf S}^2}  C_{\lambda \mu}({\bf \hat{n}}^{\prime}) \; f_{\lambda}^{(j)} \! \left( {\bf \hat{n}}^{\prime} \cdot {\bf J} \right)
 \, d{\bf \hat{n}}^{\prime}  }
\; f_{\lambda}^{(j)} \! \left( {\bf \hat{n}} \cdot {\bf J} \right)  \right] \label{raccheb2} \\
& = &  \int_{{\bf S}^2}  \frac{2\lambda +1 }{4\pi}   \; \mbox{Tr} \! \left[   f_{\lambda}^{(j)} \! \left( {\bf \hat{n}}^{\prime} \cdot {\bf J} \right) \, 
f_{\lambda}^{(j)} \! \left( {\bf \hat{n}} \cdot {\bf J} \right)  \right]  \; C_{\lambda \mu}({\bf \hat{n}}^{\prime}) \; d{\bf \hat{n}}^{\prime} \label{raccheb3}\\
& = &  \int_{{\bf S}^2} \boxed{ \frac{2\lambda +1 }{4\pi}   \; P_{\lambda}({\bf \hat{n}} \cdot {\bf \hat{n}}^{\prime}) } \; C_{\lambda \mu}({\bf \hat{n}}^{\prime}) \; d{\bf \hat{n}}^{\prime} \label{raccheb4} \\
& = &  \int_{{\bf S}^2} \delta_{C}^{(j)}({\bf \hat{n}}, {\bf \hat{n}}^{\prime})  \; C_{\lambda \mu}({\bf \hat{n}}^{\prime}) \; d{\bf \hat{n}}^{\prime} 
\label{raccheb5} \\
\mbox{where} \;\;\;\; \delta_{C}^{(j)}({\bf \hat{n}}, {\bf \hat{n}}^{\prime})   & = &  \frac{2\lambda +1 }{4\pi}   \; P_{\lambda}({\bf \hat{n}} \cdot {\bf \hat{n}}^{\prime}) 
\label{raccheb6}
\end{eqnarray} 
The trace relation of Eq.(\ref{raccheb1}) is the inversion result defining the Racah spherical harmonics $C_{\lambda \mu}({\bf \hat{n}})$  in terms of the  Chebyshev polynomial operators 
$f_{\lambda}^{(j)} \! \left( {\bf \hat{n}} \cdot {\bf J} \right)$. The spin polarization operators $\hat{T}^{(j)}_{\lambda \mu}$ of Eq.(\ref{raccheb1}) have been replaced by the ``boxed" term 
of Eq.(\ref{raccheb2}) using Eq.(\ref{spintensorcheby}). The trace in Eq.(\ref{raccheb3}) has been replaced by the ``boxed" term in Eq.(\ref{raccheb4}) using the Chebyshev polynomial operator trace relation of Eq.(\ref{traceLegendre}).   In this way, Eqs.(\ref{raccheb2}) to (\ref{raccheb5}) which follow the inversion result of Eq.(\ref{raccheb1}) lead to the definition  in Eq.(\ref{raccheb6}) of the reproducing kernel 
$\delta_{C}^{(j)}({\bf \hat{n}}, {\bf \hat{n}}^{\prime})  $, which for the spherical harmonics $C_{\lambda \mu}({\bf \hat{n}})$ acts as a delta function with 
respect to integration over ${\bf S}^2$ as shown in Eq.(\ref{raccheb5}). In Section {\bf 5.2.2}, we revisit reproducing kernels in the context of the Stratonovich-Weyl operators \cite{varilly2:moyal,heissweigert, klimov:distr}
$\Delta^{(j)}({\bf \hat{n}}) $, and the Chebyshev polynomial operators $f_{\lambda}^{(j)} \! \left( {\bf \hat{n}} \cdot {\bf J} \right)$.

\paragraph{Matrix Elements} In order to calculate matrix elements, we often follow convention by employing simulaneous eigenkets $|{\bf \hat{z}} ,m \rangle$  of both ${\bf J}^2$ and $J_z \equiv ({\bf J} \cdot {\bf \hat{z}}) $
\begin{eqnarray}
J_z \;  |{\bf \hat{z}} ,m \rangle  & = & m \,  |{\bf \hat{z}} ,m \rangle \nonumber \\
{\bf J}^2 \; |{\bf \hat{z}} ,m \rangle &  = &  j(j+1)  \, |{\bf \hat{z}} ,m \rangle \label{eigenketsz}
\end{eqnarray}
Following Sakurai's  \cite{sakurai} notation for labeling these eigenkets $|{\bf \hat{z}} ,m \rangle$, we explicitly indicate the quantization direction 
${\bf \hat{z}}$, and include the $J_z$ operator eigenvalue $m$, which for a spin-$j$ system, ranges between $-j$ and $+j$. 
For an arbitrary quantization axis ${\bf \hat{n}}$, the generalized version of Eq.(\ref{eigenketsz}) for simulaneous eigenkets $|{\bf \hat{n}} ,m \rangle$  of both ${\bf J}^2$ and $({\bf J} \cdot {\bf \hat{n}}) \equiv J_n$  would be 
\begin{eqnarray}
({\bf J} \cdot {\bf \hat{n}}) \;  |{\bf \hat{n}} ,m \rangle  & = & m \,  |{\bf \hat{n}} ,m \rangle \nonumber \\
{\bf J}^2 \; |{\bf \hat{n}} ,m \rangle &  = &  j(j+1)  \, |{\bf \hat{n}} ,m \rangle \label{eigenketsn}
\end{eqnarray}
In most contexts, ${\bf \hat{n}} \equiv {\bf \hat{z}}$, and on these occasions, we will use the shorthand notation $|jm \rangle \equiv  |{\bf \hat{z}} ,m \rangle$  to denote simultaneous eigenkets of ${\bf J}^2$ and $J_z$ when calculating matrix elements. 

From the expressions for $f_{\lambda}^{(j)} ( {\bf \hat{n}} \cdot {\bf J})  $ and $ f_{\lambda}^{(j)} (J_z)  $   in Eqs.(\ref{jn}) and (\ref{jz3}), respectively, and the matrix elements of the spin 
polarization operators in the spherical basis representation  \cite{varshal1:ang} 
\begin{equation}
\langle jm| \; \hat{T}_{\lambda \mu}^{(j)}   \;  |jm^{\prime} \rangle = 
 C_{jmj-m^{\prime}}^{\lambda \, \mu} \; (-1)^{j-m^{\prime}} 
\end{equation}
the corresponding matrix elements  are easily evaluated as
\begin{eqnarray}
\langle jm| \; f_{\lambda}^{(j)} ( {\bf \hat{n}} \cdot {\bf J})  \;  |jm^{\prime} \rangle &  =  & 
C_{\lambda \, \mu^{\prime}}^{\star}({\bf \hat{n}}) \; C_{jmj-m^{\prime}}^{\lambda \, \mu^{\prime}} \; (-1)^{j-m^{\prime}} =
C_{\lambda \, (m-m^{\prime})}^{\star}({\bf \hat{n}}) \; C_{jmj-m^{\prime}}^{\lambda \, (m-m^{\prime})} \; (-1)^{j-m^{\prime}} \;\;\;\;  \;\;\;\;  \\
\langle jm| \; f_{\lambda}^{(j)} (J_z)  \;  |jm \rangle &  \equiv   & \langle jm| \; \hat{T}^{(j)}_{\lambda 0} \;  |jm \rangle =
\; C_{jmj-m}^{\lambda \, 0} \; (-1)^{j-m} = f_{\lambda}^{(j)} (m) \label{cgeqcheb}
\end{eqnarray}
A example of how the relation in Eq.(\ref{cgeqcheb}) can be exploited is the evaluation of the  Chebyshev polynomial $ f_L^{(j)} (j) $:
\begin{eqnarray}
 f_L^{(j)} (j) & = & (-1)^{j-j} \; C^{L0}_{jjj-j}  \label{cgeval4}\\
& = & \left[ \frac{(2L+1) \, [(2j)!]^2}{(2j+L+1)! \, (2j-L)!}  \right]^{\!1/2} \label{cgeval44}
\end{eqnarray}
The Clebsch-Gordan coefficient $C^{L0}_{jjj-j} $ in Eq.(\ref{cgeval4}) was evaluated using the relation \cite{varshal1:ang}
\begin{equation}
C^{c\gamma}_{aab\beta}= \delta_{\gamma-\beta,a}
\left[ \frac{(2c+1)(2a)! (-a+b+c)!(b-\beta)!(c+\gamma)!}{(a+b+c+1)!(a-b+c)!(a+b-c)!(b+\beta)!(c-\gamma)!}  \right]^{1/2}
\end{equation}

\paragraph{Orthogonality Relations} The  relation of Eq.(\ref{cgeqcheb}) defines the Chebyshev polynomials of a discrete variable $ f_{\lambda}^{(j)} (m) $ 
\cite{meckler:angular,corio:siam, corio:ortho,filippov2:thesis}. For these polynomials, the definition of orthogonality on the uniform lattice given in 
Eq.(\ref{ortholattice}) takes the form \cite{meckler:angular,corio:siam, corio:ortho,filippov2:thesis}
\begin{equation}
\displaystyle\sum_{m=-j}^{j}  f_{\lambda}^{(j)} (m) \; f_{\lambda^{\prime}}^{(j)} (m)=  \delta_{\lambda \lambda^{\prime}} \label{ortho1}
\end{equation}
an identity first stated by Meckler \cite{meckler:angular}.
Bearing in mind the fact that the Chebyshev polynomials $ f_{\lambda}^{(j)} (m) $ are, to within a phase-factor, equivalent to 
Clebsch-Gordan coupling coefficients as shown in Eq.(\ref{cgeqcheb}), it is not surprising to find that the Chebyshev polynomial orthogonality relations given in Eq.(\ref{ortho1}) are equivalent to the Clebsch-Gordan coefficient unitarity 
relation \cite{varshal1:ang} 
\begin{equation}
\displaystyle\sum_{m=-j}^{j} \, 
C_{jmj-m}^{\lambda 0}  \; C_{jmj-m}^{\lambda^{\prime} 0} =   \delta_{\lambda \lambda^{\prime}} \label{ortho2}
\end{equation}

\subsubsection{Hermiticity}
Not only are Chebyshev polynomial operators $f_{\lambda} ^{(j)}( {\bf \hat{n}} \cdot {\bf J})$  orthonormal, but they are also Hermitian \cite{corio:siam}. Taking advantage of the 
direct product expression for $f_{\lambda}^{(j)} ( {\bf \hat{n}} \cdot {\bf J})$ discussed in  Section {\bf 6.1.2}, it is easy to demonstrate that 
$\left[ f_{\lambda}^{(j)} ( {\bf \hat{n}} \cdot {\bf J})  \right]^{\! \dagger}  =  f_{\lambda}^{(j)} ( {\bf \hat{n}} \cdot {\bf J}) $: 
\begin{eqnarray}
\left[ f_{\lambda}^{(j)} ( {\bf \hat{n}} \cdot {\bf J})  \right]^{\!  \dagger} & = &
 \left[ \; \sum_{\mu=-\lambda}^{\lambda} C_{\lambda \mu}^{\star}({\bf \hat{n}}) \; \hat{T}_{\lambda \mu}^{(j)}   \right]^{\!  \dagger} \label{her1} \\
& = & \sum_{\mu=-\lambda}^{\lambda} C_{\lambda \mu}({\bf \hat{n}}) \;  \boxed{(-1)^{\mu} \, \hat{T}_{\lambda -\mu}^{(j)}  }  \label{her2}  \\
& = &  \sum_{\mu^{\prime}=\lambda}^{-\lambda} C_{\lambda -\mu^{\prime}}({\bf \hat{n}}) \; \boxed{(-1)^{-\mu^{\prime}} } \, \hat{T}_{\lambda \mu^{\prime}}^{(j)} \label{her3}  \\
& = &  \sum_{\mu^{\prime}=\lambda}^{-\lambda} \boxed{ (-1)^{\mu^{\prime}}  \, C_{\lambda -\mu^{\prime}}({\bf \hat{n}}) }  \; \hat{T}_{\lambda \mu^{\prime}}^{(j)} \label{her4}  \\
& = & \sum_{\mu^{\prime}=\lambda}^{-\lambda} C_{\lambda \mu^{\prime}}^{\star}({\bf \hat{n}}) \; \hat{T}_{\lambda \mu^{\prime}}^{(j)} \label{her5}  \\
& = &  f_{\lambda}^{(j)} ( {\bf \hat{n}} \cdot {\bf J}) 
\end{eqnarray}
Using the properties of the spin polarization operators \cite{varshal1:ang}, the ``boxed" term in Eq.(\ref{her2}) replaces the Hermitian conjugate
 $\left[ \hat{T}_{\lambda \mu}^{(j)}\right]^{\! \dagger}$ in Eq.(\ref{her1}). A change in the dummy summation index from $ \mu \rightarrow \mu^{\prime}$ has 
been used to rewrite Eq.(\ref{her2}) as Eq.(\ref{her3}). Using the properties of the spherical harmonics
 \cite{varshal1:ang,brinksatch:ang}, the ``boxed" term of Eq.(\ref{her4}) has been replaced by the ``boxed" term in Eq.(\ref{her5}). Taking advantage of the fact that $\mu$ (or $\mu^{\prime}$) are integral, the ``boxed" term of Eq.(\ref{her3}) can  be replaced by $(-1)^{\mu^{\prime}}$ in Eq.(\ref{her4}). 

Alternatively, and more directly, given any polynomial $g(x)$, the polynomial operator $g(\hat A)$ is Hermitian if $\hat A$ is Hermitian, so 
$f_{\lambda} ^{(j)}( {\bf \hat{n}} \cdot {\bf J})$  is Hermitian since $( {\bf \hat{n}} \cdot {\bf J})$ is Hermitian.

\section{Projection operators and the calculation of transition probabilities}
Fundamental to much of the discussion in this and in subsequent sections are the actions of the rotation operator 
$\hat{{\cal D}}^{(j)} \!(\psi, {\bf \hat{n}}) = e^{-i \psi({\bf \hat{n}} \cdot {\bf J}) }$ as a change of basis operator   or   as a similarity transformation.  These two actions are defined in the next section using the angle-axis $(\psi, {\bf \hat{n}})$ parameterization \cite{Louck,siminovitch:eeht,sim:rot}, of which we make frequent but not exclusive use in this and in subsequent sections. 
\subsection{Actions of $\hat{{\cal D}}^{(j)}(\psi, {\bf \hat{n}})= e^{-i \psi ({\bf \hat{n}} \cdot {\bf J})}$}

\subsubsection{Rotation operator}

The angle-axis representation of the rotation operator $\hat{{\cal D}}^{(j)} \!(\psi, {\bf \hat{n}}) = e^{-i \psi({\bf \hat{n}} \cdot {\bf J}) }$ is parametrized by $(\psi; \Theta, \Phi)$, where $(\Theta, \Phi)$ are the polar angles of the rotation axis ${\bf \hat{n}}$, and $\psi$ is the rotation angle. The action of this unitary rotation operator on a given eigenket $ |jm \rangle$ is given by \cite{biedenharn,Louck}
\begin{equation}
\hat{{\cal D}}^{(j)}(\psi, {\bf \hat{n}})\;  |jm \rangle = e^{-i \psi ({\bf \hat{n}} \cdot {\bf J})} \; |jm \rangle =
 \sum_{m^{\prime}=-j}^j{\cal{D}}_{m^{\prime}m }^{(j)}(\psi, {\bf \hat{n}}) \; |jm^{\prime} \rangle \label{basischange}
\end{equation}
where the elements of the matrix ${\cal{D}}_{m^{\prime}m }^{(j)}(\psi, {\bf \hat{n}}) $ are defined as \cite{biedenharn}
\begin{equation}
{\cal{D}}_{m^{\prime}m }^{(j)}(\psi, {\bf \hat{n}})  = \langle jm^{\prime} |\;  e^{-i \psi ({\bf \hat{n}} \cdot {\bf J})} \;  |jm \rangle
\end{equation}
\subsubsection{Similarity transforms}
The components of the spin polarization operators $ \hat{T}_{\lambda \mu}^{(j)} $, which are irreducible tensor operators of rank $j$,  transform under the similarity action of $\hat{{\cal D}}^{(j)}(\psi, {\bf \hat{n}})$ just as 
the eigenket $ |jm \rangle$ does under the action of $\hat{{\cal D}}^{(j)}(\psi, {\bf \hat{n}})$  \cite{biedenharn}. Therefore 
the similarity transformation of the spin polarization operators $ \hat{T}_{\lambda \mu}^{(j)} $ corresponding to the unitary change of basis given by 
Eq.(\ref{basischange}) is \cite{biedenharn,Louck}
\begin{equation}
\hat{{\cal D}}^{(j)}  \!(R)  \, \hat{T}_{\lambda \mu}^{(j)}   \left [\hat{{\cal D}}^{(j)} \!(R)\right ]^{\!\dagger}  =
 e^{-i \psi ({\bf \hat{n}} \cdot {\bf J}) }\; \hat{T}_{\lambda \mu}^{(j)}  \;  e^{i \psi ({\bf \hat{n}} \cdot {\bf J}) } =
\sum_{\nu = -j}^j {\cal{D}}_{\nu \mu}^{(j)}(\psi, {\bf \hat{n}}) \;   \, \hat{T}_{\lambda \nu}^{(j)} 
\end{equation}

\subsection{Unitary transforms of $J_z$ and of related polynomial operators $g(J_z)$ }

 Since the spherical components {\large $\tau$}$_{1 \mu}$ of the angular momentum ${\bf J}$ define a rank-1 irreducible spherical tensor 
$ \mbox{\boldmath $\mathcal{T}$}\!_{1}$,  then for a rotation $R \equiv R(\theta,{\bf \hat{n}}_{\bot} )$  by an angle $\theta$ about an axis 
${\bf \hat{n}}_{\bot} = (-\sin \phi, \cos \phi, 0) $ defined by polar angles $(\Theta, \Phi) = (\frac{\pi}{2}, \phi+\frac{\pi}{2})$,  the unitary transform of 
$J_z \equiv \mbox{\large $\tau$}_{10}$ using an angle-axis parametrization can be written as \cite{Louck,siminovitch:eeht}
\begin{eqnarray}
J_z^{\prime} & = & \hat{{\cal D}}^{(j)}  \!(R)  \, J_z   \left [\hat{{\cal D}}^{(j)} \!(R)\right ]^{\!\dagger} \label{genunit} \\
& = & \hat{{\cal D}}^{(j)}(\theta, {\bf \hat{n}}_{\bot}) \, J_z   \left [\hat{{\cal D}}^{(j)} (\theta, {\bf \hat{n}}_{\bot}) \right ]^{\!\dagger}  \\
& = & e^{-i \theta ({\bf \hat{n}}_{\bot} \cdot {\bf J}) }\;  \mbox{\large $\tau$}_{10} \;  e^{i \theta ({\bf \hat{n}}_{\bot} \cdot {\bf J}) } \\
& = & \sum_{\nu = -1}^1 {\cal{D}}_{\nu 0}^{(1)}(\theta, {\bf \hat{n}}_{\bot}) \; \mbox{\large $\tau$}_{1 \nu} \label{angleaxistrans}
\end{eqnarray}
Most quantum mechanics and angular momentum textbooks use the conventional Euler angle parametrization of the rotation $R \equiv R(\alpha, \beta, \gamma)$ to define irreducible tensor operators by the unitary transformation of Eq.(\ref{genunit}). But in this case, an angle-axis parametrization of the 
rotation $R \equiv R(\theta,{\bf \hat{n}}_{\bot} )$ offers the most direct path to evaluating $J_z^{\prime}$ using Eq.(\ref{angleaxistrans}). The required 
rotation matrix 
elements ${\cal{D}}_{\nu 0}^{(1)}(\theta, {\bf \hat{n}}_{\bot}) $ are tabulated in Varshalovich et al. \cite{varshal1:ang}, and have the following values
(when $\cos \Theta = 0$ and $\sin \Theta = 1$)
\begin{eqnarray}
{\cal{D}}_{-1 0}^{(1)}(\theta, {\bf \hat{n}}_{\bot}) & = &  -\frac{i}{\sqrt{2}} \sin \theta \; e^{i(\phi + \frac{\pi}{2})} \nonumber \\
{\cal{D}}_{0 0}^{(1)}(\theta, {\bf \hat{n}}_{\bot}) & = & \cos \theta \nonumber \\
{\cal{D}}_{1 0}^{(1)}(\theta, {\bf \hat{n}}_{\bot}) & = & -\frac{i}{\sqrt{2}} \sin \theta \;  e^{-i(\phi + \frac{\pi}{2})} \label{dd1}
\end{eqnarray}
while the spherical components {\large $\tau$}$_{1 \mu}$ required are given by \cite{brinksatch:ang}
\begin{eqnarray}
\mbox{\large $\tau$}_{1 -1} & = & \frac{(J_x-iJ_y)}{\sqrt{2}}  \nonumber \\
\mbox{\large $\tau$}_{1 0} & = &  J_z \nonumber \\
\mbox{\large $\tau$}_{1 +1} & = &  - \frac{(J_x+iJ_y)}{\sqrt{2}}  \label{spherj} 
\end{eqnarray}
Using Eqs.(\ref{dd1}) and (\ref{spherj}), the unitary transform of Eq.(\ref{angleaxistrans}) is then evaluated as
\begin{eqnarray}
J_z^{\prime} & = & \hat{{\cal D}}^{(j)}(\theta, {\bf \hat{n}}_{\bot}) \, J_z   \left [\hat{{\cal D}}^{(j)} (\theta, {\bf \hat{n}}_{\bot}) \right ]^{\!\dagger}  \\
& = &  \sum_{\nu = -1}^1 {\cal{D}}_{\nu 0}^{(1)}(\theta, {\bf \hat{n}}_{\bot}) \; \mbox{\large $\tau$}_{1 \nu} \nonumber \\
 & = &  (\cos \phi \, \sin \theta ) J_x + (\sin \phi \,  \sin \theta)J_y + (\cos \theta) J_z \nonumber \\
& = & \left( {\bf \hat{n}} \cdot {\bf J}  \right) \label{simtrans}
\end{eqnarray}
where ${\bf \hat{n}} = (\cos \phi \,  \sin \theta, \sin \phi \, \sin \theta,\cos \theta)$ is a unit vector defined by polar angles $(\theta,\phi)$. The unit vector 
${\bf \hat{n}}_{\bot} =(-\sin \phi, \cos \phi, 0)$ defines a rotation axis perpendicular to the plane defined by
${\bf \hat{z}}$ and ${\bf \hat{n}}$, so that a rotation about this axis by the angle $\theta$ will transform the 
${\bf \hat{z}}$ vector into the  ${\bf \hat{n}}$ vector, just as the similarity transform of Eq.(\ref{simtrans}) transforms 
$\left( {\bf \hat{z}} \cdot {\bf J}  \right) \equiv J_z$ into $\left( {\bf \hat{n}} \cdot {\bf J}  \right)$. 

Since
\begin{equation}
({\bf \hat{n}} \cdot {\bf J})  =  \hat{{\cal D}}^{(j)}(\theta, {\bf \hat{n}}_{\bot}) \, J_z   \left [\hat{{\cal D}}^{(j)} (\theta, {\bf \hat{n}}_{\bot}) \right ]^{\!\dagger} 
 \label{JnJz}
\end{equation}
then for {\it any} polynomial function $g \equiv g(J_z)$, 
\begin{equation}
g[\left( {\bf \hat{n}} \cdot {\bf J}\right)] =  
 \hat{{\cal D}}^{(j)}(\theta, {\bf \hat{n}}_{\bot}) \, g(J_z)   \left [\hat{{\cal D}}^{(j)} (\theta, {\bf \hat{n}}_{\bot}) \right ]^{\!\dagger}    \label{polytransform}
\end{equation}
and in particular, the Chebyshev polynomial operator basis functions $f_{\lambda}^{(j)}({\bf \hat{n}} \cdot {\bf J})$ are given by \cite{filippov2:thesis}
\begin{eqnarray}
f_{\lambda}^{(j)}( {\bf \hat{n}} \cdot {\bf J}) \
&  =  &   \hat{{\cal D}}^{(j)}(\theta, {\bf \hat{n}}_{\bot}) \,  f_{\lambda}^{(j)}(J_z)    \left [\hat{{\cal D}}^{(j)} (\theta, {\bf \hat{n}}_{\bot}) \right ]^{\!\dagger} 
\label{chebytranss} \\
& = & \hat{{\cal D}}^{(j)}(\theta, {\bf \hat{n}}_{\bot}) \, \hat{T}_{\lambda 0}^{(j)}  \;   \left [\hat{{\cal D}}^{(j)} (\theta, {\bf \hat{n}}_{\bot}) \right ]^{\!\dagger}
\label{basistransform}
\end{eqnarray}
The results of Eqs.(\ref{JnJz}) and (\ref{chebytranss}) are summarized in Table V. Whereas Eq.(\ref{chebytranss}) is just a particular example of the  relation of Eq.(\ref{polytransform}) in the case 
of Chebyshev polynomials,  
Eq.(\ref{basistransform}) is a significant new relation because it exploits  the following  equivalence \cite{meckler:angular,filippov2:thesis,corio:ortho,NormRay,werb:tensor,fillipov1:qubit} between the Chebyshev polynomials $f_{\lambda}^{(j)}(J_z)$ and 
the projection-zero spin polarization operators $\hat{T}_{\lambda 0}^{(j)} $:
\begin{equation}
  f_{\lambda}^{(j)}(J_z) \equiv \hat{T}_{\lambda 0}^{(j)} \label{equivfT}
\end{equation}
In Section {\bf 6.2}, we will take advantage of this equivalence, and the similarity transformation  of the spin polarization operators 
$ \hat{T}_{\lambda \mu}^{(j)} $ as defined in Section {\bf 3.1.2}, to express the Chebyshev polynomial operators $f_{\lambda}^{(j)}( {\bf \hat{n}} \cdot {\bf J}) $
 as a direct product of spin and spatial tensors. 

The simplest way to verify the equivalence of Eq.(\ref{equivfT}) is to use the definition of Fano's state-multipole operators \cite{FanoOp,blum}, alias polarization operators \cite{varshal1:ang} or spherical coherence vectors \cite{oreg}
\begin{eqnarray}
\hat{T}_{\lambda \mu}^{(j)} & = & \sqrt{\frac{2\lambda+1}{2j+1}} \sum_{m,m^{\prime}} C^{jm^{\prime}}_{jm\lambda \mu} \; |jm^{\prime}\rangle \langle jm| 
\label{fanoop1} \\
& = & \sum_{m,m^{\prime}} (-1)^{j-m} \; C^{\lambda \mu}_{jm^{\prime}j-m} \; |jm^{\prime}\rangle \langle jm|
\end{eqnarray}
From this definition, we then obtain
\begin{eqnarray}
\hat{T}_{\lambda 0}^{(j)} & = & \sum_{m=-j}^{j} \boxed{(-1)^{j-m} \; C^{\lambda 0}_{jmj-m}} \; |jm\rangle \langle jm| \label{boxedTL0} \\
 & = &  \sum_{m=-j}^{j} f_{\lambda}^{(j)}(m) \; \mbox{{\boldmath $\Pi$}}^{(j)}(m,{\bf \hat{z}}) = f_{\lambda}^{(j)}(J_z) \label{TtoCheb} 
\end{eqnarray}
where $\mbox{{\boldmath $\Pi$}}^{(j)}(m,{\bf \hat{z}}) =  |jm\rangle \langle jm| $ is an example of a projection operator which we discuss in the next section. 
Taking advantage of the duality of the Chebyshev polynomials $ f_{\lambda}^{(j)}(m) $, which double as the Clebsch-Gordan angular momentum coupling coefficients (3$j$-symbols \cite{brinksatch:ang}) according to
\begin{equation}
 f_{\lambda}^{(j)}(m)=  (-1)^{j-m} \; C^{\lambda 0}_{jmj-m} 
\end{equation} 
the ``boxed" term of Eq.(\ref{boxedTL0}) has been replaced with the Chebyshev polynomial $ f_{\lambda}^{(j)}(m) $  in Eq.(\ref{TtoCheb}), which is just a 
statement of Sylvester's formula \cite{merzbacher2,horn}.

\subsection{Projection Operators}

\subsubsection{Chebyshev polynomial operator expansions for projection operators from Sylvester's formula}
Merzbacher \cite{merzbacher2} has discussed the use of matrix methods in quantum mechanics, with a particular emphasis on the use of Sylvester's formula \cite{horn} of Eq.(\ref{TtoCheb}),  and the associated projection operator matrices of Eq.(\ref{TtoCheb}). 
 
In general, $|{\bf \hat{n}},m \rangle$ are eigenstates of  $({\bf J} \cdot {\bf \hat{n}}) \equiv J_n$ with eigenvalue $m$, and in particular, 
 $|{\bf \hat{z}},m \rangle \equiv |jm\rangle $ are eigenstates of  $({\bf J} \cdot {\bf \hat{z}}) \equiv J_z$ with eigenvalue $m$. Then since
\begin{eqnarray}
|{\bf \hat{n}},m \rangle & = & \hat{{\cal D}} \, |{\bf \hat{z}},m \rangle \\
\langle {\bf \hat{n}},m | & = & \langle {\bf \hat{z}},m | \,  \hat{{\cal D}}^{\dagger} \\
\mbox{where}\;\;\;\; \hat{{\cal D}} & \equiv & \hat{{\cal D}}(R) =      \hat{{\cal D}}(\theta, {\bf \hat{n}}_{\bot}) = e^{-i \theta ({\bf \hat{n}}_{\bot} \cdot {\bf J}) }
\end{eqnarray}
the unitary transformation of projection operators can be expressed as
\begin{equation}
\hat{{\cal D}} \left[ \mbox{{\boldmath $\Pi$}}^{(j)}(m,{\bf \hat{z}})  \right]  \hat{{\cal D}}^{\dagger}  =
\hat{{\cal D}} \left[ \, |{\bf \hat{z}},m \rangle \langle {\bf \hat{z}},m | \, \right]  \hat{{\cal D}}^{\dagger}    \equiv 
\hat{{\cal D}}  |jm \rangle \langle jm | \hat{{\cal D}}^{\dagger} = |{\bf \hat{n}},m \rangle \langle {\bf \hat{n}},m | = \mbox{{\boldmath $\Pi$}}^{(j)}(m,{\bf \hat{n}})
\label{transprojector}
\end{equation}
Suppose we consider the unitary transformation of Sylvester's formula \cite{horn} for the Chebyshev polynomial operator 
$ f_{\lambda}^{(j)}(J_z) $:
\begin{equation}
 f_{\lambda}^{(j)}(J_z)  =  \sum_{m=-j}^{j} f_{\lambda}^{(j)}(m) \; \mbox{{\boldmath $\Pi$}}^{(j)}(m,{\bf \hat{z}}) \label{sylvest}
\end{equation}
Using Eq.(\ref{JnJz}), the transform of the left-hand side of Eq.(\ref{sylvest}) is given by 
\begin{eqnarray}
\hat{{\cal D}} \left[  f_{\lambda}^{(j)}(J_z) \right]  \hat{{\cal D}}^{\dagger}  & = & f_{\lambda}^{(j)}(\hat{{\cal D}} \, J_z  \, \hat{{\cal D}}^{\dagger}  ) \\
& = & f_{\lambda}^{(j)} \left( {\bf \hat{n}} \cdot {\bf J}\right) \label{lhsunit}
\end{eqnarray}
whereas using Eq.(\ref{transprojector}), the transform of the right-hand side of Eq.(\ref{sylvest}) is   given by 
\begin{eqnarray}
\hat{{\cal D}} \! \left[ \sum_{m=-j}^{j} f_{\lambda}^{(j)}(m) \; \mbox{{\boldmath $\Pi$}}^{(j)}(m,{\bf \hat{z}})  \right] \! \hat{{\cal D}}^{\dagger}   & = & 
 \sum_{m=-j}^{j} f_{\lambda}^{(j)}(m) \;  
\hat{{\cal D}} \!  \left[ \; \mbox{{\boldmath $\Pi$}}^{(j)}(m,{\bf \hat{z}}) \; \right] \! \hat{{\cal D}}^{\dagger}  \\
 & = &  \sum_{m=-j}^{j} f_{\lambda}^{(j)}(m) \; \mbox{{\boldmath $\Pi$}}^{(j)}(m,{\bf \hat{n}})
\end{eqnarray}
In this manner, we obtain as expected the equivalent of Eq.(\ref{sylvest}), namely Sylvester's formula \cite{merzbacher2,horn}  for the Chebyshev polynomial operator 
$ f_{\lambda}^{(j)}( {\bf \hat{n}} \cdot {\bf J} )$:
\begin{equation}
 f_{\lambda}^{(j)}( {\bf \hat{n}} \cdot {\bf J} ) =  \sum_{m=-j}^{j} f_{\lambda}^{(j)}(m) \; \mbox{{\boldmath $\Pi$}}^{(j)}(m,{\bf \hat{n}})
\label{sylvestjn}
\end{equation}
By exploiting  the Chebyshev polynomial orthogonality relation of Eq.(\ref{ortho1}), 
both Eqs.(\ref{sylvest}) and (\ref{sylvestjn}) can be inverted to develop the following Chebyshev polynomial operator expansions for the projection operators:
\begin{eqnarray}
\mbox{{\boldmath $\Pi$}}^{(j)}(m,{\bf \hat{z}}) & =  & \sum_{\lambda=0}^{2j} f_{\lambda}^{(j)}(m) \;  f_{\lambda}^{(j)}(J_z) \equiv
  \sum_{\lambda=0}^{2j} f_{\lambda}^{(j)}(m) \; \hat{T}_{\lambda 0}^{(j)} \label{projectjz}\\
\mbox{{\boldmath $\Pi$}}^{(j)}(m,{\bf \hat{n}})  & =  & \sum_{\lambda=0}^{2j} f_{\lambda}^{(j)}(m) \; f_{\lambda}^{(j)}( {\bf \hat{n}} \cdot {\bf J}) \equiv
   \sum_{\lambda=0}^{2j}\sum_{\mu=-\lambda}^{\lambda} f_{\lambda}^{(j)}(m) \; C_{\lambda \mu}^{\star}({\bf \hat{n}}) \;  \hat{T}_{\lambda \mu}^{(j)}   \;   \;\;\;\; \;\;\;\;\;\;\;\;
\label{projectJn}
\end{eqnarray}
The  novelty (and utility) of these expansions lies in the fact that all the projectors can be expressed in terms of only one special function, namely the Chebyshev polynomials, where the  scalars $f_{\lambda}^{(j)}(m)$ are the expansion coefficients and the operators $ f_{\lambda}^{(j)}(J_z)$  or $f_{\lambda}^{(j)}( {\bf \hat{n}} \cdot {\bf J} )  $ are the expansion basis. For completeness, in Eqs.(\ref{projectjz}) and 
(\ref{projectJn}) we also provide the equivalent spin polarization operator expansions. 

It is easy to verify that the 
$\mbox{{\boldmath $\Pi$}}^{(j)}(m,{\bf \hat{n}}) $ operators of Eq.(\ref{projectJn}) are actually projection operators, using  the following identity for Clebsch-Gordan coefficients 
\cite{varshal1:ang} 
\begin{equation}
\sum_{\alpha=-a}^a (-1)^{a-\alpha} \; C_{a\alpha a -\alpha}^{c0} = \sqrt{2a+1}\; \delta_{c0} \label{sumident}
\end{equation}
This relation follows from the orthogonality relation for the Clebsch-Gordan coefficients \cite{varshal1:ang}
\begin{equation}
\sum_{m_1 m_2} C_{j_1m_1j_2m_2}^{jm} \; C_{j_1m_1j_2m_2}^{j^{\prime}m^{\prime}} = \delta_{jj^{\prime}} \, \delta_{mm^{\prime}} \label{orthoga}
\end{equation}
A particular case of the orthogonality relation of Eq.(\ref{orthoga}) is the following relation
\begin{equation}
\sum_m C_{jmj-m}^{0 0} \; C_{jmj-m}^{\lambda 0} = \delta_{\lambda 0} \label{speciala}
\end{equation}
Since the first Clebsch-Gordan coefficient in Eq.(\ref{speciala}) is an example of the following special case \cite{varshal1:ang, brinksatch:ang} 
\begin{equation}
C_{a\alpha b \beta}^{00} = (-1)^{a-\alpha} \; \frac{\delta_{ab}\, \delta_{\alpha -\beta}}{\sqrt{2a+1}}
\end{equation}
it can be evaluated as 
\begin{equation}
C_{jmj-m}^{0 0}  = \frac{(-1)^{j-m}}{ \sqrt{2j+1}}
\end{equation}
Substitution of this value in Eq.(\ref{speciala}) leads to the sum identity of Eq.(\ref{sumident}). Then, taking advantage again of the Chebyshev polynomial duality, the following sum over all Chebyshev polynomials can be written as
\begin{eqnarray}
\sum_{m=-j}^j  f_{\lambda}^{(j)}(m) & = &  \sum_{m=-j}^j (-1)^{j-m} \; C_{jmj-m}^{\lambda 0} \nonumber \\
& = & \sqrt{2j+1}\; \delta_{\lambda 0} \label{chebsum}
\end{eqnarray}
If {\boldmath $\Pi$}$^{(j)}(m,{\bf \hat{n}})$ are indeed projection operators, then their sum should be the unit operator, which is easily verified using properties of the Chebyshev polynomials as follows: 
\begin{eqnarray}
\sum_{m=-j}^{j} \mbox{{\boldmath $\Pi$}}^{(j)}(m,{\bf \hat{n}}) & = & \sum_{m=-j}^{j} \left[ \sum_{\lambda=0}^{2j} f_{\lambda} ^{(j)}(m) \; 
 f_{\lambda} ^{(j)}({\bf \hat{n}} \cdot {\bf J}) \right] \\
& = & \sum_{\lambda=0}^{2j} \boxed{ \sum_{m=-j}^{j} f_{\lambda}^{(j)} (m) }\;  f_{\lambda}^{(j)} ({\bf \hat{n}} \cdot {\bf J}) \label{boxcheb} \\
& = & \sum_{\lambda=0}^{2j}  \boxed{ \sqrt{2j+1} \;  \delta_{\lambda 0} }\; f_{\lambda} ^{(j)} ({\bf \hat{n}} \cdot {\bf J}) \label{boxchebres} \\
& = & \sqrt{2j+1} \;   \; f_0 ^{(j)} ({\bf \hat{n}} \cdot {\bf J}) = \sqrt{2j+1} \; \left[  \frac{\mathds{1}}{\sqrt{2j+1}} \right] = \mathds{1}
\end{eqnarray}
The ``boxed" term in Eq.(\ref{boxcheb}) has been replaced by the ``boxed" term in Eq.(\ref{boxchebres}) using the result of Eq.(\ref{chebsum}).

\subsubsection{Coherent state projectors} In the case of the highest magnetic projection number $m=j$, Eq.(\ref{projectJn}) yields a novel, compact expression for the coherent state projector 
$ |{\bf \hat{n}},j \rangle \langle {\bf \hat{n}},j | $:
\begin{equation}
 |{\bf \hat{n}},j \rangle \langle {\bf \hat{n}},j | = \mbox{{\boldmath $\Pi$}}^{(j)}(j,{\bf \hat{n}})   =  
 \sum_{\lambda=0}^{2j} f_{\lambda}^{(j)}(j) \; f_{\lambda}^{(j)}( {\bf \hat{n}} \cdot {\bf J} ) \label{cohereproj}
\end{equation}
The novelty of this expression resides in an operator expansion whose expansion coefficients and expansion operator basis are based exclusively on Chebyshev polynomials of a discrete variable. 
Spin coherent states can be viewed as a particular state of a spin system that most closely resembles a classical spin \cite{lohkim}. In this way, the spin eigenstate with maximal $z$-angular momentum is $|{\bf \hat{z}},j \rangle $ is associated with a classical system whose angular momentum points in the 
${\bf \hat{z}}$ direction \cite{ducloy}.  By the same token, the spin state $|{\bf \hat{n}},j \rangle$, associated with a classical system whose angular momentum points in the 
${\bf \hat{n}}  \equiv (\theta, \phi)$ direction,  can be obtained by rotating  $|{\bf \hat{z}},j \rangle $ by an angle $\theta$ about the $y$-axis followed by an angle $\phi$ about the 
$z$-axis \cite{ducloy, lohkim}:
\begin{equation}
|{\bf \hat{n}},j \rangle = e^{-i \phi J_z} \, e^{-i \theta J_y} \, |{\bf \hat{z}},j \rangle
\end{equation}
Exploiting the Chebyshev polynomial operator expansion of the coherent state projector in Eq.(\ref{cohereproj}) leads to a  very simple proof of the completeness or closure relation for the spin coherent states $|{\bf \hat{n}},j \rangle$ 
\begin{eqnarray}
\frac{2j+1}{4\pi} \int_{{\bf S}^2} |{\bf \hat{n}},j \rangle \langle {\bf \hat{n}},j | \,  d{\bf \hat{n}} & = & 
 \frac{2j+1}{4\pi} \int_{{\bf S}^2}   |{\bf \hat{n}},j \rangle \langle {\bf \hat{n}},j | \,   d{\bf \hat{n}} \\
 & = & \frac{2j+1}{4\pi} \int_0^{\pi} d\theta \sin \theta \int_0^{2\pi} d\phi  \;  |{\bf \hat{n}},j \rangle \langle {\bf \hat{n}},j | \\ 
& = &\mathds{1} \label{cohereclose}  \\
\mbox{where} \;\;\; d{\bf \hat{n}} & \equiv & d\Omega=\sin \theta \, d\theta \, d\phi 
\end{eqnarray}
The  relation of Eq.(\ref{cohereclose})  also provides a resolution of the identity operator within the spin-$j$ Hilbert space \cite{lohkim}. 

Because
\begin{equation}
 f_{\lambda}^{(j)}( {\bf \hat{n}} \cdot {\bf J}  )  =   
   \sum_{\mu=-\lambda}^{\lambda} C_{\lambda \mu}^{\star}(\theta, \phi) \; \hat{T}_{\lambda \mu}^{(j)} \label{mea2} 
\end{equation}
the corresponding solid angle integral relation is given by
\begin{eqnarray}
\int_{{\bf S}^2}       f_{\lambda}^{(j)}( {\bf \hat{n}} \cdot {\bf J} ) \; d{\bf \hat{n}}  &  =  &  
   \sum_{\mu=-\lambda}^{\lambda} \left[  \int   \! \!  C_{\lambda \mu}^{\star}(\theta, \phi)\,  d\Omega \right]  \hat{T}_{\lambda \mu}^{(j)}
\end{eqnarray}

Then, making use of the Chebyshev polynomial operator expansion of the coherent state projector given in Eq.(\ref{cohereproj}), we can obtain the closure relation of Eq.(\ref{cohereclose})  as follows
\begin{eqnarray}
\frac{2j+1}{4\pi} \int_{{\bf S}^2} \;  |{\bf \hat{n}},j \rangle \langle {\bf \hat{n}},j | \,  d{\bf \hat{n}}  & = & 
     \frac{2j+1}{4\pi} \int_{{\bf S}^2}  \mbox{{\boldmath $\Pi$}}^{(j)}(j,{\bf \hat{n}})  \, d{\bf \hat{n}} \\
 & = &  \frac{2j+1}{4\pi} \;  \sum_{L=0}^{2j} f_{L}^{(j)}(j) \;  \boxed{\int_{{\bf S}^2}    f_{L}^{(j)}( {\bf \hat{n}} \cdot {\bf J}  ) \, d{\bf \hat{n}} } \label{cohderiv1} \\
& = & \frac{2j+1}{4\pi} \;    f_{0}^{(j)}(j) \,  \int_{{\bf S}^2}    f_{0} ^{(j)}( {\bf \hat{n}} \cdot {\bf J}  ) \, d{\bf \hat{n}}  \label{cohderiv2} \\
& = &\mathds{1} 
\end{eqnarray}
In the first step of this derivation, the summation over $L$ in Eq.(\ref{cohderiv1} ) can be restricted to $L=0$ because the integrals of the Chebyshev polynomial operators 
$f_{L} ^{(j)}( {\bf \hat{n}} \cdot {\bf J} )$ over all solid angles in the ``boxed" term are given by
\begin{eqnarray}
\int_{{\bf S}^2}    f_{L}^{(j)}( {\bf \hat{n}} \cdot {\bf J}  ) \, d{\bf \hat{n}}  & = & 
 \sum_{M=-L}^{L} \boxed{ \int_0^{\pi} d\theta \sin \theta \int_0^{2\pi} d\phi  \; C_{LM}^{\star}(\theta, \phi) } \; \hat{T}_{LM}^{(j)}  \label{solidangle}
\end{eqnarray}
Then using the orthogonality relations \cite{brinksatch:ang} for the Racah spherical harmonics in Eq.(\ref{racor}), and the fact that \cite{brinksatch:ang} 
\begin{equation}
C_{00}(\theta, \phi) = 1
\end{equation}
the integral of the Racah spherical harmonics function $ C_{LM}^{\star}(\theta, \phi)$ over all solid angles in the ``boxed" term of Eq.(\ref{solidangle}) is easily evaluated as 
\begin{equation}
\int C_{LM}^{\star}(\theta, \phi) \;  d\Omega= \int C_{LM}^{\star}(\theta, \phi) \; C_{00}(\theta, \phi) \;  d\Omega = 4\pi \, \delta _{L0}
\end{equation}
leading to the simplification of the ``boxed" term in Eq.(\ref{cohderiv1}))
\begin{equation}
\int_{{\bf S}^2}    f_{L}^{(j)}( {\bf J} \cdot {\bf \hat{n}} ) \, d{\bf \hat{n}} = \int_{{\bf S}^2}    f_{0}^{(j)}( {\bf J} \cdot {\bf \hat{n}} ) \, d{\bf \hat{n}} 
\end{equation}

In the second step, Eq.(\ref{cohderiv2}) can be simplified using the following properties of the Chebyshev polynomials $f_0^{(j)}  (j)  $ and Chebyshev polynomial operators $f_0^{(j)} ({\bf \hat{n}} \cdot {\bf J}) $
\begin{eqnarray}
f_0 ^{(j)} (j) & = &  \frac{1}{\sqrt{2j+1}}  \\
f_0 ^{(j)} ({\bf \hat{n}} \cdot {\bf J}) & = &   \frac{\mathds{1}}{\sqrt{2j+1}}  
\end{eqnarray} 

Using Eq.(\ref{cohereproj}), the representation of the spin polarization operators 
$ \hat{T}^{(j)}_{\lambda \mu} $ in Eq.(\ref{spintensorcheby}) as the following decomposition on the Chebyshev polynomial operators 
$f_{\lambda}^{(j)} \! \left( {\bf \hat{n}} \cdot {\bf J} \right)$ 
\begin{equation}
 \hat{T}^{(j)}_{\lambda \mu}   =  
 \frac{2\lambda +1 }{4\pi} \int_{{\bf S}^2}  C_{\lambda \mu}({\bf \hat{n}}) \; f_{\lambda}^{(j)} \! \left( {\bf \hat{n}} \cdot {\bf J} \right) \, d{\bf \hat{n}} 
\end{equation}
may easily be reexpressed as the following decomposition 
\cite{agarwal,klimovchumakov}  on the coherent state projectors 
$ |{\bf \hat{n}},j \rangle \langle {\bf \hat{n}},j | $ 
\begin{eqnarray}
 \hat{T}^{(j)}_{\lambda \mu} & = & \sqrt{\frac{2j +1 }{4\pi}} \left[C^{jj}_{jj\lambda 0} \right]^{-1} \int_{{\bf S}^2}  Y_{\lambda \mu}({\bf \hat{n}}) \;
 |{\bf \hat{n}},j \rangle \langle {\bf \hat{n}},j |  \, d{\bf \hat{n}}  \label{decomcohere} \\
 & = & \frac{2\lambda +1 }{4\pi} \left[ f^{(j)}_{\lambda}(j)\right]^{-1} \int_{{\bf S}^2}  C_{\lambda \mu}({\bf \hat{n}}) \;
 |{\bf \hat{n}},j \rangle \langle {\bf \hat{n}},j |  \, d{\bf \hat{n}}  \label{decomcohere2} 
\end{eqnarray}
In Eq.(\ref{decomcohere2}), we have rewritten the decomposition \cite{agarwal,klimovchumakov}  of  Eq.(\ref{decomcohere}) by using 
Eq.(\ref{cgeval4}) to replace the 
Clebsch-Gordan coefficient $C^{jj}_{jj\lambda 0}$ in Eq.(\ref{decomcohere})  with the Chebyshev polynomial $ f^{(j)}_{\lambda}(j)$ in 
Eq.(\ref{decomcohere2}).

\subsection{Using Projection Operators to Calculate Transition Probabilities}
A change of basis is well covered in quantum mechanics texts \cite{frenchtaylor,gottfried,merzbacher}, but rarely so in the context of projection operators, and so in this subsection, we begin by providing a brief summary of the quantum mechanics background behind a  trace relation used by Meckler \cite{meckler:majorana,meckler:angular} to calculate the spin transition probability of Eq.(\ref{modtrans}):
\begin{equation}
\mbox{ P}^{(j)}_{mm^{\prime}}(t)  =  \left| \langle {\bf \hat{b}},m^{\prime} |\,{\bf \hat{a}},m \rangle \right|^{\,2} = \mbox{Tr} \!
  \left[ \mbox{{\boldmath $\Pi$}}^{(j)}(m,{\bf \hat{a}}) \; \mbox{{\boldmath $\Pi$}}^{(j)}(m^{\prime},{\bf \hat{b}})\right] \label{meckka}
\end{equation}
Using projection operators {\boldmath $\Pi$}$^{(j)}(m,{\bf \hat{a}})$ and  {\boldmath $\Pi$}$^{(j)}(m^{\prime},{\bf \hat{b}})$, this expression  gives 
the transition probability $\mbox{ P}^{(j)}_{mm^{\prime}}(t) $ in a spin-$j$ system  that a spin, initially quantized along ${\bf \hat{a}}$ with component $m$,  will later be quantized along ${\bf \hat{b}}$ with component $m^{\prime}$. 

Before we specialize to the case of spin-$j$ systems, whose projection operators are defined by quantization axes, suppose we consider the more general case of a Hermitian operator
 ${\bf X}$  whose eigenvalues are labeled by the index $i$. Making use of Sylvester's formula \cite{merzbacher2, horn}, this operator can be expressed in terms of projection operators {\boldmath $\Pi$}$(i) $ and the eigenvalues $x_i$ of ${\bf X}$ as the spectral decomposition of 
${\bf X}$ \cite{merzbacher}
\begin{eqnarray}
{\bf X} & = &  \sum_i x_i \, \mbox{{\boldmath $\Pi$}}(i) \label{sylvester} \\
 \sum_i  \mbox{{\boldmath $\Pi$}}(i) & = & \mathds{1} \label{sumsyl}
\end{eqnarray}
Then, multiplying Eq.(\ref{sylvester}) by {\boldmath $\Pi$}$(m) $, we find
\begin{eqnarray}
\mbox{{\boldmath $\Pi$}}(m)  \, {\bf X}&  = &  \sum_i x_i  \, \boxed{ \mbox{{\boldmath $\Pi$}}(m) \, \mbox{{\boldmath $\Pi$}}(i) } \label{simplify} \\
& = &  \sum_i x_i  \,  \mbox{{\boldmath $\Pi$}}(m)  \, \delta_{mi} \\
& = & x_m  \,  \mbox{{\boldmath $\Pi$}}(m) \label{traceexpr}
\end{eqnarray}
The ``boxed" term of Eq.(\ref{simplify}) has been simplified using the idempotency of projection operators \cite{merzbacher}:
\begin{eqnarray}
 \mbox{{\boldmath $\Pi$}}(m)  \,  \mbox{{\boldmath $\Pi$}}(i)  & = & 0 \; \;\;\;\;\;\; \;\;\;\;\;\;\;\; \;\;\;\;\;\;\;\;\;\;\;\;\;\;\;\;\; (\mbox{if}\;\; m \neq i) \\
& = &  [\mbox{{\boldmath $\Pi$}}(m)]^2 =  \mbox{{\boldmath $\Pi$}}(m)\;\;\;\;\;\;\;\; (\mbox{if}\;\; m = i)
\end{eqnarray}
Taking the trace of both sides of Eq.(\ref{traceexpr} ), we obtain
\begin{equation}
\mbox{Tr}  \, [ \mbox{{\boldmath $\Pi$}}(m) \, {\bf X}] = x_m \, \boxed{\mbox{Tr} \, [ \mbox{{\boldmath $\Pi$}}(m)] } = x_m \label{ftrace}
\end{equation}
In a representation in which {\boldmath $\Pi$}$(m) $ is diagonal, the only non-zero diagonal element  is
$\left[\mbox{{\boldmath $\Pi$}}(m)  \right]_{mm}=1$, and since the trace is representation-invariant, the ``boxed" term of Eq.(\ref{ftrace}) has been simplified using the fact that $\mbox{Tr} \, [\mbox{{\boldmath $\Pi$}}(m) ] =1$.

In order to specify the basis states we have been using in more detail, let $|{\bf \hat{a}},m \rangle$ refer to basis ket states for a spin-$j$ system quantized 
along ${\bf \hat{a}}$ with component $m$, where $m=-j, -j+1, \ldots, +j$. In this notation, Eq.(\ref{ftrace}) for example, is reexpressed as
\begin{equation}
\mbox{Tr} \, [ \mbox{{\boldmath $\Pi$}}^{(j)}(m, {\bf \hat{a}})  \, {\bf X}]  = x_m   \equiv 
\langle  {\bf \hat{a}},m|\, {\bf X}\,  |{\bf \hat{a}},m \rangle \label{traceproj}
\end{equation}
while the projection operator {\boldmath $\Pi$}$^{(j)}(m, {\bf \hat{a}}) $  is given by 
\begin{eqnarray}
\mbox{{\boldmath $\Pi$}}^{(j)}(m, {\bf \hat{a}}) & =  & |{\bf \hat{a}},m \rangle  \langle {\bf \hat{a}},m| \\
& =  & \displaystyle \prod_{\stackrel{\scriptstyle{r=-j}}{\scriptstyle{r \neq m}}}^{\scriptstyle{j}}
\!  \left\{\frac{r{\bf I}-({\bf \hat{a}} \cdot {\bf J})}{m-r}\right\}  \label{explicit}
\end{eqnarray}
That the form for {\boldmath $\Pi$}$^{(j)}(m, {\bf \hat{a}}) $ given in Eq.(\ref{explicit}) is a projector can be verified by considering the following identity \cite{merzbacher2,merzbacher} for an operator ${\bf A}$ with distinct eigenvalues $a_i$ and 
corresponding eigenkets $|A_k \rangle $
\begin{eqnarray}
{\bf P}_{\!j}\,  |A_k \rangle & =  & \displaystyle \prod_{i \neq j}
\!  \left\{\frac{a_i \mathds{1}-{\bf A}}{a_i-a_j}\right\}   |A_k \rangle  = \delta_{jk}\, |A_k \rangle \\
\mbox{where}\;\;\;\; {\bf A}\,|A_k \rangle & =  & a_k\,|A_k \rangle 
\end{eqnarray}
The explicit expression given in Eq.(\ref{explicit}) for the projection operators can be used for the spectral decomposition of 
Eq.(\ref{sylvester}) if the operator ${\bf X}$ has distinct eigenvalues. 
If operator ${\bf X}$,  introduced in Eq.(\ref{sylvester}), should represent the projector for a state quantized along ${\bf \hat{b}}$, with component $m^{\prime}$
\begin{equation}
{\bf X} =  |{\bf \hat{b}},m^{\prime} \rangle  \langle {\bf \hat{b}},m^{\prime}| \equiv \mbox{{\boldmath $\Pi$}}^{(j)}(m^{\prime}, {\bf \hat{b}})
\end{equation}
then using the result of Eq.(\ref{traceproj}), we find that 
\begin{eqnarray}
\mbox{Tr} \, [\mbox{{\boldmath $\Pi$}}^{(j)}(m, {\bf \hat{a}})  \; \mbox{{\boldmath $\Pi$}}^{(j)}(m^{\prime}, {\bf \hat{b}})] & = & 
\langle {\bf \hat{a}},m| \, {\bf \hat{b}},m^{\prime} \rangle  \langle {\bf \hat{b}},m^{\prime} |\,{\bf \hat{a}},m \rangle \\
& = & \left| \langle {\bf \hat{b}},m^{\prime} |\,{\bf \hat{a}},m \rangle \right|^{2} \label{mecktrans}
\end{eqnarray}
which is just the transition probability that a spin, initially quantized along ${\bf \hat{a}}$ with component $m$,  will later be quantized along ${\bf \hat{b}}$ with component $m^{\prime}$. In the next section, this expression will be used to calculate spin transition probabilities following the elegant method originally described by Meckler. \cite{meckler:majorana,meckler:angular}

\section{Meckler's Formula for Transition Probabilities}

\subsection{Meckler's formula}

Meckler \cite{meckler:angular} cleverly eschewed the canonical form \cite{merzbacher2,bala,Zelevinsky} of the projection operator matrices given in  Eq.(\ref{explicit})  in favor of an expansion (see Eq.(\ref{projectJn})) in Chebyshev polynomials $ f_{\lambda} ^{(j)}(m) $ and Chebyshev polynomial operators $ f_{\lambda} ^{(j)} ({\bf \hat{n}} \cdot {\bf J})  $
\begin{equation}
   \mbox{{\boldmath $\Pi$}}^{(j)}(m,{\bf \hat{n}})  =  \displaystyle\sum_{\lambda=0}^{2j} f_\lambda ^{(j)}(m) \; f_\lambda  ^{(j)}({\bf \hat{n}} \cdot {\bf J})  
\end{equation}
In so doing, Meckler  \cite{meckler:angular}  avoided what would necessarily have been  a very challenging exercise in calculating expectation values for trace calculations.  Just how challenging these trace calculations might have been can be gauged by Balasubramanian's calculation \cite{bala} of the 
``spin-flip"  transition probability $\mbox{ P}^{(j)}_{j,-j}(t) $ using Sylvester's formula. Only in this special case, when the inital and final state magnetic quantum numbers differed by the maximum value of $2j$, could the matrix elements of the canonical projection operators introduced by Sylvester's formula \cite{merzbacher2, horn} be evaluated, and then summed to yield a closed-form expression \cite{bala}.

Meckler's unorthodox approach to calculating the spin transition   probability \cite{meckler:majorana,meckler:angular}  relied on the use of projection operators expanded in a Chebyshev  polynomial operator basis  $f_L  ^{(j)}({\bf \hat{n}} \cdot {\bf J})$ as described in Section {\bf 3.2.1}. The foundation of Meckler's calculation \cite{meckler:majorana} is an expression which gives 
the transition probability that a spin, initially quantized along ${\bf \hat{a}}$ with component $m$,  will later be quantized along ${\bf \hat{b}}$ with component $m^{\prime}$. This probability  was expressed in Eq.(\ref{mecktrans}) as a trace of projection operators as follows \cite{meckler:majorana}  
\begin{equation}
\mbox{ P}^{(j)}_{mm^{\prime}}(t)  = \left| \langle {\bf \hat{b}},m^{\prime} |\,{\bf \hat{a}},m \rangle \right|^{\,2} = \mbox{Tr} \!
  \left[ \mbox{{\boldmath $\Pi$}}^{(j)}(m,{\bf \hat{a}}) \; \mbox{{\boldmath $\Pi$}}^{(j)}(m^{\prime},{\bf \hat{b}})\right] \label{meckk}
\end{equation}
As we shall see, Meckler's  choice of operator basis \cite{meckler:majorana,meckler:angular} was pivotal since these 
Chebyshev  polynomial operators $f_L^{(j)}( {\bf \hat{n}} \cdot {\bf J}) $ are endowed with properties (see Eq.(\ref{traceLegendre}) for example) that render the trace calculation in Eq.(\ref{meckk})  trivial. In this section, we shall devote the first two subsections to discussing two proofs of Meckler's formula 
\cite{meckler:majorana},  the first of which is due to Meckler \cite{meckler:majorana,meckler:angular}, and the second of which is due to Schwinger \cite{schwinger:majorana}.  A discussion of the relationship between these proofs in the third subsection will lead to a novel trace relation for 
$\left|{\cal D}_{m m^{\prime}}^{(j)}(R)\right |^2$. 

\subsubsection{First proof: using projection operators expanded in terms of the $f_L ^{(j)}({\bf \hat{n}} \cdot {\bf J})$ operators}

Given a spin initially quantized along a unit vector 
${\bf \hat{a}}$ with component $m$, the probability that it is quantized along a unit vector ${\bf \hat{b}}$ with component 
$m^{\prime}$ at a later time $t$ was calculated by Meckler \cite{meckler:majorana, meckler:angular}  to be
\begin{eqnarray}
\mbox{ P}^{(j)}_{mm^{\prime}}(t) & = &
 \mbox{Tr} \!  \left[ \, \boxed{\mbox{{\boldmath $\Pi$}}^{(j)}(m,{\bf \hat{a}}) }  \; \boxed{\mbox{{\boldmath $\Pi$}}^{(j)}(m^{\prime},{\bf \hat{b}}) } \, \right  ] \label{avoid} \\
& = &  \mbox{Tr} \!  \left[ \; \boxed{\sum_{\lambda=0}^{2j}  f_{\lambda} ^{(j)}(m) \, f_{\lambda}  ^{(j)}({\bf \hat{a}}  \cdot {\bf J})} \;  
\boxed{\sum_{\lambda^{\prime}=0}^{2j} f_{\lambda^{\prime}}  ^{(j)}(m^{\prime})f_{\lambda^{\prime}}  ^{(j)}({\bf \hat{b}}  \cdot {\bf J})} \; \right] \\
& = & \sum_{\lambda,\lambda^{\prime}=0}^{2j} f_{\lambda} ^{(j)}(m) \, f_{\lambda^{\prime}} ^{(j)} (m^{\prime}) \;
 \mbox{Tr} \! \left[ f_{\lambda} ^{(j)} ({\bf \hat{a}}  \cdot {\bf J}) \, f_{\lambda^{\prime}}  ^{(j)}({\bf \hat{b}}  \cdot {\bf J}) \right]  \\
& = & \sum_{\lambda, \lambda^{\prime}=0}^{2j}  f_{\lambda} ^{(j)} (m) \, f_{\lambda^{\prime}} ^{(j)}(m^{\prime}) \; 
 \delta_{\lambda \lambda^{\prime}} \; P_{\lambda}( {\bf \hat{a}} \cdot  {\bf \hat{b}})  \\
& = & \sum_{\lambda=0}^{2j}  f_{\lambda} ^{(j)}(m) \, f_{\lambda} ^{(j)} (m^{\prime}) \; 
 P_{\lambda}( {\bf \hat{a}} \cdot  {\bf \hat{b}}) 
\end{eqnarray}
The ingenuity of Meckler's projection operator approach  \cite{meckler:majorana, meckler:angular} to calculating the spin transition probability 
$\mbox{ P}^{(j)}_{mm^{\prime}}(t)$ is evident in this calculation. By exploiting the properties of the Chebyshev polynomial operators 
$ f_{\lambda} ^{(j)} ({\bf \hat{n}}  \cdot {\bf J})$ (see Eq.(\ref{traceLegendre})), Meckler was able to circumvent the trace calculation in Eq.(\ref{avoid}).  Even more ingenious was Meckler's choice 
 \cite{meckler:majorana} for the 
quantization axis ${\bf \hat{b}}$, which up to now we have left unspecified. Meckler chose  \cite{meckler:majorana} to define  ${\bf \hat{b}} \equiv  {\bf \hat{b}}(t)$ as a moving 
instantaneous axis along which the precessing spin vector maintains its quantization. Table VI compares the relative orientations of Meckler's 
instantaneous axis ${\bf \hat{b}}(t)$  \cite{meckler:majorana} in the middle column with the corresponding relative orientations of a precessing magnetic moment ${\bf \hat{m}}(t)$ according to 
Abragam \cite{abragamtext} in the right column, and it is evident that $Z=\cos \alpha$, where $\alpha$, the angle 
between the initial orientation of the magnetic moment ${\bf \hat{m}}(0)$ along the magnetic field $\mbox{H}_0 \, {\bf \hat{z}}$  and its orientation at a later time $t$ \cite{abragamtext},  is also  just the angle between the uniform field $\mbox{H}_0 \, {\bf \hat{z}}$  and Meckler's  instantaneous axis ${\bf \hat{b}}(t)$  \cite{meckler:majorana}. By tethering his instantaneous axis to the precessing spin vector, Meckler \cite{meckler:majorana} was able to compare the results of his spin transition probability calculation with that of 
Majorana's \cite{emajorana, ramsey}. For the reminder of this article we shall use $\beta \equiv \beta(t)$ 
(and not $\alpha$) to denote the relative orientation of this instantaneous axis with respect to the magnetic field.

\subsubsection{Second proof (by Schwinger): using the Clebsch-Gordan decomposition of the direct product}

In 1959, in response to Meckler's 1958  paper \cite{meckler:majorana} on the Majorana formula \cite{emajorana}, Schwinger submitted a brief note to The Physical Review which contained an alternative  proof of Meckler's version \cite{meckler:majorana}  of the  Majorana formula  \cite{emajorana}. According to 
Schwinger  \cite{schwinger:majorana}, that note was rejected, but it appeared eighteen years later in full as an Appendix in an  article by Schwinger \cite{schwinger:majorana}. Because Schwinger \cite{schwinger:majorana} only provided the outlines of his proof, using notation that is by now quite outdated, in this section, we provide a detailed discussion of Schwinger's proof using modern notation. Schwinger \cite{schwinger:majorana}  also relied exclusively on an Euler angle $(\alpha,\beta,\gamma) $   parametrization of rotation matrices, a restriction which is of course not necessary. We demonstrate that by using Schwinger's approach 
\cite{schwinger:majorana}   with both an angle-axis $(\psi, {\bf \hat{n}}) $ and Euler angle $(\alpha,\beta,\gamma) $  parametrization to derive Meckler's version  \cite{meckler:majorana} of the Majorana formula \cite{emajorana}.

Expressed in terms of the rotation matrices, the spin transition probability is given by \cite{abragamtext}
\begin{eqnarray}
\mbox{ P}^{(j)}_{mm^{\prime}}(t) & = &\left|{\cal D}_{m m^{\prime}}^{(j)}(\psi, {\bf \hat{n}}) \right|^2 \\
 & = & 
{\cal D}_{m m^{\prime}}^{(j)}(\psi, {\bf \hat{n}})
 \boxed{\left[ {\cal D}_{m m^{\prime}}^{(j)}(\psi, {\bf \hat{n}}) \right]^{\star}} \label{conjg1} \\
 & = & {\cal D}_{m m^{\prime}}^{(j)}(\psi, {\bf \hat{n}}) \;
\boxed{{\cal D}_{m^{\prime}m }^{(j)}(-\psi, {\bf \hat{n}})} \label{conjg2}\\
& = & {\cal D}_{m m^{\prime}}^{(j)}(\psi, {\bf \hat{n}}) \;
\boxed{(-1)^{m^{\prime}-m}  \;   {\cal D}_{-m -m^{\prime}}^{(j)}(\psi, {\bf \hat{n}})    } \label{conjg3}
\end{eqnarray}
Well-known properties \cite{varshal1:ang} of the ${\cal D}_{m m^{\prime}}^{(j)}(\psi, {\bf \hat{n}})$ matrices have been used to rewrite the complex conjugated matrix element in the ``boxed" term of Eq.(\ref{conjg1}) in Eqs.(\ref{conjg2}) and (\ref{conjg3}).  

The Clebsch-Gordan decomposition of the Kronecker (or direct) product ``$\otimes$" is expressed as the reducible sum ``$\oplus$"
 \cite{gottfried,brinksatch:ang} 
\begin{equation}
{\cal D}^{(j_1)}  \otimes {\cal D}^{(j_2)}  = \sum_{j=|j_1-j_2|}^{j_1+j_2} \oplus \; {\cal D}^{(j)} \label{CGseries}
\end{equation}
In terms of a Clebsch-Gordan coefficient series, and an angle-axis $R \equiv R(\psi, {\bf \hat{n}}) $ parametrization of the ${\cal D}^{(J)}_{m m^{\prime}}(R)$ matrices, this decomposition takes the explicit form \cite{varshal1:ang} 
\begin{equation}
{\cal D}_{M_1N_1}^{(J_1)}(\psi, {\bf \hat{n}}) \; {\cal D}_{M_2N_2}^{(J_2)}(\psi, {\bf \hat{n}}) =
 \sum_{J=|J_1-J_2|}^{J_1+J_2}\; \sum_{MN}
C_{J_1M_1J_2M_2}^{JM} \; {\cal D}_{MN}^{(J)}(\psi, {\bf \hat{n}}) \; C_{J_1N_1J_2N_2}^{JN} 
\label{angleaxisseries}
\end{equation}
We can  now use the Clebsch-Gordan series of Eq.(\ref{angleaxisseries}) to reexpress the relation of 
Eq.(\ref{conjg3}) as Meckler's formula \cite{meckler:majorana, meckler:angular} for the transition probability:
\begin{eqnarray}
\mbox{ P}^{(j)}_{mm^{\prime}}(t) & = &
\left|{\cal D}_{m m^{\prime}}^{(j)}(\psi, {\bf \hat{n}}) \right|^2  \\
& = & (-1)^{m^{\prime}-m}  \; 
\boxed{ {\cal D}_{m m^{\prime}}^{(j)}(\psi, {\bf \hat{n}}) \;
  {\cal D}_{-m -m^{\prime}}^{(j)}(\psi, {\bf \hat{n}})} \label{phase} \\
& = & (-1)^{m^{\prime}-j+j-m} \; \boxed{\sum_{\lambda=0}^{2j} C_{jmj-m}^{\lambda 0} \; C_{jm^{\prime}j-m^{\prime}}^{\lambda 0}  \;
{\cal D}_{00}^{(\lambda)}(\psi, {\bf \hat{n}})} \label{phase2}\\
& = & \sum_{\lambda=0}^{2j} \boxed{(-1)^{j-m} \; C_{jmj-m}^{\lambda 0}} \;
 \boxed{(-1)^{j-m^{\prime}} \;C_{jm^{\prime}j-m^{\prime}}^{\lambda 0} } \;
{\cal D}_{00}^{(\lambda)}(\psi, {\bf \hat{n}})  \label{boxed}
\end{eqnarray}
The phase factor in Eq.(\ref{phase}) has been rewritten in Eqs.(\ref{phase2}) and (\ref{boxed}) as
\begin{equation}
 (-1)^{m^{\prime}-m} = (-1)^{m^{\prime}-j+j-m} =(-1)^{j-m} \;(-1)^{j-m^{\prime}}
\end{equation}
since $(-1)^{m-j} = (-1)^{j-m}$ for all values of $j$ (integral and half-integral). The ``boxed" 
${\cal D}^{(j)}$-matrix product term
in the same equation has been rewritten as the ``boxed" term in Eq.(\ref{phase2}) using the Clebsch-Gordan 
series of Eq.(\ref{angleaxisseries}). 
Each of the ``boxed" terms in Eq.(\ref{boxed}) is a Chebyshev polynomial ($f_{\lambda}^{(j)} (m)$ or 
$f_{\lambda}^{(j)} (m^{\prime})$ as defined in Eq.(\ref{chebyclebsch})), and as shown in Appendix B,  the rotation matrix element 
${\cal D}_{00}^{\lambda}(\psi, {\bf \hat{n}})$ can be written as an $\lambda$-th order Legendre polynomial \cite{varshal1:ang}:
\begin{equation}
{\cal D}_{00}^{(\lambda)}(\psi, {\bf \hat{n}})  = d_{00}^{(\lambda)}(\xi) \equiv P_{\lambda}(\cos \xi) = P_{\lambda}(\cos \beta) \label{legendreangleaxis}
\end{equation}

The transition probability of Eq.(\ref{boxed}) can finally then be rewritten as 
Meckler's formula \cite{meckler:angular}, a Fourier-Legendre series, whose expansion coefficients are products of Chebyshev polynomials:
\begin{equation}
\mbox{ P}^{(j)}_{mm^{\prime}} (t) = \left|{\cal D}_{m m^{\prime}}^{(j)}(\psi, {\bf \hat{n}})\right|^2 =
\sum_{\lambda=0}^{2j} f_{\lambda} ^{(j)}(m) \; f_{\lambda} ^{(j)}(m^{\prime})\;  P_{\lambda}(\cos \beta) 
\end{equation}

In the case of an  Euler angle
 $R \equiv R(\alpha,\beta,\gamma)$ parametrization of the ${\cal D}^{(j)}_{m m^{\prime}}(R)$ matrices, the 
Clebsch-Gordan series of Eq.(\ref{CGseries}) takes the same explicit form  \cite{varshal1:ang} as that of 
Eq.(\ref{angleaxisseries}) 
\begin{equation}
{\cal D}_{M_1N_1}^{(J_1)}(\alpha,\beta,\gamma) \; {\cal D}_{M_2N_2}^{(J_2)}(\alpha,\beta,\gamma) =
 \sum_{J=|J_1-J_2|}^{J_1+J_2}\; \sum_{MN}
C_{J_1M_1J_2M_2}^{JM} \; {\cal D}_{MN}^{(J)}(\alpha,\beta,\gamma) \; C_{J_1N_1J_2N_2}^{JN} 
\label{Eulerseries}
\end{equation}

A slight modification of the same argument can be used to arrive at the same result for the transition probability $\mbox{ P}^{(j)}_{mm^{\prime}}(t) $  when the ${\cal D}^{(j)}_{m m^{\prime}}(R)$ matrices are parametrized by  Euler angles
 $R \equiv R(\alpha,\beta,\gamma)$. In this case, the transition probability is given by \cite{abragamtext, schwinger:majorana}
\begin{eqnarray}
\mbox{ P}^{(j)}_{mm^{\prime}}(t) & = & 
\left|{\cal D}_{m m^{\prime}}^{(j)}(\alpha,\beta,\gamma) \right|^2 \\ 
& = & 
{\cal D}_{m m^{\prime}}^{(j)}(\alpha,\beta,\gamma) \;
 \boxed{\left[ {\cal D}_{m m^{\prime}}^{(j)}(\alpha,\beta,\gamma) \right]^{\star}} \label{conjg11} \\
 & = & {\cal D}_{m m^{\prime}}^{(j)}(\alpha,\beta,\gamma) \;
\boxed{(-1)^{m-m^{\prime}}     {\cal D}_{-m -m^{\prime}}^{(j)}(\alpha,\beta,\gamma)   } \label{conjg33} \\
\mbox{where} \;\;\;\; {\cal D}_{m m^{\prime}}^{(j)}(\alpha,\beta,\gamma) & = & e^{-im \alpha} \, d^{(j)}_{m m^{\prime}}(\beta) \, e^{-im^{\prime} \gamma}
\end{eqnarray}
As above, the corresponding well-known properties \cite{varshal1:ang} of the 
${\cal D}_{m m^{\prime}}^{(j)}(\alpha,\beta,\gamma)$ matrices have been used to rewrite the complex conjugated matrix element in the ``boxed" term of Eq.(\ref{conjg11}) in Eq.(\ref{conjg33}).  As Schwinger noted \cite{schwinger:majorana}, the net effect of the time-dependent radiofrequency field is to rotate the angular momentum vector of the magnetic moment through a definite angle, the Euler angle $\beta$. This is the same angle that Meckler used
\cite{meckler:majorana} to keep track of the angle between the uniform static field and his instantaneous axis ${\bf \hat{b}}(t)$. 

We can  now use the Clebsch-Gordan series of Eq.(\ref{Eulerseries}) to reexpress the relation of 
Eq.(\ref{conjg33}) as Meckler's formula \cite{meckler:majorana,meckler:angular} for the transition probability:
\begin{eqnarray}
\mbox{ P}^{(j)}_{mm^{\prime}}(t) & = &
\left|{\cal D}_{m m^{\prime}}^{(j)}(\alpha,\beta,\gamma) \right|^2  \\
& = & (-1)^{m-m^{\prime}}   \;
\boxed{ {\cal D}_{m m^{\prime}}^{(j)}(\alpha,\beta,\gamma) \;
    {\cal D}_{-m -m^{\prime}}^{(j)}(\alpha,\beta,\gamma) } \label{phase20} \\ 
& = & (-1)^{m-j+j-m^{\prime}} \; \boxed{ \sum_{\lambda=0}^{2j} C_{jmj-m}^{\lambda 0} \; C_{jm^{\prime}j-m^{\prime}}^{\lambda 0}  \;
{\cal D}_{00}^{(\lambda )}(\alpha,\beta,\gamma)} \label{phase22} \\
& = &  \sum_{\lambda =0}^{2j} \boxed{(-1)^{j-m} \; C_{jmj-m}^{\lambda 0}} \;
 \boxed{(-1)^{j-m^{\prime}} \;C_{jm^{\prime}j-m^{\prime}}^{\lambda 0} } \;
{\cal D}_{00}^{(\lambda)}(\alpha,\beta,\gamma) \label{boxed2}
\end{eqnarray}
The phase factor in Eq.(\ref{phase20}) has been successively rewritten in Eqs.(\ref{phase22}) and (\ref{boxed2}) as
\begin{equation}
 (-1)^{m-m^{\prime}} = (-1)^{m-j+j-m^{\prime}} =(-1)^{j-m} \;(-1)^{j-m^{\prime}}
\end{equation}
since $(-1)^{m-j} = (-1)^{j-m}$ for all values of $j$ (integral and half-integral). The ``boxed" term of Eq.(\ref{phase20}) has been reexpressed in Eq.(\ref{phase22}) using the Clebsch-Gordan series of Eq.(\ref{Eulerseries}).  
Each of the ``boxed" terms in Eq.(\ref{boxed2}) is a Chebyshev polynomial ($f_{\lambda}^{(j)} (m)$ or 
$f_{\lambda}^{(j)} (m^{\prime})$ as defined in Eq.(\ref{chebyclebsch})), and the rotation matrix element 
${\cal D}_{00}^{(\lambda)}(\alpha,\beta,\gamma)$ can be written as an $\lambda$-th order Legendre polynomial \cite{varshal1:ang}:
\begin{equation}
{\cal D}_{00}^{(\lambda)}(\alpha,\beta,\gamma)  = d_{00}^{(\lambda)}(\beta) \equiv P_{\lambda}(\cos \beta) \label{legendre}
\end{equation}
The transition probability of Eq.(\ref{boxed2}) can finally then be rewritten as Meckler's formula 
\cite{meckler:majorana,meckler:angular}, a Fourier-Legendre series whose expansion coefficients are products of Chebyshev polynomials:
\begin{equation}
\mbox{ P}^{(j)}_{mm^{\prime}}(t)  = \left|{\cal D}_{m m^{\prime}}^{(j)}(\alpha,\beta,\gamma) \right|^2 =
\sum_{\lambda=0}^{2j} f_{\lambda}^{(j)} (m) \; f_{\lambda}^{(j)} (m^{\prime})\;  P_{\lambda}(\cos \beta) \label{majoranaEuler}
\end{equation}
As Schwinger \cite{schwinger1937} first noted (without proof),  for a system initially prepared in a state with magnetic quantum number $m$, the sum 
of the transition probabilities over all possible final states labelled by  $m^{\prime}$  should be unity:
\begin{equation}
\sum_{m^{\prime}=-j}^j \!\!\! \mbox{ P}^{(j)}_{mm^{\prime}}(t)  =1
\end{equation}
Proving this relation could not be simpler with the use  of  the Majorana \cite{emajorana} formula in the version  that Meckler 
\cite{meckler:majorana, meckler:angular} first derived. By summing Eq.(\ref{majoranaEuler}) over all final states, 
we find
\begin{eqnarray}
  \sum_{m^{\prime}=-j}^j \! \!\! \mbox{ P}^{(j)}_{mm^{\prime}} (t)  & = & \sum_{m^{\prime}=-j}^j
\sum_{\lambda=0}^{2j} f_{\lambda}^{(j)} (m) \; f_{\lambda}^{(j)} (m^{\prime})\;  P_{\lambda}(\cos \beta) \\
& = & \sum_{\lambda=0}^{2j} f_{\lambda}^{(j)} (m) \; \boxed{\sum_{m^{\prime}=-j}^j  f_{\lambda}^{(j)} (m^{\prime})} \;  P_{\lambda}(\cos \beta)
\label{sumsum}
\end{eqnarray}
In order to handle the sum over Chebyshev polynomials in the ``boxed" term of Eq.(\ref{sumsum}), we exploit  the Chebyshev polynomial orthogonality relation of Eq.(\ref{ortho1}):
\begin{equation}
 \displaystyle\sum_{m=-j}^{j}f_{\lambda}^{(j)} (m) \; f_{\lambda^{\prime}}^{(j)} (m) = \delta_{\lambda \lambda^{\prime}}  
\end{equation}
If $\lambda^{\prime}=0$, then we have the following special case of this relation:
\begin{eqnarray}
\displaystyle\sum_{m=-j}^{j}f_{\lambda}^{(j)} (m) \; \boxed{f_{0}^{(j)} (m)} &  =  & \delta_{\lambda 0}  \label{sumanswer1} \\
\displaystyle\sum_{m=-j}^{j}f_{\lambda}^{(j)} (m) & = & \sqrt{2j+1} \; \,\delta_{\lambda 0} \label{sumanswer2}
\end{eqnarray}
The ``boxed" term of Eq.(\ref{sumanswer1}) has been evaluated using the relation 
\cite{filippov2:thesis, corio:ortho}
\begin{equation}
f_0^{(j)} (m) = \displaystyle\frac{1}{\sqrt{2j+1}} 
\end{equation}
The identity of Eq.(\ref{sumanswer2}) not only evaluates the sum in the ``boxed" term of Eq.(\ref{sumsum}), but it shows that 
in the sum over $\lambda$ in Eq.(\ref{sumsum}), only the $\lambda=0$ term contributes. Finally then, the sum over all final states give the following expected result for the total probability:
\begin{eqnarray}
 \sum_{m^{\prime}=-j}^j \! \!\! \! \mbox{ P}^{(j)}_{mm^{\prime}}(t)  & = &  \sqrt{2j+1} \;
f_0^{(j)} (m) \; P_0(\cos \beta) \\
&= & 1
\end{eqnarray}
using the fact that $P_0(\cos \beta) = 1$.

In the context of the Meckler formula  \cite{meckler:majorana, meckler:angular}, there are alternatives to the use of  Chebyshev polynomials 
$f_{\lambda}^{(j)}(m)=\langle jm| \; f_{\lambda}^{(j)} ( J_z)  \;  |jm \rangle  $, and in fact Schwinger \cite{schwinger:majorana}  did not choose to express the Clebsch-Gordon coefficients in Eqs.(\ref{boxed}) or (\ref{boxed2}) in terms of Chebyshev polynomials $f_{\lambda}^{(j)}(m)$, but rather in terms of matrix elements of Legendre polynomial operators $P_{\lambda}({\bf J})$ \cite{schwinger:majorana}. These operators are discussed in Appendix A.

\subsection{How are the two proofs related?}

Both methods for calculating $ \mbox{ P}^{(j)}_{mm^{\prime}}(t)$  have a foundation in angular momentum theory, either angular momentum algebra in Meckler's case  \cite{meckler:angular}, or angular momenta composition in Schwinger's case \cite{schwinger:majorana}.  No matter which method is used, either Meckler's original approach using projection 
operators \cite{meckler:majorana, meckler:angular}, or that 
adopted by Schwinger \cite{schwinger:majorana}, the final result for the Majorana  spin transition probability 
\cite{emajorana}  is 
of course the same
\begin{equation}
 \mbox{ P}^{(j)}_{mm^{\prime}}(t)= \left|{\cal D}_{m m^{\prime}}^{(j)}(R) \right|^2     =   \mbox{Tr} \!
  \left[ \mbox{{\boldmath $\Pi$}}^{(j)}(m,{\bf \hat{a}}) \; \mbox{{\boldmath $\Pi$}}^{(j)}(m^{\prime},{\bf \hat{b}}) \right] 
 =   
   \mbox{Tr} \! \left[ \mbox{{\boldmath $\Pi$}}^{(j)}(m,{\bf \hat{z}})   \; \mbox{{\boldmath $\Pi$}}^{(j)}(m^{\prime},{\bf \hat{z}}^{\prime}) \right]\,   \label{puzzle}
\end{equation} 
In this result for $\mbox{ P} ^{(j)}_{mm^{\prime}}(t)$, $R \equiv R(\alpha,\beta,\gamma)$ or $R \equiv R(\psi, {\bf \hat{n}})$ or any other  parametrization \cite{sim:rot}  of $R$ 
 for that matter, since the first equality 
of Eq.(\ref{puzzle}) does not depend on this parametrization as shown in Section {\bf 4.1.2}. On the other hand,  as shown in Section {\bf 4.1.1}, each trace is also a valid expression for the transition probability $\mbox{ P}^{(j)}_{mm^{\prime}}(t) $. But how can that be, since  Schwinger's approach \cite{schwinger:majorana} discussed in Section {\bf 4.1.2} only uses
${\cal D}_{m m^{\prime}}^{(j)}(R)$ matrix elements in the basis set $|jm \rangle  \equiv |{\bf \hat{z}},m \rangle$ (corresponding to a 
$ {\bf {\hat z}}  $ quantization axis), whereas Meckler's approach \cite{meckler:majorana, meckler:angular} discussed in Section {\bf 4.1.1} uses two distinct basis sets 
$|{\bf \hat{z}},m \rangle$  and $|{\bf \hat{z}}^{\prime},m^{\prime} \rangle$ (corresponding to two distinct quantization axes, $ {\bf {\hat a}}  \equiv {\bf {\hat z}}$  
and $ {\bf {\hat b}}  \equiv {\bf {\hat z}}^{\prime}$, respectively) 
 to define the projection operators  {\boldmath $\Pi$}$^{(j)}(m,{\bf \hat{z}}) $  and  {\boldmath $\Pi$}$^{(j)}(m^{\prime},{\bf \hat{z}}^{\prime}) $?

There is nothing physical about the quantization axes, since they are just a way of labeling states, and certainly the final result for the transition probability cannot depend on the choice of quantization axes for the initial and final states. In order to demonstrate this independence, a careful consideration of basis set transformations is required. Following Brink and Satchler \cite{brinksatch:ang},
 we note  that if a set of 
axes $(x^{\prime},y^{\prime},z^{\prime})$ is obtained by a rotation $R$ from a set $(x,y,z)$, then the eigenstates
    $|{\bf \hat{z}}^{\prime},n \rangle$ of $({\bf J} \cdot {\bf \hat{z}}^{\prime}) \equiv J_{z^{\prime}}$ are determined by rotating the corresponding eigenstates 
$|{\bf \hat{z}}, n \rangle$ of $J_{z}$ along with the axes. In this way, the 
state $|{\bf \hat{z}}, n \rangle$ is transformed by the rotation operator $\hat{{\cal D}}^{(j)} \! (R)$ as follows
\begin{eqnarray}
 |{\bf \hat{z}}^{\prime},n \rangle & = & \hat{{\cal D}}^{(j)} \! (R) \, |{\bf \hat{z}}, n \rangle  \label{rot} \\
& = & \boxed{ \sum_{m=-j}^j |{\bf \hat{z}}, m \rangle \langle {\bf \hat{z}}, m | }\,  \hat{{\cal D}}^{(j)} \! (R) \, |{\bf \hat{z}}, n \rangle \label{completeness} \\
& = &  \sum_{m=-j}^j |{\bf \hat{z}}, m \rangle \, {\cal D}_{mn}^{(j)}(R)
\end{eqnarray} 
The ``boxed" term of Eq.(\ref{completeness}) is  a representation of the identity operator $\mathds{1}$, using the completeness relation for the 
eigenstates $ |{\bf \hat{z}}, m\rangle$:
\begin{equation}
\mathds{1} =  \sum_{m=-j}^j |{\bf \hat{z}}, m\rangle \langle {\bf \hat{z}}, m| 
\end{equation}
The states $ \langle {\bf \hat{z}}, n | $ conjugate to those rotated in Eq.(\ref{rot}) are transformed by the adjoint (transpose conjugate) 
rotation operator $\left[\hat{{\cal D}}^{(j)} \!(R)\right]^{\!\dagger} $  as follows \cite{brinksatch:ang}
\begin{eqnarray}
  \langle {\bf \hat{z}}^{\prime},n  | = \langle  {\bf \hat{z}}, n  | \, \left [\hat{{\cal D}}^{(j)} \!(R)  \right ]^{\!\dagger}
& =  & \sum_{m=-j}^{j} \, \left[  {\cal D}_{mn}^{(j)} \!(R)  \right]^{\!\star}  \langle {\bf \hat{z}}, m  | \label{rotprime}  \\
\mbox{where} \;\;\;  \left[  {\cal D}_{mn}^{(j)} (R)  \right]^{\!\star}  & =  & 
\langle  {\bf \hat{z}}, m | \, \hat{{\cal D}}^{(j)}  \! (R)  \,  | {\bf \hat{z}}, n \rangle ^{\star} =
 \langle  {\bf \hat{z}}, n| \;   \left [\hat{{\cal D}}^{(j)}\!(R)  \right ]^{\!\dagger}      \,  |{\bf \hat{z}}, m  \rangle
\end{eqnarray}
Then, referring to Eq.(\ref{puzzle})
\begin{eqnarray}
\left|{\cal D}_{m m^{\prime}}^{(j)}(R) \right|^2 & = & {\cal D}_{m m^{\prime}}^{(j)}(R)
 \left[  {\cal D}_{m m^{\prime}}^{(j)}(R)   \right]^{\!\star} \label{modulus} \\
 & = &  \langle {\bf \hat{z}}, m | \, \boxed{\hat{{\cal D}}^{(j)}(R) |{\bf \hat{z}}, m^{\prime}\rangle} \; 
\boxed{ \langle {\bf \hat{z}}, m^{\prime}| \! \left [\hat{{\cal D}}^{(j)} \!(R)\right ]^{\!\dagger} }       \;  |{\bf \hat{z}}, m \rangle  \label{transf}\\
& = &  \langle {\bf \hat{z}}, m | \, \boxed{  |{\bf \hat{z}}^{\prime},m^{\prime}\rangle \; 
  \! \langle {\bf \hat{z}}^{\prime},m^{\prime}|} \,  |{\bf \hat{z}}, m  \rangle \label{project} \\
& = &  \langle {\bf \hat{z}}, m | \;  \mbox{{\boldmath $\Pi$}}^{(j)} (m^{\prime},{\bf \hat{z}}^{\prime})  \; |{\bf \hat{z}}, m  \rangle \label{expect} \\
& = &  \mbox{Tr} \! \left[ \,  \mbox{{\boldmath $\Pi$}}^{(j)}(m,{\bf \hat{z}})   \; \mbox{{\boldmath $\Pi$}}^{(j)}(m^{\prime},{\bf \hat{z}}^{\prime}) \right] \label{finaltrace}
\end{eqnarray}
We recognize the ``boxed" terms of Eq.(\ref{transf}) as the state transformation equations of 
Eqs.(\ref{rot}) and (\ref{rotprime}), and identify the ``boxed" term of Eq.(\ref{project}) as the projection operator 
{\boldmath $\Pi$}$^{(j)}(m^{\prime},{\bf \hat{z}}^{\prime}) $:
\begin{equation}
 \mbox{{\boldmath $\Pi$}}^{(j)}(m^{\prime},{\bf \hat{z}}^{\prime})  = |{\bf \hat{z}}^{\prime},m^{\prime}\rangle \; 
  \! \langle {\bf \hat{z}}^{\prime},m^{\prime}|
\end{equation}
Finally, Eq.(\ref{finaltrace}) was obtained from Eq.(\ref{expect})  by using the following form of Eq.(\ref{traceproj}) expressed 
in the notation of Eqs.(\ref{modulus} - \ref{finaltrace}):
\begin{eqnarray}
\langle {\bf \hat{z}}, m  |\, {\bf X}\,  | {\bf \hat{z}}, m \rangle & = & \mbox{Tr}\!  \left[ \,  \mbox{{\boldmath $\Pi$}}^{(j)}(m,{\bf \hat{z}}) \; {\bf X} \right] \\
\mbox{where} \;\;\;  {\bf X} & = & \mbox{{\boldmath $\Pi$}}^{(j)}(m^{\prime},{\bf \hat{z}}^{\prime}) 
\end{eqnarray}

We conclude by stating the result in Eq.(\ref{finaltrace}) in more general terms. For a spin-$j$ system, 
whose magnetic quantum numbers are chosen from the set $\Big\{m \Big\}_{\!\!-j}^{\,\,j}$,  let us consider two quantization axes, defined  by unit vectors 
${\bf \hat{z}}$ and $ {\bf \hat{z}}^{\prime}$, where the quantization axis 
$ {\bf \hat{z}}^{\prime}$ is obtained from ${\bf \hat{z}}$ by a rotation $R$. Associated with these axes are  projection operators 
$ \mbox{{\boldmath $\Pi$}}^{(j)}(m,{\bf \hat{z}})$ and $ \mbox{{\boldmath $\Pi$}}^{(j)}(m^{\prime},{\bf \hat{z}}^{\prime})$, each of which is 
also a function of a magnetic quantum number ($m$ or $m^{\prime}$).   Acting on an arbitrary superposition of multiplet states $|{\bf \hat{z}}, n\rangle$,  
$ \mbox{{\boldmath $\Pi$}}^{(j)}(m,{\bf \hat{z}}) = |{\bf \hat{z}}, m \rangle \;  \! \langle {\bf \hat{z}}, m | $
singles out the component with the projection $ ({\bf J} \cdot {\bf \hat{z}}) =m$, whereas 
$ \mbox{{\boldmath $\Pi$}}^{(j)}(m^{\prime},{\bf \hat{z}}^{\prime}) = |{\bf \hat{z}}^{\prime},m^{\prime}\rangle \; 
  \! \langle {\bf \hat{z}}^{\prime},m^{\prime}|$, 
 acting on an arbitrary superposition of multiplet states $|{\bf \hat{z}}^{\prime},n\rangle $, 
singles out the component with the projection $ ({\bf J} \cdot {\bf \hat{z}}^{\prime}) = m^{\prime}$ \cite{Zelevinsky}.
Then the modulus squared of the Wigner rotation matrix element  ${\cal D}_{m m^{\prime}}^{(j)}(R) $ is given by the trace of the product of these projection 
operators as follows:
\begin{equation}
\boxed{\left|{\cal D}_{m m^{\prime}}^{(j)}(R)\right |^2 = 
 \mbox{Tr} \! \left[ \,  \mbox{{\boldmath $\Pi$}}^{(j)}(m,{\bf \hat{z}})   \; \mbox{{\boldmath $\Pi$}}^{(j)}(m^{\prime},{\bf \hat{z}}^{\prime}) \, \right] 
=  \mbox{ P}^{(j)}_{mm^{\prime}}(t) } \label{traceidid}
\end{equation}
As we noted at the outset, and as we emphasize once again, this result for the transition probability $ \mbox{ P}^{(j)}_{mm^{\prime}}(t) $ does not depend upon the parametrization of $R$. What it does very much depend upon is the relative orientation of the quantization axes axes ${\bf {\hat z}}$  
and ${\bf {\hat z}}^{\prime}$ as shown in Section {\bf 4.1}.

\subsection{Applications of the Meckler formula and related expressions}

Although Schwinger \cite{schwinger:majorana} took notice of Meckler's formula \cite{meckler:majorana, meckler:angular}, and Biedenharn and Louck mention it in their discussion of the Majorana formula  \cite{emajorana}, 
Meckler's formula \cite{meckler:majorana} for the Majorana \cite{emajorana} spin transition probability has remained relatively obscure. In this section,  some practical applications of the Meckler formula \cite{meckler:majorana} are discussed. These applications include the calculation of the ``spin-flip" probability $\mbox{P}^{(j)}_{j,-j} (t) $, and the elucidation of some properties of the Wigner rotation matrix elements.  We also demonstrate the rediscovery of the Meckler formula in  a recent solution for  the multi-level Landau-Zener  transition 
probability $\mbox{ P}_{mm^{\prime}}^{\mbox{\tiny{LZ}}}(t)  $ by Fai et al. \cite{fai}. 

\subsubsection{Spin-flip transition probabilities expressed as a  Fourier-Legendre series}

Meckler's formula \cite{meckler:majorana}  provides a straightforward answer to the calculation of the ``spin-flip" probability $\mbox{P}^{(j)}_{j,-j} (t) $, the probability for a  radiofrequency-induced transition in a spin-$j$ system between the state $| j,j \rangle$ with the highest magnetic projection number,  and the state 
$|j,-j \rangle$ with the lowest magnetic projection number. Using Meckler's formula \cite{meckler:majorana}    for $ \mbox{P}^{(j)}_{mm^{\prime}}(t) $ given in Eq.(\ref{majoranaEuler}), 
\begin{eqnarray}
\mbox{ P}^{(j)}_{j,-j} (t)  & = & 
\sum_{L=0}^{2j} \boxed{ f_L^{(j)} (j) \; f_L^{(j)} (-j) }\;  P_L(\cos \beta) \label{spinflip1} \\
& = &  \sum_{L=0}^{2j} c(j,L) \, P_L(\cos \beta) \label{legserr} \\
& = & \left[(2j)! \right]^2 \sum_{L=0}^{2j} \frac{(-1)^L \, (2L+1) }{(2j-L)!(2j+L+1)!} \, P_L(\cos \beta) \label{leg} \\
& = & \left[ \frac {1- \cos \beta}{2} \right]^{\!2j} = \left[ \sin (\beta / 2) \right]^{4j} \label{leg3}\\
& = &  (\sin  \Theta)^{4j} \left( \sin \psi/2 \right) ^{4j} \label{jjth}\\
& = & \left[   \frac{\omega_1}{\omega_e}  \right]^{\!4j} 
 \sin^{4j} \! \left\{   \omega_e t/2 \right\} \label{spinflip} \\
\mbox{where} \;\;\;\; \omega_e & \equiv & \left[\omega_1^2+(\omega_0-\omega)^2 \right]^{\!1/2} \label{rfdef}
\end{eqnarray}
In Eq.(\ref{rfdef}), $\omega_e$ is the effective radiofrequency field strength, defined in terms of the applied radiofrequency field $\omega_1=\gamma H_1$ and the resonance offset $\Delta = \omega_0-\omega$ where $\omega_0 = \gamma H_0$ is the Larmor frequency. 
The  results given in Eqs.(\ref{jjth}) and (\ref{spinflip})  agree with analogous expressions obtained by Balasubramanian \cite{bala} using Sylvester's formula \cite{merzbacher2,horn}, and by Siemens et al. \cite{sim:beyond} using the Chebyshev polynomial operator expansion of the rotation operator as given in Eq.(\ref{coriorotfirst}) below (see Section {\bf 5.1}) . Each step leading to the expressions for the spin-flip probability given in  Eqs.(\ref{jjth}) and (\ref{spinflip})   is now justified. 

 The first Chebyshev polynomial $f_L^{(j)} (j)$ in the ``boxed" term of Eq.(\ref{spinflip1}) can be evaluated using Eq.(\ref{cgeval44}). 
After the second Chebyshev polynomial $f_L^{(j)} (-j)$ in the ``boxed" term of Eq.(\ref{spinflip1}) is evaluated using the parity relation of Eq.(\ref{parity}) (see Section {\bf 2.1.3}), 
the coefficients $c(j,L)= f_L^{(j)} (j) \; f_L^{(j)} (-j) $ in the Fourier-Legendre series expansion of  the spin-flip transition probability $ \mbox{P}^{(j)}_{j,-j}(t) $ in Eq.(\ref{legserr}) are easily obtained. This expansion can be summed to yield the very simple and closed-form expression of  Eq.(\ref{leg3}),  as described in Appendix C.

Using the same approach, Meckler's formula \cite{meckler:majorana}  also provides a straightforward answer to the calculation of the ``spin-flip" probability $\mbox{P}^{(j)}_{j-1,-(j-1)} (t) $, the probability in a spin-$j$ system for a  radiofrequency-induced  transition between the state $| j,j-1 \rangle$ with the next to highest magnetic projection number,  and the state  $|j,-(j-1) \rangle$ with the next to  lowest magnetic projection number. Using Meckler's formula \cite{meckler:majorana}    for $ \mbox{P}^{(j)}_{mm^{\prime}}(t) $ given in Eq.(\ref{majoranaEuler}), 
\begin{eqnarray}
  &  & \mbox{ P}^{(j)}_{j-1,-(j-1)} (t) \\
& = & \sum_{L=0}^{2j} \boxed{ f_L^{(j)} (j-1) \; f_L^{(j)} (-(j-1)) }\;  P_L(\cos \beta) \label{spinflip2} \\
& = &  \sum_{L=0}^{2j} c^{\prime}(j,L) \, P_L(\cos \beta) \label{legser} \\
& = &  \sum_{L=0}^{2j} \frac{(-1)^L \, [L(L+1)-2j]^2 \; (2L+1) \left[(2j-1)! \right]^2}{(2j-L)!(2j+L+1)!} \, P_L(\cos \beta) \label{legser22} \\
& = & \left[ \frac {1- \cos \beta}{2} \right]^{\!2(j-1)} \left[2j \cos^2 (\beta / 2)  -1 \right]^{\!2}= \left[ \sin (\beta / 2) \right]^{4(j-1)} 
 \left[2j \cos^2 (\beta / 2)  -1 \right]^{\!2}
\label{legser23} \\
& = &  (\sin  \Theta)^{4(j-1)}    \left(\sin \psi/2  \right)^{4(j-1)} \left[2j \cos^2 \Theta \, \sin^2 \psi/2 -1 \right]^{\!2} \\
& = & \left[ \frac{\omega_1}{\omega_e} \right]^{\!4(j-1)} 
 \sin^{4(j-1)} \! \left\{ \omega_e t/2 \right\}  \; \left[2j \frac{(\omega_0-\omega)^2 }{\omega_e^2} 
 \sin^{2} \left\{  \omega_e t/2  \right\} -1 \right]^{\!2}  \label{spinflip3} 
\end{eqnarray}

 The first Chebyshev polynomial $f_L^{(j-1)} (j)$ in the ``boxed" term of Eq.(\ref{spinflip2}) can be evaluated using the relation of Eq.(\ref{cgeqcheb}) between the Chebyshev polynomials $ f_L^{(j)}(m)$ and the Clebsch-Gordan coupling coefficients $ C^{L0}_{jmj-m} $,  from which we obtain
\begin{eqnarray}
 f_L^{(j)} (j-1) & = & (-1)^{j-(j-1)} \; C^{L0}_{j(j-1)j-(j-1)}  \label{cgeval2}\\
& = & -[L(L+1)-2j]\left[ \frac{(2L+1) \, [(2j-1)!]^2}{(2j+L+1)! \, (2j-L)!}  \right]^{\!1/2}
\end{eqnarray}
The Clebsch-Gordan coefficient $C^{L0}_{j(j-1)j-(j-1)} $ in Eq.(\ref{cgeval2}) was evaluated using the relation \cite{varshal1:ang}
\begin{eqnarray}
C^{c\gamma}_{aa-1b\beta}& = & \delta_{\gamma-\beta,a-1} \left\{(c-\gamma)(c+\gamma +1)-(b+\beta)(b-\beta +1)\right\} \\
& & \;\;\;\times \; \left[ \frac{(2c+1)(2a-1)! (-a+b+c)!(b-\beta)!(c+\gamma)!}{(a+b+c+1)!(a-b+c)!(a+b-c)!(b+\beta)!(c-\gamma)!}  \right]^{1/2}
\end{eqnarray}
After the second Chebyshev polynomial $f_L^{(j)} (-(j-1))$ in the ``boxed" term of Eq.(\ref{spinflip2}) is evaluated the parity relation of Eq.(\ref{parity}) (see Section {\bf 2.1.3}), 
the coefficients $c^{\prime}(j,L)= f_L^{(j)} (j) \; f_L^{(j)} (-j) $ in the Fourier-Legendre series expansion of  the spin-flip transition probability $ \mbox{P}^{(j)}_{j-1,-(j-1)}(t) $ in Eq.(\ref{legser22}) are easily obtained, and used to write the explicit form of this expansion in Eq.(\ref{legser23}).

\subsubsection{Multi-level Landau-Zener transition probability}

Recently, Tchouobiap et al. \cite{fai} have solved the multi-level Landau-Zener problem to obtain the following exact analytical expression for the transition 
probability $\mbox{ P}_{mm^{\prime}}^{\mbox{\tiny{LZ}}}(t)  $ between two Zeeman levels of an arbitrary spin $S$: 
\begin{eqnarray}
\mbox{ P}_{mm^{\prime}}^{\mbox{\tiny{LZ}}}(t)   & =  & \sum_{L=0}^{2S} \sqrt{\displaystyle\frac{2L+1}{2S+1}} \left[ \hat{T}_{L0}^{(S)}   \right]_{\!mm}
C_{Sm^{\prime}L0}^{Sm^{\prime}} \; \;_2F_1[-L,L+1,1;1-p(t)] \label{LZtransprob}\;\;\;\;\;\;\; \\
\mbox{where} \;\;\;\; 
\left[ \hat{T}_{L0}^{(S)}   \right]_{\!mm} & \equiv & \langle Sm |\;  \hat{T}_{L0}^{(S)}  \; |Sm \rangle \\
\hat{T}_{LM}^{(S)} & \equiv & \sqrt{\displaystyle\frac{2L+1}{2S+1}} \sum_{m,m^{\prime}} 
C_{SmLM }^{Sm^{\prime}} \; |Sm^{\prime} \rangle \, \langle Sm |
\end{eqnarray}
At first sight, the transition probability expression in Eq.(\ref{LZtransprob}) seems to be quite unrelated in form to the Meckler formula \cite{meckler:majorana}, especially with the appearance of the hypergeometric function \cite{olver} $_2F_1[-L,L+1,1;1-p(t)] \equiv \; _2F_1[a,b,c;d]$, 
matrix elements of the spin polarization operators $\hat{T}_{L0}^{(S)} $, and Clebsch-Gordan coefficients $C_{Sm^{\prime}L0}^{Sm^{\prime}} $. However, it is easy to show that the expression of Eq.(\ref{LZtransprob}) for 
$\mbox{ P}_{mm^{\prime}}^{\mbox{\tiny{LZ}}}(t)  $ is actually identical to Meckler's formula \cite{meckler:majorana}. First, using well-documented \cite{varshal1:ang} symmetry properties of the Clebsch-Gordan coefficients, and 
definitions of the hypergeometric functions \cite{arken,olver},  we can rewrite the summands in Eq.(\ref{LZtransprob}) in terms of either Chebyshev polynomials 
$f_L^{(S)} (m) $ or  Legendre polynomials  $P_L(x)$ as follows:
\begin{eqnarray}
\left[ \hat{T}_{L0}^{(S)}   \right]_{\!mm} & = & \langle Sm| \; f_L^{(S)} (J_z) \;  |Sm \rangle  = f_L^{(S)} (m) \label{summand1}\\
 \sqrt{\displaystyle\frac{2L+1}{2S+1}} \; C_{Sm^{\prime}L0}^{Sm^{\prime}}  & \equiv & 
(-1)^{m^{\prime}-S} \; C_{Sm^{\prime}S-m^{\prime}}^{L0}  = f_L^{(S)} (m^{\prime}) \\
_2F_1[-L,L+1,1;1-p(t)] & \equiv  & P_L[2p(t)-1] \label{summand3}
\end{eqnarray}
In Eq.(\ref{summand3}),  $P_L(x)$ is an $L$-th order Legendre polynomial, and  the two-level Landau-Zener transition probability \cite{fai} $p(t)$ is given by
\begin{equation}
p(t) \equiv \mbox{ P}_{\frac{1}{2}, -\frac{1}{2} } (t)   
\end{equation}
With the results of Eqs.(\ref{summand1}) to  (\ref{summand3}) in hand, the multi-level Landau-Zener transition probability formula derived by 
 Tchouobiap et al. \cite{fai} can now  be written in condensed form as
\begin{equation}
\mbox{ P}_{mm^{\prime}}^{\mbox{\tiny{LZ}}} (t)   =   \sum_{L=0}^{2S} \;  \boxed{f_L^{(S)} (m)} \; \boxed{f_L^{(S)} (m^{\prime})} \; P_L[2\, p(t)-1] \label{LZChebya}
\end{equation}
However, given that the two-level, spin-1/2 transition probability from Meckler's formula \cite{meckler:majorana} is 
\begin{equation}
\mbox{ P}_{\frac{1}{2}, -\frac{1}{2} } (t)  = \frac{1}{2} (1+\cos \beta) \equiv p(t)
\end{equation}
so that 
\begin{equation}
\cos \beta = 2\, p(t) -1
\end{equation}
Meckler's formula  \cite{meckler:majorana} for the multi-level transition probability can be expressed as
\begin{equation}
\mbox{ P}_{mm^{\prime}} (t)   =   \sum_{L=0}^{2S} \;  \boxed{f_L^{(S)} (m)} \; \boxed{f_L^{(S)} (m^{\prime})} \;  \;P_L[2\, p(t)-1] \label{LZCheby}
\end{equation}
which is identical in form to the multi-level Landau-Zener transition probability formula derived by Tchouobiap et al. \cite{fai}.

\subsubsection{Squared Wigner rotation matrix elements}

Varshalovich et al. \cite{varshal1:ang} tabulate (without proof) the following expression for the squares of the ${\cal D}^{(j)}$ matrix elements when
$\beta= \pi/2$:
\begin{equation}
\left[ {\cal D}_{M M^{\prime}}^{(j)}(\alpha,\tfrac{\pi}{2},\gamma)   \right]^2 =e^{-i 2M\alpha -i2M^{\prime}\gamma} \;
(-1)^{M-M^{\prime}} \!\! \sum_{L=0,2,4, \dots} \! \!(-1)^{L/2} \; \displaystyle\frac{(L-1)!!}{L!!} \;
C_{JMJ-M}^{L 0} \;  C_{JM^{\prime}J-M^{\prime}}^{L0} \label{varshalsquare}
\end{equation}
Proving this expression is  trivial using the Majorana formula expressions we have just discussed in Section 
{\bf 4.1}. To begin, note that for any complex number $z \equiv |z| \, e^{i\zeta}$, 
\begin{equation}
z^2 =\left|z\right|^2 e^{2i\zeta} \label{polar}
\end{equation}
so that in particular, 
\begin{equation}
\left[ {\cal D}_{M M^{\prime}}^{(j)}(\alpha,\beta,\gamma)   \right]^2 = 
\left|{\cal D}_{M M^{\prime}}^{(j)}(\alpha,\beta,\gamma)\right|^2  e^{-i 2(M\alpha + M^{\prime}\gamma)} \label{sqmod}
\end{equation}
where  the phase angle of Eq.(\ref{polar}) $\zeta = -(M\alpha +M^{\prime}\gamma)$. To verify the expression for 
$ \left[ {\cal D}_{M M^{\prime}}^{(j)}(\alpha,\tfrac{\pi}{2},\gamma)   \right]^{\!2}$ given in Eq.(\ref{varshalsquare}), it remains to determine $\left|{\cal D}_{M M^{\prime}}^{(j)}(\alpha,\tfrac{\pi}{2},\gamma)\right|^2$, and this can easily be done from the Majorana expansion of Eq.(\ref{majoranaEuler})  evaluated when $\beta = \tfrac{\pi}{2}$:
\begin{eqnarray}
\left|{\cal D}_{m m^{\prime}}^{(j)}(\alpha,\tfrac{\pi}{2},\gamma) \right|^2 & = & 
\sum_{L=0}^{2j} \boxed{f_L^{(j)} (m)}\;  \boxed{f_L^{(j)} (m^{\prime})} \;  P_L(\cos \tfrac{\pi}{2}) \label{boxedCheby} \\
& = & \sum_{L=0}^{2j} \boxed{(-1)^{j-m} \, C^{L0}_{jmj-m}} \; 
\boxed{(-1)^{j-m^{\prime}}\,  C^{L0}_{jm^{\prime}j-m^{\prime}}} \; P_L(0) \label{Peval} \\
& = & (-1)^{m-m^{\prime}}  \sum_{L=0,2,4, \ldots} (-1)^{L/2} \; \displaystyle\frac{(L-1)!!}{L!!} \;
C^{L0}_{jmj-m} \; C^{L0}_{jm^{\prime}j-m^{\prime}} \label{restrict}
\end{eqnarray}

The parity of the Legendre polynomials $P_L(\cos \beta)$ is even or odd, depending on whether $L$ is even or odd. Those polynomials with odd parity will vanish at the origin, and therefore the only non-vanishing values of the Legendre polynomials $P_L(\cos \beta)$  evaluated at the origin (when 
$\cos \beta \equiv \cos ( \pi/2) =0$) are given by \cite{arken}
\begin{equation}
P_{2n}(0) = (-1)^n \;  \frac{(2n-1)!!}{(2n)!!}
\end{equation}
This identity restricts the sum in Eq.(\ref{restrict}) to even values of $L=2n$, and has also been used to evaluate 
$P_L(0)$ in Eq.(\ref{Peval}). The ``boxed" Chebyshev polynomial terms in  Eq.(\ref{boxedCheby}) have been replaced by their Clebsch-Gordan coefficient equivalents in Eq.(\ref{Peval}). After substitution of the expression 
for $\left|{\cal D}_{m m^{\prime}}^{(j)}(\alpha,\tfrac{\pi}{2},\gamma)\right |^2$ given in Eq.(\ref{restrict}) in 
Eq.(\ref{sqmod}), we obtain the relation of Eq.(\ref{varshalsquare}). 

\subsubsection{Squared reduced Wigner rotation matrix elements}

\paragraph{Using Meckler's formula to expand $ \left[d^{(j)}_{m m^{\prime}}(\beta)\right]^{\!2}$ in a Fourier-Legendre series.  } In the Condon and Shortley phase convention \cite{condonshortley}, the reduced rotation matrices $ d^{(j)}_{m m^{\prime}}(\beta)$ are 
real, so that there is no distinction between the modulus squared of these matrices and their square:
\begin{equation}
\left|{\cal D}_{m m^{\prime}}^{(j)}(\alpha,\beta,\gamma) \right|^2 = \left|d^{(j)}_{m m^{\prime}}(\beta)\right|^{2} \equiv 
  \left[d^{(j)}_{m m^{\prime}}(\beta)\right]^{\!2} \label{equivalence}
\end{equation}
Therefore, Meckler's expression \cite{meckler:majorana,meckler:angular} for the Majorana transition probability \cite{emajorana} given in Eq.(\ref{majoranaEuler})  can be rewritten as a Fourier-Legendre series for the squared reduced rotation matrix elements
\begin{eqnarray}
 \left[d^{(j)}_{m m^{\prime}}(\beta)\right]^{\!2} & = &
\sum_{L=0}^{2j} \boxed{f_L^{(j)} (m)} \;  \boxed{f_L^{(j)} (m^{\prime})} \;  P_L(\cos \beta) \label{relaxationa} \\
& = & (-1)^{m-m^{\prime}} \sum_{L=0}^{2j} 
 \boxed{ C_{jmj-m}^{L0}} \;
 \boxed{C_{jm^{\prime}j-m^{\prime}}^{L0} }
\;  P_L(\cos \beta) \label{relaxation}
\end{eqnarray}
Just as we simplified the expansion of $\left|{\cal D}_{m m^{\prime}}^{(j)}(\alpha,\tfrac{\pi}{2},\gamma)\right |^2$   in
 Eq.(\ref{boxedCheby}), the ``boxed" Chebyshev polynomial terms in  Eq.(\ref{relaxationa}) have been replaced by their Clebsch-Gordan coefficient equivalents in Eq.(\ref{relaxation}). A version of this latter relation,  adapted to accomodate inversion symmetry of coordinate frame rotations (so that $L=0,2,4,\ldots$), proved  indispensable in a theoretical analysis \cite{brown} of  NMR  spin-lattice relaxation of $^2$H and $^{14}$N nuclei in lipid bilayers and membrane systems.  
\paragraph{Inverting Meckler's formula}
If we were to view Eq.(\ref{relaxationa}) as a system of equations for the unknowns $P_L(\cos \beta)$, then we could use standard matrix techniques to solve for these unknowns. However, a more direct approach is possible using the properties of the basis functions $f_L ^{(j)} ({\bf \hat{n}}  \cdot {\bf J})$. The trace relation of Eq.(\ref{traceLegendre}) 
\begin{equation}
 \mbox{Tr} \! \left[  f_L ^{(j)} ({\bf \hat{n}}  \cdot {\bf J}) \;  
f_{L^{\prime}} ^{(j)} ({\bf \hat{n}}^{\prime}  \cdot {\bf J}) \right] =  \delta_{LL^{\prime}} \; P_L( {\bf \hat{n}} \cdot  {\bf \hat{n}}^{\prime}) 
\end{equation}
can be rewritten as follows: 
\begin{eqnarray}
&  & \delta_{LL^{\prime}} \; P_L( {\bf \hat{n}} \cdot  {\bf \hat{n}}^{\prime}) \nonumber \\
& = & \! \!\!\mbox{Tr} \! \left[ \; \boxed{f_L ^{(j)} ({\bf \hat{n}}  \cdot {\bf J})} \;  
\boxed{f_{L^{\prime}} ^{(j)} ({\bf \hat{n}}^{\prime}  \cdot {\bf J})} \; \right] \label{inverse1} \\
& = &
\! \!\!\mbox{Tr} \! \left[ \; \boxed{\displaystyle\sum_{m=-j}^{j} f_L^{(j)} (m)  \;   \mbox{{\boldmath $\Pi$}}^{(j)}(m,{\bf \hat{n}})         } \;
 \boxed{\displaystyle\sum_{m^{\prime}=-j}^{j} f_{L^{\prime}}^{(j)} (m^{\prime})  \;   
\mbox{{\boldmath $\Pi$}}^{(j)}(m^{\prime},{\bf \hat{n}}^{\prime})          } \; \right] \label{inverse2} \\
& = & \sum_{m, \, m^{\prime}=-j}^j \,  f_L^{(j)} (m) \,  f_{L^{\prime}}^{(j)}(m^{\prime}) 
\; \mbox{Tr} \! \left[  \mbox{{\boldmath $\Pi$}}^{(j)}(m,{\bf \hat{n}})   \; \mbox{{\boldmath $\Pi$}}^{(j)}(m^{\prime},{\bf \hat{n}}^{\prime})    \right] \label{inverse3} \\
& = & \sum_{m,\, m^{\prime}=-j}^j \,  f_L^{(j)} (m) \,  f_{L^{\prime}}^{(j)}(m^{\prime})  
\left[d^{(j)}_{m m^{\prime}}(\beta)\right]^{\!2} \label{inverse4}
\end{eqnarray} 
Each of the ``boxed" basis functions $f_L ^{(j)} ({\bf \hat{n}}  \cdot {\bf J})$ in Eq.(\ref{inverse1}) has been replaced in Eq.(\ref{inverse2}) by the corresponding Sylvester's formula \cite{merzbacher2, horn} expansions from Section {\bf 3.2.1}, while the trace of the projection operator product in Eq.(\ref{inverse3}) has been reduced to the square of reduced matrix elements 
$\left[d^{(j)}_{m m^{\prime}}(\beta)\right]^{\!2}$ 
using Eqs.(\ref{puzzle}) and (\ref{equivalence}). Taking advantage of the delta function on the left-hand side of Eq.(\ref{inverse1}), we arrive at the final result for the unknowns $P_L(\cos \beta)$:
\begin{equation}
P_L(\cos \beta) = \sum_{m, \, m^{\prime}=-j}^j \,  f_L^{(j)} (m) \,  f_{L}^{(j)}(m^{\prime})  
\left[d^{(j)}_{m m^{\prime}}(\beta)\right]^{\!2} \label{intriguing}
\end{equation}
This inverse of Meckler's \cite{meckler:majorana} Majorana formula for the spin transition probability is an intriguing expression, because the Legendre polynomial $P_L(\cos \beta) $ is clearly $j$-independent, whereas every expansion term  is $j$-dependent.   Once $L$ is fixed, the elements of any reduced matrix
 $d^{(j)}_{m m^{\prime}}(\beta)$  (arbitrary $j$) suffice to calculate the expansion, with expansion coefficients given by the product of appropriate $j$-dependent  Chebyshev polynomials $ f_L^{(j)} (m) $. The summation of Eq.(\ref{intriguing}), as well as other closely related summations \cite{lai} involving 
$\left[d^{(j)}_{m m^{\prime}}(\beta)\right]^{\!2} $ and $P_L(\cos \beta) $, may be derived using the Clebsch-Gordan series discussed in Section {\bf 4.1.2} and the orthogonality condition of the Clebsch-Gordan coefficients \cite{varshal1:ang}. 

 Beyond the well-known symmetry properties of the reduced Wigner rotation matrix elements $d^{(j)}_{m m^{\prime}}(\beta)$
 \cite{varshal1:ang}, the relation 
of Eq.(\ref{intriguing}) puts an additional  constraint on the values of the $(2j+1) \times (2j+1)$ matrix $d^{(j)}_{m m^{\prime}}(\beta)$. When $L=0$, the simplest version of this constraint is
\begin{eqnarray}
P_0(\cos \beta) \equiv 1  & = & \sum_{m , \, m^{\prime} =-j}^j \,  f_0^{(j)} (m) \,  f_{0}^{(j)}(m^{\prime})  
\left[d^{(j)}_{m m^{\prime}}(\beta)\right]^{\!2} \label{intrigue2} \\
 & = & \displaystyle\frac{1}{2j+1} \;  \sum_{m , \, m^{\prime} =-j}^j \,  \left[d^{(j)}_{m m^{\prime}}(\beta)\right]^{\!2} \label{intrigue3} \\
\mbox{since \cite{corio:ortho, filippov2:thesis}} \;\;\;\; f_0^{(j)} (m) & =  & \displaystyle\frac{1}{\sqrt{2j+1}}
\end{eqnarray} 
From the constraint of Eq.(\ref{intrigue3}), it follows that for any reduced matrix $d^{(j)}_{m m^{\prime}}(\beta)$, the sum of all of its squared elements is $2j+1$:
\begin{equation}
\sum_{m,\, m^{\prime}=-j}^j \,  \left[d^{(j)}_{m m^{\prime}}(\beta)\right]^{\!2}  = 2j+1 \label{nosurprise}
\end{equation}
But this relation  is actually just a consequence of the unitarity of the rotation 
operator $\hat{{\cal D}}^{(j)}(R)$. A special case of the unitarity sum of the rotation matrices ${\cal D}_{m m^{\prime}}^{(j)}(\alpha,\beta,\gamma)$ is the following orthogonality sum of the (real) reduced rotation matrix elements \cite{thompson}
\begin{equation}
\sum_{m^{\prime \prime}=-j}^j d^{(j)}_{m^{\prime} m^{\prime \prime}}(\beta) \; d^{(j)}_{m m^{\prime \prime}}(\beta) = \delta_{m^{\prime} m} \label{dorthog}
\end{equation}
Setting $m^{\prime} = m$ in Eq.(\ref{dorthog}) leads to the relation
\begin{equation}
\sum_{m^{\prime \prime}=-j}^j d^{(j)}_{m^{\prime} m^{\prime \prime}}(\beta) \; d^{(j)}_{m^{\prime}  m^{\prime \prime}}(\beta) \equiv \sum_{m^{\prime \prime}=-j}^j \left[d^{(j)}_{m^{\prime} m^{\prime \prime}}(\beta) \right]^{\!2} = 1
\end{equation}
which when summed over both sides, leads again to the identity of Eq.(\ref{nosurprise}) 
\begin{equation}
\sum_{m^{\prime} \!, \, m^{\prime \prime} =-j}^j \left[d^{(j)}_{m^{\prime} m^{\prime \prime}}(\beta) \right]^{\!2} = 2j+1
\end{equation}

\section{Operator Expansions}

\subsection{Rotation Operator}

Developing polynomial operator expressions in the variable $({\bf \hat{n}} \cdot {\bf J})$  for the rotation operator 
 $\hat{{\cal D}}^{(j)} \! (\psi, {\bf \hat{n}}) = e^{i \psi({\bf \hat{n}} \cdot {\bf J}) }$ is a challenging problem. Over the last five decades, it has been solved by a variety of ingenious methods \cite{corio:siam,wageningen:rotoppoly, webwill:spinmatrixpol, albert:rotgroup, happer, varshal2:expansion,torruella:rotgroup,  
kusnezov:sun, curtright}, but the solution presented by Corio \cite{corio:siam,corio:correction}  
\begin{eqnarray}
\hat{{\cal D}} ^{(j)}\! (\psi, {\bf \hat{n}}) \equiv e^{i \psi({\bf \hat{n}} \cdot {\bf J}) }  
& =  & \sum_{\lambda=0}^{2j}  \mbox{Tr} \!  \left[ f^{(j)}_{\lambda}({\bf \hat{n}} \cdot {\bf J})  \; \hat{{\cal D}}^{(j)} \! (\psi, {\bf \hat{n}}) \right]  
   f_{\lambda}^{(j)}({\bf \hat{n}} \cdot {\bf J}) \label{coriotracrel} \\
& =  & \sum_{\lambda=0}^{2j} a_{\lambda}^{(j)}(\psi) \;  f_{\lambda}^{(j)}({\bf \hat{n}} \cdot {\bf J}) \label{coriover}\\
& = & \sum_{\lambda=0}^{2j} i^{\lambda} \sqrt{\displaystyle\frac{2\lambda+1}{2j+1}}\, \mbox{{\large $\chi$}}_{\lambda}^{(j)}(\psi)   \; f_{\lambda}^{(j)}({\bf \hat{n}} \cdot {\bf J}) \label{coriorotfirst}
\end{eqnarray} 
is unique because the basis operator polynomials $ f_{\lambda}^{(j)}({\bf \hat{n}} \cdot {\bf J})$ used by Corio 
\cite{corio:siam,corio:correction}  define an orthonormal set \cite{corio:siam} 
\begin{equation}
\mbox{Tr} \! \left[ f^{(j)}_{\lambda}({\bf \hat{n}} \cdot {\bf J}) \; f^{(j)}_{{\lambda}^{\prime}}({\bf \hat{n}} \cdot {\bf J}) \right] = 
\delta_{\lambda \lambda^{\prime}} 
\end{equation}
and because these polynomials first introduced in a physics application  by Meckler \cite{meckler:angular} are the Chebyshev polynomial operators  
$ f_{\lambda}^{(j)}({\bf \hat{n}} \cdot {\bf J})$.  
The general form of Corio's Chebyshev polynomial operator expansion \cite{corio:siam} given in Eq.(\ref{coriover})  I have modified in Eq.({\ref{coriorotfirst}) in order to introduce the  generalized characters $\mbox{{\large $\chi$}}_{\lambda}^{(j)}(\psi) $ \cite{varshal1:ang}. The canonical differential relation \cite{varshal1:ang} which defines the generalized characters $\mbox{{\large $\chi$}}_{\lambda}^{(j)}(\psi) $ in terms of the Gegenbauer polynomials 
\cite{ tem:bk, askey} $\mbox{ C}_{2j}^{1}(c)$   is
\begin{eqnarray}
\mbox{{\large $\chi$}}_{\lambda}^{(j)}(\psi)  & = & \sqrt{2j+1} \; \sqrt{\displaystyle\frac{(2j-\lambda)!}{(2j+\lambda+1)!}}\; s^{\lambda}
\left( \displaystyle\frac{d}{dc  }\right)^{\!\!\lambda}  \mbox{{\large $\chi$}}^{(j)}(\psi) \label{diffchar} \\
\mbox{where} \;\;\;\;\mbox{{\large $\chi$}}^{(j)}(\psi)  & = & \mbox{ C}_{2j}^{1}(c) \\
s & = &  \sin(\psi/2) \\
c & = & \cos(\psi/2)
\end{eqnarray}
Well-documented properties of the Gegenbauer polynomials \cite{ tem:bk, askey} can then be used to derive the following Gegenbauer polynomial 
definition  \cite{varshal1:ang}  of the  generalized characters $\mbox{{\large $\chi$}}_{\lambda}^{(j)}(\psi) $ which we will use in Section {\bf 6.1.1}:
\begin{equation}
\mbox{{\large $\chi$}}_{\lambda}^{(j)}(\psi) =  (2\lambda)!!\sqrt{2j+1}\sqrt{\frac{(2j-\lambda)!}
{(2j+\lambda+1)!}}
\;\; s^{\lambda}\mbox{ C}_{2j-\lambda}^{\lambda+1}(c)
\end{equation}
It is instructive to compare the  differential relation of  Eq.(\ref{diffchar}) with the trace relation obtained from Eq.(\ref{coriotracrel}) that  defines the generalized characters $\mbox{{\large $\chi$}}_{\lambda}^{(j)}(\psi) $ in terms of the Chebyshev polynomial operators 
$f^{(j)}_{\lambda}({\bf \hat{n}} \cdot {\bf J})$ and the rotation operator $\hat{{\cal D}}^{(j)} \! (\psi, {\bf \hat{n}}) $
\begin{equation}
 \mbox{{\large $\chi$}}_{\lambda}^{(j)}(\psi)  =    i^{3 \lambda}  \sqrt{\displaystyle\frac{2j+1}{2\lambda+1}} \; 
\mbox{Tr}\!  \left[ f^{(j)}_{\lambda}({\bf \hat{n}} \cdot {\bf J})  \; \hat{{\cal D}}^{(j)}\! (\psi, {\bf \hat{n}}) \right] 
\end{equation}
Neither definition is more fundamental, but it goes without saying that calculating traces is far simpler than calculating derivatives.

The important role that the Chebyshev polynomial operators 
$f^{(j)}_{\lambda}({\bf \hat{n}} \cdot {\bf J})$ play in representations of the rotation group is evident in the expansion coefficients as defined by Corio 
\cite{corio:siam}
\begin{equation}
a_{\lambda}^{(j)}(\psi)= \boxed{\mbox{Tr} \! \left[ f^{(j)}_{\lambda}({\bf \hat{n}} \cdot {\bf J})  \; \hat{{\cal D}}^{(j)} \! (\psi, {\bf \hat{n}}) \right] =
 i^{\lambda} \sqrt{\displaystyle\frac{2\lambda+1}{2j+1}}\, \mbox{{\large $\chi$}}_{\lambda}^{(j)}(\psi) } \label{charcheby}
\end{equation}
which are proportional to the generalized characters $\mbox{{\large $\chi$}}_{\lambda}^{(j)}(\psi) $ \cite{varshal1:ang} of the rotation group. The first equality of Eq.(\ref{charcheby}) is due to Corio 
\cite{corio:siam}, but the 
``boxed" relation of Eq.(\ref{charcheby}) is new, and, as we illustrate with an example in Appendix D,  can be used as another definition of the generalized characters $ \mbox{{\large $\chi$}}_{\lambda}^{(j)}(\psi)$. Several other definitions of $ \mbox{{\large $\chi$}}_{\lambda}^{(j)}(\psi)$ are tabulated in Varshalovich et al. \cite{varshal1:ang}.   Putting 
$\lambda=0$ in this ``boxed" relation, we recover the character $\mbox{{\large $\chi$}}^{(j)}(\psi) $ of the irreducible representation $\hat{{\cal D}}^{(j)}(\psi, {\bf \hat{n}})$
\begin{equation}
\mbox{Tr} \! \left[  \hat{{\cal D}}^{(j)} \! (\psi, {\bf \hat{n}}) \right] =
 \mbox{{\large $\chi$}}^{(j)}(\psi) 
\end{equation}
using the facts that \cite{ varshal1:ang, corio:siam} 
\begin{eqnarray}
f_0^{(j)}({\bf \hat{n}} \cdot {\bf J}) &  =  & \displaystyle\frac{ \mathds{1} }{\sqrt{2j+1} } \\
\mbox{{\large $\chi$}}_{0}^{(j)}(\psi) & \equiv & \mbox{{\large $\chi$}}^{(j)}(\psi) 
\end{eqnarray}

By taking the trace of the rotation operator $\hat{{\cal D}}^{(j)} \! (\psi, {\bf \hat{n}})$ expanded in the Chebyshev polynomial operator basis 
$f_{\lambda}^{(j)} \! \left( {\bf \hat{n}} \cdot {\bf J} \right)$, the character $ \mbox{{\large $\chi$}}^{(j)}(\psi) $ of irreducible representations of the rotation group can be obtained.  Using Corio's expansion \cite{corio:siam} of the rotation operator $\hat{{\cal D}}^{(j)}  \!(\psi, {\bf \hat{n}}) = e^{i \psi({\bf \hat{n}} \cdot {\bf J}) }$, 
this trace can be evaluated as
\begin{eqnarray}
\mbox{Tr} \! \left[ \hat{{\cal D}}^{(j)} \! (\psi, {\bf \hat{n}}) \right]  
& = & \sum_{\lambda=0}^{2j} i^{\lambda} \sqrt{\displaystyle\frac{2\lambda+1}{2j+1}}\, \mbox{{\large $\chi$}}_{\lambda}^{(j)} (\psi)   \; 
\boxed{\mbox{Tr}\! \left[  f_{\lambda}^{(j)} ({\bf \hat{n}} \cdot {\bf J}) \right] } \label{chartrace}\\
& = & \sum_{\lambda=0}^{2j} i^{\lambda} \sqrt{\displaystyle\frac{2\lambda+1}{2j+1}}\, \mbox{{\large $\chi$}}_{\lambda}^{(j)} (\psi)   \; 
\boxed{\sqrt{2j+1} \; \delta_{\lambda 0}} \label{chartraceb} \\
& = & \frac{1}{\sqrt{2j+1}} \;  \mbox{{\large $\chi$}}_{0}^{(j)} (\psi) \;  \sqrt{2j+1} \\
& = & \mbox{{\large $\chi$}}^{(j)} (\psi) \\
 \mbox{where \cite{varshal1:ang}}\;\;\;\;   \mbox{{\large $\chi$}}_{0}^{(j)} (\psi) & \equiv &  \mbox{{\large $\chi$}}^{(j)} (\psi)
\end{eqnarray}
The ``boxed" trace term in Eq.(\ref{chartrace}) has been replaced by the ``boxed" term of Eq.(\ref{chartraceb}) by calculating the trace in  a representation where $({\bf \hat{n}} \cdot {\bf J}) $ is diagonal. Then we find
\begin{eqnarray}
\mbox{Tr} \! \left[  f_{\lambda}^{(j)} ({\bf \hat{n}} \cdot {\bf J}) \right] & = & \sum_m \langle m |\, f_{\lambda}^{(j)} (J_z) \, |m \rangle \\
& = & \sum_m f_{\lambda}^{(j)} (m) \\
& = & \sqrt{2j+1} \; \delta_{\lambda 0} \;\;\; \;\;(\mbox{using Eq.(\ref{chebsum})}
\end{eqnarray}

The Chebyshev polynomial expansions of the projection operators (see Eqs.(\ref{projectjz}) and (\ref{projectJn}) of Section {\bf 3.3.1})  are particularly effective when they are used with Sylvester's formula 
 \cite{merzbacher2, horn}.  We illustrate this effectiveness with a simple alternative derivation of the Chebyshev polynomial operator expansion of the rotation operator \cite{corio:siam} in the version given in Eq.(\ref{coriorotfirst}).  We begin by exploiting Sylvester's formula \cite{merzbacher2, horn} to write the exponential matrix operator as a sum of projection operators in Eq.(\ref{projj}), and then use the Chebyshev polynomial expansion 
(see Eq.(\ref{projectJn}) of Section {\bf 3.3.1})  of the projection operator {\boldmath $\Pi$}$^{(j)}(m,{\bf \hat{n}}) $ 
\begin{eqnarray}
e^{-i \psi({\bf \hat{n}} \cdot {\bf J}) } & = & \sum_{m=-j}^j e^{-im\psi} \; \boxed{\mbox{{\boldmath $\Pi$}}^{(j)}(m,{\bf \hat{n}})  } \label{projj}\\
& = &  \sum_{m=-j}^j  e^{-im\psi} \,  \boxed{\displaystyle\sum_{\lambda=0}^{2j} f_{\lambda}^{(j)} (m) \; f_{\lambda} ^{(j)} ({\bf \hat{n}} \cdot {\bf J})} \label{projj2} \\
& = &  \displaystyle\sum_{\lambda=0}^{2j} \boxed{ \sum_{m=-j}^j e^{-im\psi} \,  f_{\lambda}^{(j)} (m)} \;
 f_{\lambda} ^{(j)} ({\bf \hat{n}} \cdot {\bf J}) \label{coriosylvester}  \\
& = &   \displaystyle\sum_{\lambda=0}^{2j} \boxed{ i^{-\lambda}  \, \sqrt{\displaystyle\frac{2\lambda+1}{2j+1}}\; \mbox{{\large $\chi$}}_{\lambda}^{(j)}(\psi)} \; 
f_{\lambda} ^{(j)} ({\bf \hat{n}} \cdot {\bf J})  \label{coriorotproj}
\end{eqnarray}
The projection operator {\boldmath $\Pi$}$^{(j)}(m,{\bf \hat{n}}) $ in the ``boxed" term of Eq.(\ref{projj}) has been replaced by its Chebyshev polynomial expansion in the ``boxed" term of Eq.(\ref{projj2}). The sum in the ``boxed" term of Eq.(\ref{coriosylvester}) has been reexpressed as the ``boxed" term of Eq.(\ref{coriorotproj}) by converting a  trigonometric series identity \cite{varshal1:ang}  for the generalized characters of the rotation group 
$\mbox{{\large $\chi$}}_{\lambda}^{(j)}(\psi) $ into a Chebyshev polynomial series as follows:
\begin{eqnarray}
\mbox{{\large $\chi$}}_{\lambda}^{(j)}(\psi) & = & i^{\lambda}  \sum_{m=-j}^j \; e^{-im\psi} \,  \boxed{C_{jm\lambda 0}^{jm}} \label{charCG}\\
& = &  i^{\lambda}  \sum_{m=-j}^j \; e^{-im\psi} \, \boxed{\sqrt{\displaystyle\frac{2j+1}{2\lambda+1}}(-1)^{j-m}\;  C_{jmj-m}^{\lambda 0}}
\label{charCG2} \\
& = &  i^{\lambda}  \sum_{m=-j}^j \; e^{-im\psi} \, \sqrt{\displaystyle\frac{2j+1}{2\lambda+1}} \; f_{\lambda}^{(j)} (m) 
\label{charCG3} 
\end{eqnarray}
Well-documented \cite{varshal1:ang} symmetry properties of the Clebsch-Gordan coefficients have been used to reexpress the 
``boxed" term in Eq.(\ref{charCG}) as the ``boxed" term in Eq.(\ref{charCG2}). The Clebsch-Gordan coefficient and phase factor in the 
``boxed" term of Eq.(\ref{charCG2}) have been replaced by their Chebyshev polynomial equivalent $ f_{\lambda}^{(j)} (m)$ in Eq.(\ref{charCG3}). Aside from the use of Sylvester's formula \cite{merzbacher2, horn}, which is not particularly unique or novel, the novelty of this derivation is that it exploits some of the most felicitous properties  of the Chebyshev polynomials, namely the Chebyshev polynomial operator expansion of the projection operator, and the duality of the Chebyshev polynomials as Clebsch-Gordan coefficients. 

Remarkably, it is clear in retrospect that  in 1967, Albert \cite{albert:rotgroup}  had anticipated Corio's \cite{corio:siam} Chebyshev polynomial operator expansion of the rotation operator (see Eqs.(\ref{coriover} and (\ref{coriorotfirst})). In modern notation, Albert's expansion of $\exp[i \psi J_z]$ as a finite sum of irreducible tensor components 
$\hat{T}_{\lambda 0}^{(J)} $ can be expressed as 
\begin{eqnarray}
e^{i \psi J_z } & = & \sum_{\lambda=0}^{2j} i^{\lambda} \sqrt{\displaystyle\frac{2\lambda+1}{2j+1}}\, \mbox{{\large $\chi$}}_{\lambda}^{(j)}(\psi)   \; \boxed{ \hat{T}_{\lambda 0}^{(j)} } \label{albert1} \\
& = & \sum_{\lambda=0}^{2j} i^{\lambda} \sqrt{\displaystyle\frac{2\lambda+1}{2j+1}}\, \mbox{{\large $\chi$}}_{\lambda}^{(j)}(\psi)   \; \boxed{ f_{\lambda}^{(j)} \! (J_z)} 
\label{corioalbert1}
\end{eqnarray}
The ``boxed" irreducible tensor operator term in 
Eq.(\ref{albert1}) has been reexpressed as the ``boxed" Chebyshev polynomial operator term in Eq.(\ref{corioalbert1}) using the operator equivalence of Eq.(\ref{equivfT}). By considering the unitary transform of both sides of Eq.({\ref{corioalbert1}), and using the results of similarity transforms summarized in Table V,  we then obtain 
Corio's \cite{corio:siam} Chebyshev polynomial operator expansion of the rotation operator 
\begin{eqnarray}
& & \hat{{\cal D}}^{(j)}(\theta, {\bf \hat{n}}_{\bot}) \, e^{i \psi J_z }  \left [\hat{{\cal D}}^{(j)} (\theta, {\bf \hat{n}}_{\bot}) \right ]^{\!\dagger} \\
& = & \exp \! \left\{i \psi\; \hat{{\cal D}}^{(j)}(\theta, {\bf \hat{n}}_{\bot}) \, J_z   \left [\hat{{\cal D}}^{(j)} (\theta, {\bf \hat{n}}_{\bot}) \right ]^{\!\dagger} \right\} \\
& = & e^{i \psi({\bf \hat{n}} \cdot {\bf J}) } \\
& = & \sum_{\lambda=0}^{2j} i^{\lambda} \sqrt{\displaystyle\frac{2\lambda+1}{2j+1}}\, \mbox{{\large $\chi$}}_{\lambda}^{(j)}(\psi)   
\left\{\hat{{\cal D}}^{(j)}(\theta, {\bf \hat{n}}_{\bot}) \, \hat{T}_{\lambda 0}^{(j)}  \left [\hat{{\cal D}}^{(j)} (\theta, {\bf \hat{n}}_{\bot}) \right ]^{\!\dagger} \right\}
 \label{albert11} \\
& = & \sum_{\lambda=0}^{2j} i^{\lambda} \sqrt{\displaystyle\frac{2\lambda+1}{2j+1}}\, \mbox{{\large $\chi$}}_{\lambda}^{(j)}(\psi) \;
\left\{ \hat{{\cal D}}^{(j)}(\theta, {\bf \hat{n}}_{\bot})  \;  f_{\lambda}^{(j)} \! (J_z) \; \left [\hat{{\cal D}}^{(j)} (\theta, {\bf \hat{n}}_{\bot}) \right ]^{\!\dagger} \right\} \\
& = & \sum_{\lambda=0}^{2j} i^{\lambda} \sqrt{\displaystyle\frac{2\lambda+1}{2j+1}}\, \mbox{{\large $\chi$}}_{\lambda}^{(j)}(\psi) \;  f_{\lambda}^{(j)} ({\bf \hat{n}} \cdot {\bf J} ) 
\label{corioalbert11}
\end{eqnarray}

It is astonishing that Corio's \cite{corio:siam,corio:correction}  Chebyshev polynomial operator $f_{\lambda}^{(j)} ({\bf \hat{n}} \cdot {\bf J})$  expansion of the rotation operator 
$\hat{{\cal D}} ^{(j)}\! (\psi, {\bf \hat{n}})$ in Eq.(\ref{coriorotfirst}) is hardly known at all, due in no small part to the fact that this expansion is not mentioned or discussed in any angular momentum or quantum mechanics textbook. And yet this expansion emphasizes the fact the Chebyshev polynomial operators 
$f_{\lambda}^{(j)} ({\bf \hat{n}} \cdot {\bf J})$ play a unique role in irreducible representations of the rotation group. In Section {\bf 5.3}, we will exploit this expansion to derive a tomographic reconstruction relation for the density operator $\hat \rho$ using the Chebyshev polynomial operators $f_{\lambda}^{(j)} ({\bf \hat{n}} \cdot {\bf J})$.

\subsection{Stratonovich-Weyl Operator}

In the phase-space approach to spin, the conventional quantum mechanical operators are replaced by functions on the classical phase-space of 
the unit sphere ${\bf S}^2$. Central to this correspondence between Hilbert space operators and functions on the phase-space is the Stratonovich-Weyl operator  \cite{varilly2:moyal,heissweigert}, also known as the Wigner-Stratonovich-Agarwal operator \cite{klimov:distr,agarwal,klimovchumakov}. 
Beginning with the conventional representation \cite{varilly2:moyal, klimov:distr, klimovchumakov}, this operator (kernel)  $\Delta^{(j)}({\bf \hat{n}})$   can be expanded in terms of the Chebyshev polynomial operator basis $f_{\lambda}^{(j)} ({\bf \hat{n}} \cdot {\bf J})$ as follows
\begin{eqnarray}
\Delta^{(j)}({\bf \hat{n}}) & = & \sum_{\lambda=0}^{2j} \sum_{\mu = -\lambda}^{\lambda} \sqrt{\displaystyle\frac{4\pi}{2j+1}}
 \; \boxed{\left[\hat{T}^{(j)}_{\lambda\mu} \right]^{\!\dagger}}\;  Y_{\lambda\mu}({\bf \hat{n}}) \label{dagrep} \\
 & = & \sum_{\lambda=0}^{2j} \sum_{\mu = -\lambda}^{\lambda} \sqrt{\displaystyle\frac{4\pi}{2j+1}}
 \; \hat{T}^{(j)}_{\lambda\mu}  \; Y^{\star}_{\lambda\mu}({\bf \hat{n}}) \\
& = & \frac{1}{\sqrt{2j+1}} \sum_{\lambda=0}^{2j} \sqrt{2\lambda+1}\;\, \boxed{ {\bf C}_{\lambda}({\bf \hat{n}}) \cdot {\bf T}_{\lambda}({\bf J})   } \label{stratopexpa}\\
& = & \frac{1}{\sqrt{2j+1}} \sum_{\lambda=0}^{2j} \sqrt{2\lambda+1}\; \, \boxed{f_{\lambda}^{(j)} ({\bf \hat{n}} \cdot {\bf J})} \label{stratopexp}
\end{eqnarray}
 where the ``boxed" term of Eq.(\ref{dagrep}) has been reexpressed using the following relation \cite{varshal1:ang}  for the adjoint of the polarization operators
\begin{equation}
\left[\hat{T}^{(j)}_{\lambda\mu} \right]^{\!\dagger} = (-1)^{\mu} \; \hat{T}^{(j)}_{\lambda -\mu}
\end{equation}
To arrive at the result of Eq.(\ref{stratopexp}), we have reexpressed the canonical version of the  Stratonovich-Weyl operator  
\cite{varilly2:moyal, klimov:distr, klimovchumakov}  in terms of a  direct product of spin and spatial tensors in Eq.(\ref{stratopexpa}). The ``boxed" tensor direct product term in 
  Eq.(\ref{stratopexpa}) has been reexpressed as the ``boxed" Chebyshev polynomial operator term in Eq.(\ref{stratopexp}) using a recoupling relation which we derive in Section {\bf 6.1}. 

\subsubsection{$\Delta^{(j)}({\bf \hat{n}})$ Operator Trace}

Using the Chebyshev polynomial operator expansion of Eq.(\ref{stratopexp}), the  Stratonovich-Weyl operator  $\Delta^{(j)}({\bf \hat{n}})$ \cite{varilly2:moyal,heissweigert} trace is easly evaluated using the properties of the Chebyshev polynomials. We find 
  
\begin{eqnarray}
\mbox{Tr} \left[   \Delta^{(j)}({\bf \hat{n}}) \right] & = & 
\frac{1}{\sqrt{2j+1}} \sum_{\lambda=0}^{2j} \sqrt{2\lambda+1}\; \,  \mbox{Tr}  \left[   f_{\lambda}^{(j)} ({\bf \hat{n}} \cdot {\bf J}) \right] \\
& = & 
\frac{1}{\sqrt{2j+1}} \sum_{\lambda=0}^{2j} \sqrt{2\lambda+1}\; \,  \mbox{Tr}  \left[   f_{\lambda}^{(j)}(J_z) \right] \\
& = & 
\frac{1}{\sqrt{2j+1}} \sum_{\lambda=0}^{2j} \sqrt{2\lambda+1}\; \,  \sum_m \langle jm| \,  f_{\lambda}^{(j)}(J_z)  \, | jm \rangle \\
& = & 
\frac{1}{\sqrt{2j+1}} \sum_{\lambda=0}^{2j} \sqrt{2\lambda+1}\; \, \boxed{ \sum_m  f_{\lambda}^{(j)}(m) } \label{sumdeltaltrace}\\
  & = & 
\frac{1}{\sqrt{2j+1}} \sum_{\lambda=0}^{2j} \sqrt{2\lambda+1}\; \, \sqrt{2j+1} \, \delta_{\lambda 0} \\
& = & 1
\end{eqnarray}
The sum over the Chebyshev polynomials  in the ``boxed" term of Eq.(\ref{sumdeltaltrace}) was evaluated using Eq.(\ref{chebsum}) of Section {\bf 3.2.1}. 
\subsubsection{The Reproducing Kernel for the $\Delta^{(j)}({\bf \hat{n}})$ Operators }
Among the most important properties of the Stratonovich-Weyl operator kernel is the traciality condition \cite{varilly2:moyal, heissweigert}
\begin{eqnarray}
 \Delta^{(j)}({\bf \hat{n}}) & = & \frac{(2j+1)}{4\pi} \int_{{\bf S}^2} \mbox{Tr} \! \left[  \Delta^{(j)}({\bf \hat{n}}) \; \Delta^{(j)}({\bf \hat{n}}^{\prime})  \right] 
 \Delta^{(j)}({\bf \hat{n}}^{\prime}) \, d{\bf \hat{n}}^{\prime} \label{traciality} \\
 & = &  \int_{{\bf S}^2}\;  K^{(j)}({\bf \hat{n}}, {\bf \hat{n}}^{\prime}) \;  \Delta^{(j)}({\bf \hat{n}}^{\prime})  \; d{\bf \hat{n}}^{\prime} \label{deltakernela} \\
  & = &  \int_{{\bf S}^2}\;  \delta_{\Delta}^{(j)}({\bf \hat{n}}, {\bf \hat{n}}^{\prime}) \;  \Delta^{(j)}({\bf \hat{n}}^{\prime})  \; d{\bf \hat{n}}^{\prime} \label{deltakernel} \\
 & & \nonumber \\
 \mbox{where} \;\;\;\; K^{(j)}({\bf \hat{n}}, {\bf \hat{n}}^{\prime}) & = &   \frac{(2j+1)}{4\pi} \; \mbox{Tr} \! \left[  \Delta^{(j)}({\bf \hat{n}}) \;
 \Delta^{(j)}({\bf \hat{n}}^{\prime})  \right]  \\
  d{\bf \hat{n}} &  = & \sin \theta \, d\theta \, d\phi
\end{eqnarray}
Eqs.(\ref{deltakernela}) and (\ref{deltakernel}) show that for a certain subset of $(2j+1)^2$ functions on the sphere ${\bf S}^2$ \cite{varilly2:moyal, heissweigert}, $K^{(j)}({\bf \hat{n}}, {\bf \hat{n}}^{\prime}) \equiv 
 \delta_{\Delta}^{(j)}({\bf \hat{n}}, {\bf \hat{n}}^{\prime})$ is the reproducing kernel  \cite{varilly2:moyal, heissweigert}, acting as a delta function with respect to integration over ${\bf S}^2$. 
  Armed with the Chebyshev polynomial operator expansion of Eq.(\ref{stratopexp}), this delta function can be  simply evaluated using 
the Chebyshev polynomial operator $f_{\lambda}^{(j)}({\bf \hat{n}} \cdot {\bf J})$  trace identity discussed in Section {\bf 2.2.1} as follows:
\begin{eqnarray}
 \delta_{\Delta}^{(j)}({\bf \hat{n}}, {\bf \hat{n}}^{\prime})
 & = & \frac{1}{4\pi} \; \sum_{\lambda=0}^{2j}\; \sum_{\lambda^{\prime}=0}^{2j}\; \sqrt{2\lambda+1}\,\sqrt{2\lambda^{\prime}+1} \;\,
\mbox{Tr} \! \left[  f_{\lambda}^{(j)} ({\bf \hat{n}} \cdot {\bf J})\, f_{\lambda^{\prime}}^{(j)} ({\bf \hat{n}}^{\prime}\cdot {\bf J})  \right]\;\;\;\;\;\; \\
& = & \frac{1}{4\pi} \; \sum_{\lambda=0}^{2j}\; \sum_{\lambda^{\prime}=0}^{2j}\; \sqrt{2\lambda+1}\,\sqrt{2\lambda^{\prime}+1} \;\,
\delta_{\lambda \lambda^{\prime}} \; P_{\lambda}({\bf \hat{n}}  \cdot {\bf \hat{n}}^{\prime})\\
& = &  \; \sum_{\lambda=0}^{2j}\; \boxed{\frac{2\lambda+1}{4\pi}\, P_{\lambda}({\bf \hat{n}}  \cdot {\bf \hat{n}}^{\prime}) } \label{legend} \\
& = &  \; \sum_{\lambda=0}^{2j} \boxed{\sum_{\mu=-\lambda}^{\lambda} Y_{\lambda \mu}({\bf \hat{n}}) \, Y_{\lambda \mu}^{\star}({\bf \hat{n}}^{\prime}) }
\label{addsphere}
\end{eqnarray}
The ``boxed" term of Eq.(\ref{legend}) has been reexpressed as the ``boxed" term of Eq.(\ref{addsphere}) using the spherical harmonics 
addition theorem \cite{gottfried}:
\begin{equation}
P_{\lambda}({\bf \hat{n}}  \cdot {\bf \hat{n}}^{\prime}) = \frac{4 \pi}{2\lambda+1} \; \sum_{\mu=-\lambda}^{\lambda} Y_{\lambda \mu}({\bf \hat{n}}) \, Y_{\lambda \mu}^{\star}({\bf \hat{n}}^{\prime}) \label{gottadd}
\end{equation}
Certainly there are other methods for evaluating $\delta_{\Delta}^{(j)}({\bf \hat{n}}, {\bf \hat{n}}^{\prime})$  \cite{varilly2:moyal, heissweigert}, but this method using 
the Chebyshev polynomial operators $f_{\lambda}^{(j)} ({\bf \hat{n}} \cdot {\bf J})$ is among the most direct. Note that the closure relation  \cite{gottfried} for the spherical harmonics $\left\{ Y_{\lambda \mu}({\bf \hat{n}}) \right\} $ could be obtained 
from Eq.(\ref{addsphere}) in the limit that $j \rightarrow \infty$ to give
\begin{eqnarray}
\delta({\bf \hat{n}} - {\bf \hat{n}}^{\prime}) & \equiv &  \frac{\delta(\theta - \theta^{\prime}) \;  \delta(\phi - \phi^{\prime} )}{\sin \, \theta} \\
& & \\
& = &  \; \sum_{\lambda=0}^{\infty} \sum_{\mu=-\lambda}^{\lambda} Y_{\lambda \mu}({\bf \hat{n}}) \, Y_{\lambda \mu}^{\star}({\bf \hat{n}}^{\prime}) 
\label{closure}
\end{eqnarray}
Substituting the expression for the reproducing kernel $ \delta_{\Delta}^{(j)}({\bf \hat{n}}, {\bf \hat{n}}^{\prime})$ from Eq.(\ref{legend}) in Eq.(\ref{deltakernel}), and using the Chebyshev polynomial operator $f_{\lambda}^{(j)} ({\bf \hat{n}} \cdot {\bf J})$ expansion of the  Stratonovich-Weyl operator  given in Eq.(\ref{stratopexp}), we easily obtain the reproducing kernel $ \delta_{f}^{(j)}({\bf \hat{n}}, {\bf \hat{n}}^{\prime})$
\begin{equation}
\delta_{f}^{(j)}({\bf \hat{n}}, {\bf \hat{n}}^{\prime}) = \frac{2\lambda+1}{4\pi}\, P_{\lambda}({\bf \hat{n}}  \cdot {\bf \hat{n}}^{\prime}) 
\end{equation}
which for the Chebyshev polynomial operators $f_{\lambda}^{(j)} ({\bf \hat{n}} \cdot {\bf J})$,  acts as a delta function with respect to integration over ${\bf S}^2$:
\begin{eqnarray}
f_{\lambda}^{(j)} ({\bf \hat{n}} \cdot {\bf J}) & = &  \int_{{\bf S}^2}\;  \delta_{f}^{(j)}({\bf \hat{n}}, {\bf \hat{n}}^{\prime}) \; 
f_{\lambda}^{(j)} ({\bf \hat{n}}^{\prime} \cdot {\bf J})  \; d{\bf \hat{n}}^{\prime} \\
& = &  \frac{2\lambda+1}{4\pi}\, \int_{{\bf S}^2}\;   P_{\lambda}({\bf \hat{n}}  \cdot {\bf \hat{n}}^{\prime}) \; 
f_{\lambda}^{(j)} ({\bf \hat{n}}^{\prime} \cdot {\bf J})  \; d{\bf \hat{n}}^{\prime} \label{reprokernelcheby} \\
& = &  \frac{2\lambda+1}{4\pi}\, \int_{{\bf S}^2}\; 
\mbox{Tr} \!  \left[ f^{(j)}_{\lambda}({\bf \hat{n}} \cdot {\bf J}) \; f^{(j)} _{{\lambda}}({\bf \hat{n}}^{\prime} \! \cdot {\bf J}) \right]   \; 
f_{\lambda}^{(j)} ({\bf \hat{n}}^{\prime} \cdot {\bf J})  \; d{\bf \hat{n}}^{\prime} \label{reprokernelcheby2} 
\end{eqnarray}
The reproducing kernel in Eq.(\ref{reprokernelcheby}) has been rewritten as a trace in Eq.(\ref{reprokernelcheby2}) to elicit the analogy with the 
traciality condition \cite{varilly2:moyal, heissweigert} of Eq.(\ref{traciality}) for the Stratonovich-Weyl operators. 
If in Eq.(\ref{reprokernelcheby}), we put ${\bf \hat{n}}  = {\bf \hat{z}}$ and ${\bf \hat{n}}^{\prime}  =  {\bf \hat{n}}$, so that 
${\bf \hat{n}} \cdot {\bf \hat{n}}^{\prime} \equiv {\bf \hat{z}} \cdot {\bf \hat{n}} = \cos \theta$ then what at first sight is certainly not a familiar relation becomes  
\begin{eqnarray}
f^{(j)} _{{\lambda}}({\bf \hat{z}} \! \cdot {\bf J}) \equiv f^{(j)} _{{\lambda}}(J_z) & = & 
\frac{2\lambda+1}{4\pi}\, \int_{{\bf S}^2}\;   P_{\lambda}(\cos \theta) \; 
f_{\lambda}^{(j)} ({\bf \hat{n}} \cdot {\bf J})  \; d{\bf \hat{n}} \label{familiar}
\end{eqnarray}
But since $ f^{(j)} _{{\lambda}}(J_z) \equiv  \hat{T}^{(j)}_{\lambda 0}$, and $ P_{\lambda}(\cos \theta)  \equiv C_{\lambda 0}({\bf \hat{n}})$, Eq.(\ref{familiar}) can be reexpressed as 
\begin{equation}
 \hat{T}^{(j)}_{\lambda 0}  = \frac{2\lambda+1}{4\pi}\, \int_{{\bf S}^2}\; C_{\lambda 0}({\bf \hat{n}})   \; 
f_{\lambda}^{(j)} ({\bf \hat{n}} \cdot {\bf J})  \; d{\bf \hat{n}} 
\end{equation}
which for the spin polarization operators 
$ \hat{T}^{(j)}_{\lambda \mu} $ is a particular version $(\mu=0)$ of a decomposition on the Chebyshev polynomial operators 
$f_{\lambda}^{(j)} \! \left( {\bf \hat{n}} \cdot {\bf J} \right)$   (see Eq.(\ref{spintensorcheby}) in Section {\bf 2.2.2}).

In the next section, we will use this reproducing kernel $ \delta_{f}^{(j)}({\bf \hat{n}}, {\bf \hat{n}}^{\prime}) $ to derive a tomographic reconstruction formula for the density operator from the conventional statistical tensor expansion of the density operator. 

\subsection{Tomographic Reconstruction of the Density Operator}

As an alternative to the phase-space approach to spin, the tomographic map of spin states onto a probability distribution 
\cite{filippov4, filippov2:thesis, fillipov1:qubit, filippov3, manko:spintomo,klimovchumakov} represents another approach to mapping spin operators onto functions. In the first two parts of this section, specific examples of these distributions are considered, and in each case, these mappings are shown to lead to a novel tomographic reconstruction  formula for the density operator $\hat \rho$ expressed exclusively in terms of Chebyshev polynomial operators 
$f_{\lambda}^{(j)}({\bf \hat{n}} \cdot {\bf J})$. In the concluding part,  without considering a specific probability distribution, this tomographic reconstruction formula is recovered using more fundamental approaches.

\subsubsection{Spin Tomogram Distributions}
One example of tomographic mapping is the approach developed by Man'ko and colleagues  \cite{filippov4, filippov2:thesis, fillipov1:qubit, filippov3, manko:spintomo}, who have made very effective use of the following Chebyshev polynomial operator $f_{\lambda}^{(j)}({\bf \hat{n}} \cdot {\bf J})$ expansions of the dequantizer (alias projection) operators $\mbox{{\boldmath $\Pi$}}^{(j)}(m,{\bf \hat{n}})$   and quantizer operators
 $\mbox{{\boldmath $\Xi$}}^{(j)}(m,{\bf \hat{n}}) $
\begin{eqnarray}
\mbox{{\boldmath $\Pi$}}^{(j)}(m,{\bf \hat{n}})  & =  & \sum_{\lambda=0}^{2j} f_{\lambda}^{(j)}(m) \; f_{\lambda}^{(j)}({\bf \hat{n}} \cdot {\bf J}) \nonumber \\
\mbox{{\boldmath $\Xi$}}^{(j)}(m,{\bf \hat{n}})  & =  & \sum_{\lambda=0}^{2j} (2\lambda +1) \, f_{\lambda}^{(j)}(m) \; f_{\lambda}^{(j)}({\bf \hat{n}} \cdot {\bf J})
\label{chebexp}
\end{eqnarray}

The spin tomogram $w^{(j)}(m,{\bf \hat{n}}) $ of a state determined by the density operator $\hat \rho$ is \cite{filippov4, filippov2:thesis, fillipov1:qubit, filippov3, manko:spintomo}
\begin{equation}
w^{(j)}(m,{\bf \hat{n}}) =  \mbox{Tr} \! \left[  \hat \rho\; \mbox{{\boldmath $\Pi$}}^{(j)}(m,{\bf \hat{n}}) \right] \label{tomtrace}
\end{equation}
whereas the inverse mapping of the spin tomogram $w^{(j)}(m,{\bf \hat{n}}) $ onto the density operator $\hat \rho$  was expressed through the  quantizer operator $\mbox{{\boldmath $\Xi$}}^{(j)}(m,{\bf \hat{n}}) $ as the tomographic reconstruction \cite{filippov4, filippov2:thesis, fillipov1:qubit, filippov3, manko:spintomo}
\begin{eqnarray}
\hat \rho & = & \sum_{m^{\prime}=-j}^j \; \frac{1}{4\pi} \int _{{\bf S}^2}  \, w^{(j)}(m^{\prime},{\bf \hat{n}}^{\prime}) \; \,
\mbox{{\boldmath $\Xi$}}^{(j)}(m^{\prime},{\bf \hat{n}}^{\prime})\;  d{\bf \hat{n}}^{\prime}  \label{densityrecon} \\
\mbox{where} \;\;\;\;  
d{\bf \hat{n}}^{\prime} & \equiv &d\Omega= \sin \theta \, d\theta \, d\phi \nonumber
\end{eqnarray}
Upon substituting this  integral representation for the density operator in Eq.(\ref{tomtrace}), we obtain
\begin{eqnarray}
w^{(j)}(m,{\bf \hat{n}}) & = &  \mbox{Tr} \! \left[  \hat \rho\; \mbox{{\boldmath $\Pi$}}^{(j)}(m,{\bf \hat{n}}) \right]\\
& = & \sum_{m=-j}^j \; \frac{1}{4\pi}  \int_{{\bf S}^2} \, \; \mbox{Tr}  \! \left[ \mbox{{\boldmath $\Xi$}}^{(j)}(m^{\prime},{\bf \hat{n}}^{\prime})  \; 
\mbox{{\boldmath $\Pi$}}^{(j)}(m,{\bf \hat{n}})  \right]   w^{(j)}(m^{\prime},{\bf \hat{n}}^{\prime})
  \, d{\bf \hat{n}}^{\prime} \label{trdq}
\end{eqnarray}
This implies that for the set of tomograms $ w^{(j)}(m,{\bf \hat{n}}) $ on the sphere ${\bf S}^2$, the trace in Eq.(\ref{trdq}) must act like a delta function with respect to integration over ${\bf S}^2$. This delta function can easily be evaluated using the Chebyshev polynomial operator $f_{\lambda}^{(j)}({\bf \hat{n}} \cdot {\bf J})$  trace identity discussed in Section {\bf 2.2.1} as follows:
\begin{eqnarray}
&  & \delta_w^{(j)}(m, {\bf \hat{n}};m^{\prime},{\bf \hat{n}}^{\prime}) \\
 & = &  \delta_{m,m^{\prime}} \;
\mbox{Tr}  \! \left[ \mbox{{\boldmath $\Xi$}}^{(j)}(m^{\prime},{\bf \hat{n}}^{\prime})  \; 
\mbox{{\boldmath $\Pi$}}^{(j)}(m,{\bf \hat{n}})  \right]  \\
& = & \delta_{m,m^{\prime}} \; \sum_{\lambda=0}^{2j}   \; \sum_{\lambda^{\prime}=0}^{2j} \; (2\lambda{^\prime}+1) \,
 f_{\lambda}^{(j)}(m) \;  f_{\lambda^{\prime}}^{(j)}(m^{\prime}) \; 
\mbox{Tr} \! \left[  f_{\lambda}^{(j)} ({\bf \hat{n}} \cdot {\bf J})\, f_{\lambda^{\prime}}^{(j)} ({\bf \hat{n}}^{\prime}\cdot {\bf J})  \right]  \\
 & = & \delta_{m,m^{\prime}} \; \sum_{\lambda=0}^{2j}(2\lambda+1) \,  f_{\lambda}^{(j)}(m) \;  f_{\lambda}^{(j)}(m^{\prime}) \; 
P_{\lambda}({\bf \hat{n}} \cdot{\bf \hat{n}}^{\prime}) 
\end{eqnarray}
As with the evaluation of $ \delta_{\Delta}^{(j)}({\bf \hat{n}}, {\bf \hat{n}}^{\prime})$ in Section {\bf 5.2}, evaluating 
$\delta_w^{(j)}(m, {\bf \hat{n}};m^{\prime},{\bf \hat{n}}^{\prime}) $ is straightforward using the 
the Chebyshev polynomial operators $f_{\lambda}^{(j)} ({\bf \hat{n}} \cdot {\bf J})$.

A much simpler expression for the tomographic reconstruction of the density operator can be obtained from Eq.(\ref{densityrecon}) just by replacing the quantizer and dequantizer operators with their corresponding Chebyshev polynomial operator $f_{\lambda}^{(j)}({\bf \hat{n}} \cdot {\bf J})$ expansions given in Eq.(\ref{chebexp}). By means of these replacements, the density operator tomographic reconstruction formula can be  expressed exclusively in terms of Chebyshev polynomial operators 
$f_{\lambda}^{(j)}({\bf \hat{n}} \cdot {\bf J})$
\begin{eqnarray}
\hat \rho & = &  \sum_{m^{\prime}=-j}^j \; \frac{1}{4\pi} \int _{{\bf S}^2} \mbox{Tr} \! \left[  \hat \rho \;\mbox{{\boldmath $\Pi$}}^{(j)}(m,{\bf \hat{n}})  \right] \; 
\mbox{{\boldmath $\Xi$}}^{(j)}(m,{\bf \hat{n}}) \; d{\bf \hat{n}} \\
& = &  \frac{1}{4\pi} \sum_{\lambda,\lambda^{\prime}=0}^{2j} (2\lambda+1) \; 
\boxed{ \sum_{m^{\prime}=-j}^j   f_{\lambda^{\prime}}^{(j)}(m) \;   f_{\lambda}^{(j)}(m) }  \int _{{\bf S}^2}  f_{\lambda}^{(j)}({\bf \hat{n}} \cdot {\bf J} ) \; 
\; \mbox{Tr} \! \left[ \hat \rho \,
 f_{\lambda^{\prime}}^{(j)}( {\bf \hat{n}} \cdot {\bf J}) \right]  d{\bf \hat{n}} \;\;\;\; \label{boxorthog} \\
& = &  \sum_{\lambda=0}^{2j} \frac{(2\lambda+1)}{4\pi}  \int _{{\bf S}^2}  f_{\lambda}^{(j)}({\bf \hat{n}} \cdot {\bf J}) \;  
\mbox{Tr} \! \left[ \hat \rho \,  f_{\lambda}^{(j)}({\bf \hat{n}} \cdot {\bf J} ) \right] d{\bf \hat{n}} \label{singlesum}
\end{eqnarray}
The ``boxed" term in Eq.(\ref{boxorthog}) has been simplified using the Chebyshev polynomial orthogonality relation of Eq.(\ref{ortho1}), which collapses the double sum 
in Eq.(\ref{boxorthog})  to a single sum in Eq.(\ref{singlesum}). 

\subsubsection{Other Distributions}
Other examples of probability distributions are Husimi's \cite{husimi} $Q({\bf \hat{n}})$ function \cite{klimovchumakov} and the Stratonovich-Weyl distribution $W({\bf \hat{n}})$  \cite{klimovchumakov}, defined as 
\begin{eqnarray}
Q({\bf \hat{n}}) &  =  &  \langle {\bf \hat{n}},j |   \; \hat \rho \;    |{\bf \hat{n}},j \rangle = 
 \mbox{Tr} \! \left[ \hat \rho \;  \mbox{{\boldmath $\Pi$}}^{(j)}(j,{\bf \hat{n}})  \right] \label{Hist} \\
W({\bf \hat{n}}) & = &  \mbox{Tr} \! \left[ \hat \rho \;  \Delta^{\!(j)}({\bf \hat{n}})  \right] \label{Wdist}
\end{eqnarray}
For a spin state of well-defined angular momentum,  Husimi's \cite{husimi} $Q({\bf \hat{n}})$ function \cite{klimovchumakov}  is defined as the average value of the density matrix in the coherent state $|{\bf \hat{n}},j \rangle $, while the Stratonovich-Weyl distribution $W({\bf \hat{n}})$ is defined in terms of the 
 Stratonovich-Weyl operator  \cite{varilly2:moyal,heissweigert} $ \Delta^{\!(j)}({\bf \hat{n}})$ of Section {\bf 5.2} by the trace of Eq.(\ref{Wdist}). Table VII compares these probability distributions and their tomographic reconstruction relations with those of the 
spin tomogram distributions $w^{(j)}(m,{\bf \hat{n}}) $ \cite{filippov4, filippov2:thesis, fillipov1:qubit, filippov3, manko:spintomo}. All the distributions and tomographic reconstruction relations in this table are expressed only in terms of Chebyshev polynomials $ f_{\lambda}^{(j)}(m) $ or Chebyshev polynomial operators 
$ f_{\lambda}^{(j)}({\bf \hat{n}} \cdot {\bf J} )$, so that substitution of each distribution in the corresponding tomographic reconstruction relation immediately yields the tomographic reconstruction relation of Eq.(\ref{singlesum}). All the distributions $D({\bf \hat{n}}) \;\;[\equiv w^{(j)}(m,{\bf \hat{n}}) , Q({\bf \hat{n}})$ or $W({\bf \hat{n}})]$ of Table VI  are normalized \cite{klimovchumakov}  so that 
\begin{equation}
\frac{2j+1}{4\pi} \int _{{\bf S}^2}  D({\bf \hat{n}}) \; d{\bf \hat{n}} = 1
\end{equation}
Just as measurements of the spin tomograms $w^{(j)}(m,{\bf \hat{n}}) $ enable the reconstruction of the density operator $\hat \rho $ via 
the tomographic reconstruction relation of Eq.(\ref{densityrecon}) \cite{filippov4, filippov2:thesis, fillipov1:qubit, filippov3, manko:spintomo}, 
so do the corresponding tomographic reconstruction relations for $\hat \rho $ in Table VII  also permit the reconstruction of all the density matrix elements simply by measuring the corresponding $Q({\bf \hat{n}})$  or  $W({\bf \hat{n}})$ function distributions \cite{klimovchumakov}.

\subsubsection{More Fundamental Perspectives}

In the previous two sections, specific examples of tomographic reconstruction formulae for the density operator were considered. Regardless of the probability distribution under consideration, the tomographic map of spin states lead to a novel tomographic reconstruction formula expressed exclusively in terms of  Chebyshev polynomial operators $ f_{\lambda}^{(j)}({\bf \hat{n}} \cdot {\bf J} )$.  From a more fundamental perspective, can this formula be derived without considering a specific probability distribution? Two approaches are now considered which demonstrate that such a derivation is possible.

\paragraph{Statistical tensor expansion of the density operator}
In this approach, we begin with a consideration of the statistical tensor expansion of the density operator given in Table IV
\begin{equation}
\hat \rho =   \displaystyle\sum_{\lambda=0}^{2j}\sum_{\mu=-\lambda}^{\lambda}  \mbox{Tr} \! \left[ \hat \rho \,   
\left[ \hat{T}_{\lambda \mu }^{(j)} \right]^{\! \dagger}  \right]  \hat{T}_{\lambda \mu }^{(j)}  \label{stat}
\end{equation}
Viewing the  spin polarization operators $\hat{T}^{(j)}_{\lambda \mu} $ in Eq.(\ref{stat}}) as the following integral transformation of the Chebyshev polynomial operators 
$f_{\lambda}^{(j)} \! \left( {\bf \hat{n}} \cdot {\bf J} \right)$ (see Eq.(\ref{spintensorcheby}) in Section {\bf 2.2.2})
\begin{eqnarray}
 \hat{T}^{(j)}_{\lambda \mu} &  =  &
 \frac{2\lambda +1 }{4\pi} \int C_{\lambda \mu}({\bf \hat{n}}) \; f_{\lambda}^{(j)} \! \left( {\bf \hat{n}} \cdot {\bf J} \right) \, d{\bf \hat{n}}  \label{altspintensorcheby} 
\end{eqnarray}
the density operator can be written as 
\begin{eqnarray}
&  &  \hat \rho  \nonumber  \\
&\!\! \!=  &  \displaystyle\sum_{\lambda=0}^{2j}\sum_{\mu=-\lambda}^{\lambda} \! \mbox{Tr} \! \left[ \hat \rho \,   
\left[  \frac{2\lambda +1 }{4\pi} \int_{{\bf S}^2}  \!C_{\lambda \mu}({\bf \hat{n}}) \; f_{\lambda}^{(j)} \! \left( {\bf \hat{n}} \cdot {\bf J} \right) \, d{\bf \hat{n}}  \right]^{\! \dagger}  \right]   \frac{2\lambda +1 }{4\pi} \! \int_{{\bf S}^2}  \! C_{\lambda \mu}({\bf \hat{n}}^{\prime}) \; f_{\lambda}^{(j)} \! \left( {\bf \hat{n}}^{\prime} \cdot {\bf J} \right) \, d{\bf \hat{n}} ^{\prime}  \;\;\;\;\;\;\;\;\;\;\;\\
&\!\! \!=  &  \displaystyle\sum_{\lambda=0}^{2j} \frac{2\lambda +1 }{4\pi}  \int \! \mbox{Tr} \! \left[ \hat \rho  \;
 f_{\lambda}^{(j)} \! \left( {\bf \hat{n}} \cdot {\bf J} \right) \right]   \int_{{\bf S}^2}  \frac{2\lambda +1 }{4\pi}   \boxed{\sum_{\mu=-\lambda}^{\lambda}
 C_{\lambda \mu}^{\star}({\bf \hat{n}}) \; C_{\lambda \mu}({\bf \hat{n}}^{\prime}) }  \; f_{\lambda}^{(j)} \! \left( {\bf \hat{n}}^{\prime} \cdot {\bf J} \right) \,
d{\bf \hat{n}} ^{\prime} \; d{\bf \hat{n}}  \; \label{addtheorem}  \\
&\!\! \!=  &  \displaystyle\sum_{\lambda=0}^{2j} \frac{2\lambda +1 }{4\pi}  \int_{{\bf S}^2}  \! \mbox{Tr} \! \left[ \hat \rho  \;
 f_{\lambda}^{(j)} \! \left( {\bf \hat{n}} \cdot {\bf J} \right) \right]  \boxed{ \int_{{\bf S}^2}  \frac{2\lambda +1 }{4\pi}  P_{\lambda}( {\bf \hat{n}} \cdot {\bf \hat{n}}^{\prime}) \; f_{\lambda}^{(j)} \! \left( {\bf \hat{n}}^{\prime} \cdot {\bf J} \right) \, d{\bf \hat{n}} ^{\prime}  }\; d{\bf \hat{n}}  \; \label{kernelsimp} \\
&\!\! \!=  &  \displaystyle\sum_{\lambda=0}^{2j} \frac{2\lambda +1 }{4\pi}  \int_{{\bf S}^2}  \! \mbox{Tr} \! \left[ \hat \rho  \;
 f_{\lambda}^{(j)} \! \left( {\bf \hat{n}} \cdot {\bf J} \right) \right]    f_{\lambda}^{(j)} \! \left( {\bf \hat{n}} \cdot {\bf J} \right) d{\bf \hat{n}}  \label{statalt}
\end{eqnarray}
The ``boxed" term of Eq.(\ref{addtheorem}) has been simplified using the spherical harmonics 
addition theorem \cite{gottfried} of Eq.(\ref{gottadd}), and the ``boxed" term of Eq.(\ref{kernelsimp}) has been simplified using Eq.(\ref{reprokernelcheby}), recognizing $\left[(2\lambda+1)/4\pi \right] \; P_{\lambda}( {\bf \hat{n}} \cdot {\bf \hat{n}}^{\prime})$ as the reproducing kernel for the Chebyshev polynomial operators 
$ f_{\lambda}^{(j)} \! \left( {\bf \hat{n}} \cdot {\bf J}\right)$.

\paragraph{Group theory operator identity} In this altenative approach,  we demonstrate that the Chebyshev polynomial operator tomographic reconstruction formula of Eqs.(\ref{singlesum}) and (\ref{statalt}) is a consequence of much deeper and more general result in the form of a single operator identity based on group theory \cite{ariano}. 
Using group theory, D'Ariano et al. \cite{ariano} have derived the following fundamental tomographic reconstruction formula for the density operator
\begin{equation}
\hat \rho  = \int \mbox{Tr} \! \left[ \hat \rho \; {\cal R}(g) \right] {\cal R}^{\dagger}(g) \, dg \label{Ariano}
\end{equation}
valid for an irreducible unitary representation ${\cal R}(g)$ on the Hilbert space ${\cal H}$ of the physical system. 
The appropriate operators of such a  unitary irreducible representation in the case of a single spin physical system are  \cite{ariano} 
\begin{equation}
{\cal R}(g) \equiv {\cal R}({\bf \hat{n}}, \psi) =  e^{i \psi ({\bf \hat{n}} \cdot {\bf J}) }
\end{equation}
In this parametrization, the invariant measure is given by \cite{ariano} 
\begin{equation}
dg({\bf \hat{n}}, \psi ) =\frac{(2j+1)}{4\pi^2} \sin^2  \bfrac{\psi }{2}  \sin \theta \,  d\theta  \, d\phi \, d\psi 
\end{equation}
so that the density operator according to Eq.(\ref{Ariano}) can be written \cite{ariano} 
\begin{equation}
\hat \rho  = \frac{(2j+1)}{4\pi^2}  \int_0^{2\pi} d\psi \,  \sin^2  \bfrac{\psi }{2}  \;
\int_{{\bf S}^2} \mbox{Tr} \! \left[ \hat \rho \, e^{i \psi  ({\bf \hat{n}} \cdot {\bf J}) } \right] e^{-i \psi  ({\bf \hat{n}} \cdot {\bf J}) } \, d{\bf \hat{n}}  \label{arianoconstruct}
\end{equation}
D'Ariano et al. \cite{ariano} have also described an experimental apparatus designed to reconstruct the density operator according to Eq.(\ref{arianoconstruct}). 

Taking advantage of Corio's  \cite{corio:siam,corio:correction}  Chebyshev polynomial operator expansion of the operator $e^{i \psi ({\bf \hat{n}} \cdot {\bf J}) }$ given in Eq.(\ref{coriorotfirst}) of Section {\bf 5.1}, and using the orthogonality relation \cite{varshal1:ang} for the generalized character functions $\mbox{{\large $\chi$}}_{\lambda}^{(j)}(\psi ) $
\begin{equation}
\int_0^{2\pi}  d\psi  \,  \sin^2  \bfrac{\psi }{2}  \; \mbox{{\large $\chi$}}_{\lambda_1}^{(j_1)}\!(\psi  )  \; \mbox{{\large $\chi$}}_{\lambda_2}^{(j_2)}\!(\psi )  = \pi \, 
\delta_{j_1j_2} \, \delta_{\lambda_1 \lambda_2}
\end{equation}
which define the expansion coefficients, 
the spin tomographic reconstruction formula of Eq.(\ref{Ariano}) derived by D'Ariano et al. from group theory 
\cite{ariano} can be expressed exclusively in terms of the Chebyshev polynomial operators $f_{\lambda}^{(j)}({\bf \hat{n}} \cdot {\bf J} ) $ 
\begin{equation}
  \hat \rho =   \sum_{\lambda=0}^{2j} \frac{(2\lambda+1)}{4\pi}  \int _{{\bf S}^2}  f_{\lambda}^{(j)}({\bf \hat{n}} \cdot {\bf J} ) \;  
\mbox{Tr} \! \left[ \hat \rho \,  f_{\lambda}^{(j)}({\bf \hat{n}} \cdot {\bf J} ) \right]  d \Omega
\end{equation}
This is the tomographic reconstruction formula for the density operator obtained in Sections {\bf 5.3.1} and {\bf 5.3.2} from specific examples of 
tomographic reconstruction formulae using different spin tomography probability distributions. Recovering this formula from the group theoretical operator identity derived by D'Ariano et al. \cite{ariano} indicates that for the angle-axis 
$(\psi,{\bf \hat{n}} )$ parameterization of the $SU(2)$  group, the Chebyshev polynomial operators $f_{\lambda}^{(j)}({\bf \hat{n}} \cdot {\bf J} )$ have a unique role to play in 
 tomographic reconstruction of the density operator.

\section{ The recoupling of spin and spatial tensors via $f_{\lambda}^{(j)}({\bf \hat{n}} \cdot {\bf J})$ }
From a comparison of two different operator expansions \cite{corio:siam, happer} for both  the rotation operator 
$\hat{{\cal D}}^{(j)} \!(\psi, {\bf \hat{n}})=e^{i \psi({\bf \hat{n}} \cdot {\bf J}) } $ in Section {\bf 6.1.1}, and for the coherent state projector 
$ |{\bf \hat{n}},j \rangle \langle {\bf \hat{n}},j | = \mbox{{\boldmath $\Pi$}}^{(j)}(j,{\bf \hat{n}})  $  in Section {\bf 6.1.2}, we derive a novel recoupling expression for the  Chebyshev polynomial operators $f_{\lambda}^{(j)}({\bf \hat{n}} \cdot {\bf J})$.    In Section {\bf 6.2}, we discuss a distinctly different approach  
 \cite{filippov2:thesis}  to deriving the same recoupling expression. We conclude with    specific examples of this recoupling expression in the case of first- and second-rank tensors in 
Section {\bf 6.3}. 
\subsection{Operator expansion comparisons} 
\subsubsection{Rotation operator expansions}
Using the orthonormal basis functions $f_{\lambda}^{(j)}({\bf \hat{n}} \cdot {\bf J})$, Corio \cite{corio:siam} derived the following expansion of the rotation operator $\hat{{\cal D}}^{(j)} \!(\psi, {\bf \hat{n}}) = e^{i \psi({\bf \hat{n}} \cdot {\bf J}) }$
\begin{eqnarray}
e^{i \psi({\bf \hat{n}} \cdot {\bf J}) } & = & \sum_{\lambda=0}^{2j} a_{\lambda}^{(j)} \! (\psi) \; f_{\lambda}^{(j)}({\bf \hat{n}} \cdot {\bf J})\\
& = &  \sum_{\lambda=0}^{2J} A(\lambda,j) \,  i^{\lambda}  \,  s^{\lambda} \,  
\mbox{ C}_{2j-\lambda}^{\lambda+1}(c)\; f_{\lambda}^{(j)}({\bf \hat{n}} \cdot {\bf J}) \label{Afac}\\
& = & \sum_{\lambda=0}^{2j} i^{\lambda} \sqrt{\displaystyle\frac{2\lambda+1}{2j+1}}\, \mbox{{\large $\chi$}}_{\lambda}^{(j)}(\psi)   \; f_{\lambda}^{(j)}({\bf \hat{n}} \cdot {\bf J}) \label{coriorot}\\
\mbox{where} \;\;A(\lambda, j) & = & (2\lambda)!! \sqrt{2\lambda+1} \sqrt{\displaystyle\frac{(2j-\lambda )!}{(2j+\lambda+1)!}} \\
s & = &  \sin(\psi/2) \\
c & = & \cos(\psi/2)\\
 \mbox{{\large $\chi$}}_{\lambda}^{(j)}(\psi)  & = & \sqrt{\displaystyle\frac{2j+1}{2\lambda+1}} \,  A(\lambda, j) \, 
s^{\lambda} \,  
\mbox{ C}_{2j-\lambda}^{\lambda+1}(c)
\end{eqnarray}

In Eq.(\ref{coriorot}), we have introduced the generalized character functions \cite{varshal1:ang}
$\mbox{{\large $\chi$}}_{\lambda}^{(j)}(\psi)$  to reexpress the operator expansion in Eq.(\ref{Afac}) that Corio \cite{corio:siam} first presented.  These generalized 
character functions  are defined in terms of 
the Gegenbauer polynomials \cite{tem:bk, askey} 
$\mbox{ C}_{2j-\lambda}^{\lambda+1}(c)$ as
\begin{equation}
\mbox{{\large $\chi$}}_{\lambda}^{(j)}(\psi) =  (2\lambda)!!\sqrt{2j+1}\sqrt{\frac{(2j-\lambda)!}
{(2j+\lambda+1)!}}
\;\;s^{\lambda}\mbox{ C}_{2j-\lambda}^{\lambda+1}
(c) \label{xgegen} 
\end{equation}

Using the direct product  of spin and spatial tensors expressed in terms of the  spherical harmonics  \cite{brinksatch:ang} 
$Y_{\lambda \mu}(\theta,\phi)$ and the spin  polarization tensor operators \cite{varshal1:ang}  $\hat{T}_{\lambda \mu}^{(j)}$ 
as
\begin{eqnarray}
{\bf Y}_{\lambda}({\bf \hat{n}}) \cdot  {\bf T}_{\lambda}({\bf J}) & \equiv & \sum_{\mu=-\lambda}^{\lambda}(-1)^{\mu}\; Y_{\lambda -\mu}({\bf \hat{n}})\;
\hat{T}_{\lambda \mu}^{(j)}=
 \sum_{\mu=-\lambda}^{\lambda}Y_{\lambda \mu}^{\star}({\bf \hat{n}}) \; \hat{T}_{\lambda \mu}^{(j)} \label{dir} 
\end{eqnarray}
Happer  \cite{happer} derived an alternative rotation operator expression.
Making use of the generalized character functions  \cite{varshal1:ang}, Varshalovich et al. \cite{ varshal1:ang,varshal2:expansion} subsequently reexpressed Happer's  \cite{happer} 
partial-wave expansion of the rotation operator $\hat{{\cal D}}^{(j)} \!(\psi, {\bf \hat{n}}) = e^{i \psi({\bf \hat{n}} \cdot {\bf J}) }$ as  
\begin{eqnarray}
e^{i \psi({\bf \hat{n}} \cdot {\bf J}) } & = & \sum_{\lambda=0}^{2j} b_{\lambda}^{(j)} \! (\psi) \;  
{\bf Y}_{\lambda}({\bf \hat{n}}) \cdot  {\bf T}_{\lambda}({\bf J})  \label{directpro} \\
& = & \sum_{\lambda=0}^{2j} B(\lambda,j)\;  i^{\lambda}  \,  s^{\lambda}  
\mbox{ C}_{2j-\lambda}^{\lambda + 1}(c) \;  {\bf Y}_{\lambda}({\bf \hat{n}}) \cdot  {\bf T}_{\lambda}({\bf J}) 
\label{Bfac} \\
& = & \sum_{\lambda=0}^{2j} i^{\lambda} \sqrt{\displaystyle\frac{2\lambda+1}{2j+1}}\, \mbox{{\large $\chi$}}_{\lambda}^{(j)}(\psi)   \;
{\bf C}_{\lambda}({\bf \hat{n}}) \cdot  {\bf T}_{\lambda}({\bf J})  \label{varshalrot} \\
\mbox{where} \;\;B(\lambda, j) & = & \sqrt{4 \pi}\;  (2\lambda)!! \; \sqrt{\displaystyle\frac{(2j-\lambda )!}{(2j+\lambda+1)!}}\\
C_{\lambda \mu}({\bf \hat{n}}) & = & \sqrt{\displaystyle\frac{4 \pi}{2\lambda+1}} \;  Y_{\lambda \mu}({\bf \hat{n}}) \label{RachSH}
\end{eqnarray}

Comparing Eq.(\ref{Afac}) with Eq.(\ref{Bfac}), we see that 
\begin{equation}
A(\lambda,j) \;  f_{\lambda}^{(j)}({\bf \hat{n}} \cdot {\bf J}) = B(\lambda, j)\;  {\bf Y}_{\lambda}({\bf \hat{n}}) \cdot  {\bf T}_{\lambda}({\bf J}) 
\end{equation}
so that 
\begin{eqnarray}
f_{\lambda}^{(j)}({\bf \hat{n}} \cdot {\bf J}) & = & \displaystyle\frac{B(\lambda, j)}{A(\lambda,j)} \; 
 {\bf Y}_{\lambda}({\bf \hat{n}}) \cdot  {\bf T}_{\lambda}({\bf J})   \\
& = & \sqrt{\displaystyle\frac{4 \pi}{2\lambda+1}} \;   {\bf Y}_{\lambda}({\bf \hat{n}}) \cdot  {\bf T}_{\lambda}({\bf J}) \\
& = &   {\bf C}_{\lambda}({\bf \hat{n}}) \cdot  {\bf T}_{\lambda}({\bf J})  \label{directcorio}  \\
& = &  \sum_{\mu=-{\lambda}}^{{\lambda}}(-1)^{\mu} \, C_{{\lambda}-{\mu}}({\bf \hat{n}})\;
\hat{T}^{(j)}_{{\lambda}{\mu}}\\
& = &  \sum_{\mu=-{\lambda}}^{\lambda}C^{\;\star}_{\lambda\mu}({\bf \hat{n}})\;
\hat{T}^{(j)}_{\lambda\mu}  \label{recoupleverb} \\
& = &  (-1)^{\lambda} \sqrt{2\lambda +1} \, \Big\{  {\bf T}_{\lambda}({\bf J}) \circledast {\bf C}_{\lambda}({\bf \hat{n}}) \Big\}_{\!0}^{\!0}    \label{recouplevera}
\end{eqnarray}
where in Eq.(\ref{recouplevera}), we have introduced the rank-zero composite product tensor (see Appendix E) defined in terms of 
Clebsch-Gordan coefficients as 
\begin{equation}
  \Big\{{\bf R}^{(k)}   \circledast {\bf S}^{(k)}  \Big\}_{\!0}^{\!0}  =\sum_{\substack{q,q^{\prime} \\ q+q^{\prime}=0}} {\bf R}^{(k)}_q  \;   {\bf S}^{(k)}_{q^{\prime}} \; 
C^{00}_{kqkq^{\prime}}
 \end{equation}
In his expansion of the rotation operator using the Chebyshev polynomial operators $f_{\lambda}^{(j)}({\bf \hat{n}} \cdot {\bf J}) $,  Corio \cite{corio:siam} did not express these basis operators as  the direct product in Eq.(\ref{directcorio}) of rank-$\lambda$ spatial and spin tensors. On the other hand, Happer \cite{happer} did not identify or recognize this direct product  
$ {\bf C}_{\lambda}({\bf \hat{n}}) \cdot  {\bf T}_{\lambda}({\bf J}) $  in his expansion of the rotation operator as the Chebyshev polynomial operator 
$f_{\lambda}^{(j)}({\bf \hat{n}} \cdot {\bf J}) $. The essential conclusion of this section, embodied in Eq.(\ref{directcorio}), is that the 
 Chebyshev polynomial operators $f_{\lambda}^{(j)}({\bf \hat{n}} \cdot {\bf J}) $ and the direct product $ {\bf C}_{\lambda}({\bf \hat{n}}) \cdot  {\bf T}_{\lambda}({\bf J})  $ of rank-$\lambda$ spatial and spin tensors are identical, and this identity leads to the recoupling expression of Eq.(\ref{recouplevera}). In Section 
{\bf 6.2},  our discussion is centred around an alternative approach taken by 
Filippov  \cite{filippov2:thesis} which leads to the same recoupling expression.

\subsubsection{Coherent state projection operator expansions}
The coherent state projector $ |{\bf \hat{n}},j \rangle \langle {\bf \hat{n}},j |$  has the following Chebyshev polynomial operator 
$ f_{\lambda}^{(j)}({\bf \hat{n}} \cdot {\bf J}) $ expansion 
(see Eq.(\ref{cohereproj}) of Section {\bf 3.3.2}):
\begin{equation}
 |{\bf \hat{n}},j \rangle \langle {\bf \hat{n}},j | = \mbox{{\boldmath $\Pi$}}^{(j)}(j,{\bf \hat{n}})   =  
 \sum_{\lambda=0}^{2j} f_{\lambda}^{(j)}(j) \; f_{\lambda}^{(j)}( {\bf \hat{n}} \cdot {\bf J} ) \label{cohereproj2}
\end{equation}
In the notation of this article, Ducloy \cite{ducloy} had already obtained the following operator expansion for the coherent state projector
\begin{eqnarray}
 |{\bf \hat{n}},j \rangle \langle {\bf \hat{n}},j | & = & \sqrt{4\pi} \,  \sum_{\lambda=0}^{2j} \sum_{\mu=-\lambda}^{\lambda}
\frac{(2j)!}{\sqrt{(2j+\lambda+1)!(2j-\lambda)!}} \; Y^{\;\star}_{\lambda\mu}({\bf \hat{n}})\; \hat{T}^{(j)}_{\lambda\mu}  \label{duc1}\\
 & = &  \sum_{\lambda=0}^{2j}
\boxed{\frac{(2j)! \sqrt{2\lambda+1}}{\sqrt{(2j+\lambda+1)!(2j-\lambda)!}}} \;  \sum_{\mu=-\lambda}^{\lambda} C^{\;\star}_{\lambda\mu}({\bf \hat{n}})\; \hat{T}^{(j)}_{\lambda\mu}  \label{duc2} \\
 & = &  \sum_{\lambda=0}^{2j} f_{\lambda}^{(j)}(j) \; \boxed{ \sum_{\mu=-\lambda}^{\lambda} C^{\;\star}_{\lambda\mu}({\bf \hat{n}})\;
 \hat{T}^{(j)}_{\lambda\mu} } \label{duc3}
\end{eqnarray}
The original expansion obtained by Ducloy \cite{ducloy}  in Eq.(\ref{duc1}) has been rewritten in Eqs.(\ref{duc2}) and (\ref{duc3}) in order to facilitate a comparison with  Eq.(\ref{cohereproj2}). After using Eq.(\ref{cgeval44}) to  replace the ``boxed" coefficient term in Eq.(\ref{duc2}) with the Chebyshev polynomial $ f_{\lambda}^{(j)}(j) $  in 
Eq.(\ref{duc3}), it is clear from a comparison of the operator expansions in Eq.(\ref{cohereproj2}) and Eq.(\ref{duc3}) that the 
Chebyshev polynomial operator 
$ f_{\lambda}^{(j)}({\bf \hat{n}} \cdot {\bf J}) $  can be expressed as the following  direct product or recoupling expressions 
\begin{eqnarray}
 f_{\lambda}^{(j)}({\bf \hat{n}} \cdot {\bf J}) & = &  \sum_{\mu=-{\lambda}}^{\lambda}C^{\;\star}_{\lambda\mu}({\bf \hat{n}})\;
\hat{T}^{(j)}_{\lambda\mu}  \\
& = &  {\bf C}_{\lambda}({\bf \hat{n}}) \cdot  {\bf T}_{\lambda}({\bf J})  \\
& = &  (-1)^{\lambda} \sqrt{2\lambda +1} \, \Big\{  {\bf T}_{\lambda}({\bf J})   \circledast {\bf C}_{\lambda}({\bf \hat{n}}) \Big\}_{\!0}^{\!0}    \label{recouplevera2}
\end{eqnarray}
in agreement with  Eqs.(\ref{recoupleverb}) and (\ref{recouplevera}) in
Section {\bf 6.1.1}. 

\subsection{Exploiting the transformation properties of the Chebyshev polynomial operators $ f_{\lambda}^{(j)}(J_z) $}

Independently established by many workers \cite{meckler:angular,filippov2:thesis,corio:ortho, NormRay, werb:tensor, fillipov1:qubit} over the last 50 years, the relation of Eq.(\ref{jz3}) 
\begin{equation}
f_{\lambda}^{(j)}(J_z)  = \hat{T}_{\lambda 0}^{(j)}
\end{equation}

 is an important  and useful operator equivalent. In this section, we show how it can provide another independent proof of the operator equivalence between 
the Chebyshev polynomials $f_{\lambda}^{(j)}({\bf \hat{n}} \cdot {\bf J})$,  and 
$ {\bf C}_{\lambda}({\bf \hat{n}}) \cdot  {\bf T}_{\lambda}({\bf J}) = \sum_{\mu=-{\lambda}}^{\lambda}C^{\;\star}_{\lambda\mu}({\bf \hat{n}})\;
\hat{T}^{(j)}_{\lambda\mu}$,  the direct product of rank-$\lambda$ spatial and spin tensors, namely the (renormalized) Racah spherical harmonics 
 \cite{brinksatch:ang} $C_{\lambda\mu}({\bf \hat{n}})$ and the spin polarization tensor operators \cite{varshal1:ang} $\hat{T}^{(j)}_{\lambda\mu}$, 
respectively.

As  irreducible tensor operators, the spin polarization  operators $ \hat{T}^{(j)}_{\lambda\nu}$ transform as \cite{varshal1:ang}
\begin{equation}
\hat{{\cal D}}^{(j)}(\theta, {\bf \hat{n}}_{\bot}) \,  \hat{T}^{(j)}_{\lambda\nu}   \left [\hat{{\cal D}}^{(j)} (\theta, {\bf \hat{n}}_{\bot}) \right ]^{\!\dagger} 
= \sum_{\mu=-\lambda}^{\lambda} {\cal{D}}_{\mu \nu}^{(\lambda)}(\theta, {\bf \hat{n}}_{\bot}) \; \hat{T}^{(j)}_{\lambda\mu}
\end{equation}
and in particular, the basis functions of Eq.(\ref{basistransform}) can be rewritten as \cite{filippov2:thesis}
\begin{eqnarray}
 f_{\lambda}^{(j)} \! \left( {\bf \hat{n}} \cdot {\bf J} \right)  & = &  
\hat{{\cal D}}^{(j)}(\theta, {\bf \hat{n}}_{\bot}) \,  f_{\lambda}^{(j)}(J_z)    \left [\hat{{\cal D}}^{(j)} (\theta, {\bf \hat{n}}_{\bot}) \right ]^{\!\dagger} \\
& = &  \hat{{\cal D}}^{(j)}(\theta, {\bf \hat{n}}_{\bot}) \,  \hat{T}^{(j)}_{\lambda 0}   \left [\hat{{\cal D}}^{(j)} (\theta, {\bf \hat{n}}_{\bot}) \right ]^{\!\dagger}  \label{filla}  \\
 & = & \sum_{\mu=-\lambda}^{\lambda} {\cal D}^{(\lambda)}_{\mu 0}(\theta, {\bf \hat{n}}_{\bot})\; \hat{T}^{(j)}_{\lambda \mu} \label{fillb} \\
& = & \sum_{\mu=-\lambda}^{\lambda} C_{\lambda \mu}^{\star}(\theta, \phi) \; \hat{T}^{(j)}_{\lambda \mu} \label{mea} \\
& = & \sum_{\mu=-\lambda}^{\lambda} (-1)^{\mu} \; C_{\lambda-\mu}(\theta, \phi) \; \hat{T}^{(j)}_{\lambda \mu} \label{me2a}  \\
& = &     {\bf C}_{\lambda}({\bf \hat{n}}) \cdot  {\bf T}_{\lambda}({\bf J})  \label{dotprod} \label{me3} \\
& = &  (-1)^{\lambda} \sqrt{2\lambda +1} \, \Big\{  {\bf T}_{\lambda}({\bf J})  \circledast  {\bf C}_{\lambda}({\bf \hat{n}}) \Big\}_{\!0}^{\!0}  \label{me4}
\label{recouplever}
\end{eqnarray}
where as shown in Appendix B, the rotation matrix elements  
${\cal D}^{(\lambda)}_{\mu 0}(\theta, {\bf \hat{n}}_{\bot}) \equiv C_{\lambda \mu}^{\, \star}(\theta, \phi)$ 
\cite{brinksatch:ang}  of Eq.(\ref{fillb}) are the Racah spherical harmonics of Eq.(\ref{RachSH}). Eqs.(\ref{filla}) and (\ref{fillb}), originally derived by Filippov 
 \cite{filippov2:thesis} using a Euler angle parametrization, have been reexpressed using an angle-axis parametrization. In Eqs.(\ref{mea} - \ref{me4}), we demonstrate that Filippov's \cite{filippov2:thesis}  transformation equation for the tensor operators in Eq.(\ref{fillb}) can then be expressed as a recoupling of spin and spatial tensors.  The final expressions for the basis functions $
  f_{\lambda}^{(j)} \! \left( {\bf \hat{n}} \cdot {\bf J} \right)  $ in Eqs.(\ref{recouplever}) and (\ref{recouplevera}) demonstrate that 
a Chebyshev polynomial of degree $\lambda$ in the variable $  ({\bf \hat{n}} \cdot {\bf J})$ can be recoupled as a 
rank-zero irreducible composite tensor defined by the product of two rank-$\lambda$ tensors, one the spin tensor $ {\bf T}_{\lambda}({\bf J}) $,  and the other, the spatial Racah tensor ${\bf C}_{\lambda}({\bf \hat{n}})$:
\begin{equation}
 f_{\lambda}^{(j)} \! \left({\bf \hat{n}} \cdot {\bf J} \right)= {\bf C}_{\lambda}({\bf \hat{n}}) \cdot  {\bf T}_{\lambda}({\bf J}) =
(-1)^{\lambda} \sqrt{2\lambda +1} \, \Big\{  {\bf T}_{\lambda}({\bf J})  \circledast  {\bf C}_{\lambda}({\bf \hat{n}}) \Big\}_{\!0}^{\!0}  \label{chebrecoup}
\end{equation}

Alternatively, by determining the Euler angles $(\alpha, \beta, \gamma)$ of the  rotation  $R \equiv R(\theta,{\bf \hat{n}}_{\bot} ) \equiv 
R(\alpha, \beta, \gamma)$ discussed in Section {\bf 3.3.1}, an Euler angle parametrization of the rotation operator 
$ \hat{{\cal D}}^{(J)}\!(\alpha, \beta, \gamma) $ can be used to verify Eq.(\ref{me3}).  If ${\bf \hat{n}}_{\bot} =(-\sin \phi, \cos \phi, 0)$, then the angle-axis parameters are $(\psi; \Theta, \Phi) = 
\left(\theta; \displaystyle\frac{\pi}{2}, \phi + \displaystyle\frac{\pi}{2}\right)$. Since the Euler angles $(\alpha,\beta,\gamma)$ can be expressed in terms of the 
angle-axis parameters $(\psi; \Theta, \Phi)$ using the following relations \cite{varshal1:ang}
\begin{eqnarray}
\sin \frac{\beta}{2} & = & \sin \Theta\; \sin \frac{\psi}{2} \nonumber \\
\tan \frac{\alpha + \gamma}{2} & = & \cos \Theta \; \tan \frac{\psi}{2} \nonumber \\
\frac{\alpha - \gamma}{2} & = & \Phi -\frac{\pi}{2}
\label{ieuanglea}
\end{eqnarray}
the Euler angles corresponding to the rotation operator $R(\theta, {\bf \hat{n}}_{\bot}) = e^{-i \theta ({\bf \hat{n}}_{\bot} \cdot {\bf J}) }$ are therefore
$(\phi, \theta, -\phi)$,  determined as a solution of  the following relations
\begin{eqnarray}
\sin \frac{\beta}{2} & = & \sin \frac{\theta}{2} \rightarrow \boxed{\beta = \theta}\nonumber \\
\tan \frac{\alpha + \gamma}{2} & = & 0   \rightarrow \boxed{\alpha + \gamma =0} \nonumber \\
\frac{\alpha - \gamma}{2} & = & \phi \rightarrow \boxed{\alpha - \gamma =2\phi} 
\label{ieuangleb}
\end{eqnarray}

Since irreducible spherical tensors $ \hat{T}_{LM}$ transform as
\begin{equation}
 \left[\hat{{\cal D}}^{(J)}(\phi, \theta, -\phi) \right ]^{\!\dagger} \; \hat{T}_{LM} \; \hat{{\cal D}}^{(J)}\!(\phi, \theta, -\phi) \;  = \sum_m {\cal{D}}_{mM}^{(L)}(\phi, \theta, -\phi) \; \hat{T}_{LM}
\end{equation}
then in particular
\begin{eqnarray}
f_{\lambda}^{(j)} \! \left( {\bf \hat{n}} \cdot {\bf J}  \right) 
 & = &  
  \hat{{\cal D}}^{(j)}\!(\phi, \theta, -\phi)  \; \hat{T}_{\lambda 0}^{(j)}  \;   \left[\hat{{\cal D}}^{(j)}(\phi, \theta, -\phi) \right ]^{\!\dagger}  \label{fil1}  \nonumber  \\
 & = & \sum_{\mu=-\lambda}^{\lambda} {\cal D}^{(\lambda)}_{\mu 0}(\phi, \theta, -\phi)\; \hat{T}^{(j)}_{\lambda \mu} \label{fil2}  \nonumber  \\
& = & \sum_{\mu=-\lambda}^{\lambda} C_{\lambda \mu}^{\star}(\theta, \phi) \; \hat{T}^{(j)}_{\lambda \mu} \label{me1} 
 \;\;(\mbox{using \cite{brinksatch:ang}}\;\; {\cal{D}}_{\mu 0}^{(\lambda)}(\alpha, \beta, \gamma) = C_{\lambda \mu}^{\star}(\beta, \alpha))  \nonumber 
\\
& = & \sum_{\mu=-\lambda}^{\lambda} (-1)^{\mu} \; C_{\lambda-\mu}(\theta, \phi) \; \hat{T}^{(j)}_{\lambda \mu} \label{me2b}  \nonumber  \\
& = &    {\bf C}_{\lambda}({\bf \hat{n}}) \cdot  {\bf T}_{\lambda}({\bf J})  \nonumber  \\
& = &  (-1)^{\lambda} \sqrt{2\lambda +1} \;  \Big\{  {\bf T}_{\lambda}({\bf J})  \circledast  {\bf C}_{\lambda}({\bf \hat{n}}) \Big\}_{\!0}^{\!0} \label{recoop}
\end{eqnarray}
where ${\bf C}_{\lambda}({\bf \hat{n}})$ is the renormalized Racah spherical harmonics (tensor) 
\cite{brinksatch:ang}.

\subsection{Specific recoupling examples: first- and second-rank tensors}

In order to elicit the recoupling feature of the relation in Eq.(\ref{recoop}), we consider the more familiar and commonly encountered cases of first-rank 
$(\lambda=1)$ and second-rank $(\lambda=2)$ spherical tensors. 

\subsubsection{First-rank tensors} To illustrate our approach, we begin with the trivial case of first-rank spherical tensors. The spin polarization operators 
$ {\bf T}_{1}({\bf J})  $ in Eq.(\ref{dotprod}) can be replaced with their spherical operator equivalents 
{\boldmath $\mathcal{T}$}$\!_1 =  {\bf J}   $  using the following relation \cite{varshal1:ang,ambler:traces}
\begin{eqnarray}
 {\bf T}_{1}({\bf J})   & = & a_1(j)\; \mbox{\boldmath $\mathcal{T}$}\!_{1}  = a_1(j) \;  {\bf J}  \label{firstrankrel} \\
\mbox{where} \;\;\;\; a_1(j) & = & \sqrt{\displaystyle\frac{3}{j(j+1)(2j+1)}}
\end{eqnarray}
Component-wise, the relation of Eq.(\ref{firstrankrel}) can be expressed as the following proportionality between the spin polarization operators 
$ \hat{T}^{(j)}_{1M}$ and the spherical components of the spin operator $J_M$ as 
\begin{equation}
 \hat{T}^{(j)}_{1M} = a_1(j) \,  J_M \;\;\;\;\;(M= \pm 1,0)
 \end{equation}
The Chebyshev polynomial operator $f_{1}^{(j)}({\bf \hat{n}} \cdot {\bf J}) $ can be replaced with $ ({\bf \hat{n}} \cdot {\bf J})$ using the following relation \cite{corio:ortho}
\begin{eqnarray}
f_{1}^{(j)}({\bf \hat{n}} \cdot {\bf J}) & =  & a_1(j)  \; ({\bf \hat{n}}  \cdot {\bf J}) 
\end{eqnarray}
After these replacements, in this special case of first-rank spherical tensors, Eq.(\ref{chebrecoup}) can then be written as 
\begin{eqnarray}
 ({\bf \hat{n}}  \cdot {\bf J})  =   {\bf J} \cdot {\bf C}_{1}({\bf \hat{n}})   & = &  ({\bf \hat{n}}  \cdot {\bf J})  \;\;\;\;\;\;\;\;\label{rereb} \\
  \mbox{where \cite{brinksatch:ang}} \;\;\;\; {\bf C}_{1 }({\bf \hat{n}})  & = & {\bf \hat{n}}
  \end{eqnarray}
\subsubsection{Second-rank tensors} While the result of Eq.(\ref{rereb}) is self-evident, the case of second-rank tensors is much more revealing. The spin  polarization operators $ {\bf T}_{2}({\bf J})  $ in Eq.(\ref{dotprod}) can be replaced with their spherical operator equivalents 
{\boldmath $\mathcal{T}$}$\!_2 = \sqrt{6} \;  \left\{ {\bf J} \circledast {\bf J}  \right\}^2   $  using the following relation \cite{ambler:traces}
\begin{eqnarray}
 {\bf T}_{2}({\bf J})  & = & a_2(j)\; \mbox{\boldmath $\mathcal{T}$}\!_{2}  = a_2(j) \sqrt{6}  \left\{ {\bf J} \circledast {\bf J}  \right\}^2 \label{secondrankrel} \\
\mbox{where} \;\;\;\; a_2(j) & = & \displaystyle\frac{\sqrt{5}}
{\sqrt{j(j+1)(2j+3)(2j-1)(2j+1)}} \label{tensorcoeff}
\end{eqnarray}
Component-wise, the relation of Eq.(\ref{secondrankrel}) can be expressed as the following proportionalities between the spin polarization operators 
$ \hat{T}^{(j)}_{2M}$ and the spherical components of the spin operator $ \left\{ {\bf J}  \circledast {\bf J}  \right\}_{\!M}^2 $ as 
\begin{eqnarray}
\hat{T}^{(j)}_{20} & = &  \sqrt{6}   \left\{ {\bf J} \circledast {\bf J}  \right\}_0^2  \\
\hat{T}^{(j)}_{2\pm 1} & = &  \sqrt{6}   \left\{ {\bf J} \circledast {\bf J}  \right\}_{\pm 1}^2  \\
\hat{T}^{(j)}_{2 \pm 2} & = &  \sqrt{6}  \left\{ {\bf J} \circledast {\bf J}  \right\}_{\pm 2}^2  
\end{eqnarray}
The Chebyshev polynomial operator $f_{2}^{(j)}({\bf \hat{n}} \cdot {\bf J}) $ can be replaced with $ \left[ 3 ({\bf \hat{n}} \cdot {\bf J})^2 -  ({\bf J} \cdot {\bf J}) \right]$ using the following relation \cite{corio:ortho}
\begin{eqnarray}
f_{2}^{(j)}({\bf \hat{n}} \cdot {\bf J}) & =  & a_2(j) \! \left[ 3 ({\bf \hat{n}}  \cdot {\bf J})^2 -  ({\bf J} \cdot {\bf J}) \right]  \label{f2dip} \\
\mbox{where} \;\;\;\; ({\bf J} \cdot {\bf J})  & = & \kappa \, \mathds{1}  =j(j+1) \, \mathds{1} 
\end{eqnarray}
After these replacements, in this special case of second-rank spherical tensors, Eq.(\ref{chebrecoup}) can be written as 
\begin{eqnarray}
\left[ 3 ({\bf \hat{n}}  \cdot {\bf J})^2 -  ({\bf J} \cdot {\bf J}) \right]  =  \sqrt{6}  \left\{ {\bf J} \circledast {\bf J}  \right\}^2 
\cdot {\bf C}_{2 }({\bf \hat{n}})   & = &  3\sqrt{5} \;  \Big\{  \left\{ {\bf J} \circledast {\bf J}  \right\}^2 \circledast
 \left\{ {\bf \hat{n}} \circledast {\bf \hat{n}}  \right\}^2  \Big\}_{\!0}^{\!0}  \;\;\;\;\;\;\;\;\label{rere} \\
  \mbox{where \cite{brinksatch:ang}} \;\;\;\; {\bf C}_{2 }({\bf \hat{n}})  & = & \sqrt{\displaystyle\frac{3}{2}}  \left\{ {\bf \hat{n}} \circledast {\bf \hat{n}}  \right\}^2 
\end{eqnarray}
This recoupling of two irreducible second-rank spherical tensors in Eq.(\ref{rere}) is exactly analogous to the recoupling of the magnetic dipolar 
interaction \cite{abragamtext} or the tensor force \cite{brinksatch:ang}
\begin{equation}
\left[ 3 ({\bf \hat{n}}  \cdot {\bf J}_1)({\bf \hat{n}}  \cdot {\bf J}_2)  -  ({\bf J}_1 \cdot {\bf J}_2) \right] = 
3\sqrt{5} \;  \Big\{  \left\{ {\bf J}_1 \circledast {\bf J}_2  \right\}^2 \circledast 
 \left\{ {\bf \hat{n}} \circledast {\bf \hat{n}}  \right\}^2  \Big\}_{\!0}^{\!0}  \label{rere2}
\end{equation}
Normally, the recoupling relationships of Eqs.(\ref{rere}) or (\ref{rere2})  can only be established by a laborious expansion of both sides of these equations, or by using the formalism of angular momentum theory \cite{brinksatch:ang} to recouple the four rank-1 tensors on the right-hand side of these equations  via a 
9$j$-symbol as outlined in Appendix E. Instead, the recoupling relation of Eq.(\ref{rere}) arises very simply and directly just from the definition of the Chebyshev polynomial operator 
$f_{2}^{(j)} \!  \left( {\bf \hat{n}} \cdot  {\bf J}  \right)  $, examples of which  are provided in Table I for $j=1/2,1$ and $3/2$.

\subsubsection{Matrix elements of the   truncated homonuclear dipolar Hamiltonian} From the recoupling expression in Eq.(\ref{mea}) applied to 
$f_{2}^{(j)} \!  \left( {\bf \hat{n}} \cdot  {\bf J}  \right)  $, the general matrix elements are 
\begin{eqnarray}
  \langle jm| \,  f_{\lambda}^{(j)} \!  \left( {\bf \hat{n}} \cdot  {\bf J}  \right) |jm^{\prime} \rangle & = & 
\sum_{\mu=-\lambda}^{\lambda} C_{\lambda \mu}^{\star}(\theta, \phi) \;  \langle jm| \, \hat{T}^{(j)}_{\lambda \mu}  |jm^{\prime} \rangle \\
& = & 
\sum_{\mu=-\lambda}^{\lambda} C_{\lambda \mu}^{\star}(\theta, \phi) \; \sqrt{\frac{2\lambda+1}{2j+1}} \; C_{jm^{\prime}\lambda \mu}^{jm} \\
& = & C_{\lambda \, (m-m^{\prime})}^{\star}(\theta, \phi) \; C_{jmj-m^{\prime}}^{\lambda \, (m-m^{\prime})} \; (-1)^{j-m^{\prime}} \label{f2matrix}
\end{eqnarray}
The diagonal matrix elements $(m= m^{\prime})$ for $f_{2}^{(j)} \!  \left( {\bf \hat{n}} \cdot  {\bf J}  \right)  $ are therefore 
\begin{eqnarray}
 \langle jm| \,  f_{2}^{(j)} \!  \left( {\bf \hat{n}} \cdot  {\bf J}  \right) |jm \rangle & = & 
C_{20}^{\star}(\theta, \phi) \; C_{jmj-m}^{20} \; (-1)^{j-m} \\
& = & P_2(\cos \theta) \; f_2^{(j)}(m) \\
& = & P_2(\cos \theta) \; \frac{3m^2-2}{\sqrt{6}}  \;\;\; \;\;\;   (j=1) 
\end{eqnarray}
But then using Eqs.(\ref{tensorcoeff}) and (\ref{f2dip}), 
\begin{eqnarray}
f_{2}^{(j)}({\bf \hat{n}} \cdot {\bf J}) & =  & a_2(j) \! \left[ 3 ({\bf \hat{n}}  \cdot {\bf J})^2 -  ({\bf J} \cdot {\bf J}) \right]  \label{f2dipb} \\
& = & \frac{1}{\sqrt{6}} \; \left[ 3 ({\bf \hat{n}}  \cdot {\bf J})^2 -  ({\bf J} \cdot {\bf J}) \right]  \;\;\; \;\;\;   (j=1) 
\end{eqnarray}
we find that the diagonal matrix elements of $ \left[ 3 ({\bf \hat{n}}  \cdot {\bf J})^2 -  ({\bf J} \cdot {\bf J}) \right] $ for $j=1$ are given by
\begin{equation}
        \langle m| \,  \left[ 3 ({\bf \hat{n}}  \cdot {\bf J})^2 -  ({\bf J} \cdot {\bf J}) \right]  \,  |m \rangle 
= P_2(\cos \theta)  \left[ 3m^2-2  \right] \label{ddiag}
\end{equation}
The homonuclear dipolar Hamiltonian ${\cal H}_{D}^{\mbox{intra}}$ \cite{abragamtext,keller}, which describes the through-space interaction between the magnetic moments of two adjacent protons separated by a distance $r$ is given by 
\begin{eqnarray}
{\cal H}_{D}^{\mbox{intra}} & = & \frac{\mu_0 \gamma^2 \hbar}{4\pi r^3} \;  \left[ 3 ({\bf \hat{n}}  \cdot {\bf J}_1)({\bf \hat{n}}  \cdot {\bf J}_2) -  ({\bf J}_1 \cdot {\bf J}_2) \right] \label{fulldip} \\
\mbox{where} \;\;\;\; {\bf \hat{n}} & = & {\bf \hat{r}}
\end{eqnarray}
This interaction Hamiltonian can also be expressed in terms of the total spin operator \cite{keller} ${\bf J} = {\bf J}_1 + {\bf J}_2 $ for each pair of protons as
\begin{equation}
{\cal H}_{D}^{\mbox{intra}} = \frac{\mu_0 \gamma^2 \hbar}{4\pi r^3} \;  \left[ 3 ({\bf \hat{n}}  \cdot {\bf J})^2 -  ({\bf J} \cdot {\bf J}) \right] 
\end{equation}
High-field approximation of this Hamiltonian leaves only the truncated part \cite{abragamtext}
\begin{eqnarray}
\left[{\cal H}_{D}^{\mbox{intra}}\right]^{\prime} = \frac{\mu_0 \gamma^2 \hbar}{4\pi r^3} \; P_2(\cos \theta)   \left[ 3 J_z^2 -  ({\bf J} \cdot {\bf J}) \right] 
\label{truncatedip}
\end{eqnarray}
where $\theta$ is the angle between the internuclear vector ${\bf r}$ and the magnetic field ${\bf B}_0$. 
The form of $\left[{\cal H}_{D}^{\mbox{intra}}\right]^{\prime} $ is identical to that of spin-1 systems, provided the ``quadrupole frequency" $\nu_Q$  is redefined as \cite{keller}
\begin{equation}
\nu_Q = \frac{3 \hbar \gamma^2}{4 \pi r^3}
\end{equation}
If a basis of the total angular momentum eigenstates of ${\bf J} = {\bf J}_1 + {\bf J}_2 $ is used, it is evident from Eq.(\ref{ddiag}) that the diagonal elements of the truncated Hamiltonian $\left[{\cal H}_{D}^{\mbox{intra}}\right]^{\prime} $ of Eq.(\ref{truncatedip}) match the diagonal elements of the diagonal elements of 
the dipolar Hamiltonian ${\cal H}_{D}^{\mbox{intra}} $ of Eq.(\ref{fulldip}). The latter elements are just what we expect from the secular 
approximation \cite{abragamtext}, which retains only that part of the dipolar Hamiltonian $\left[{\cal H}_{D}^{\mbox{intra}}\right]^{\prime} $  which commutes with the Zeeman Hamiltonian ${\cal H}_Z $
\begin{equation}
\left[ \left[{\cal H}_{D}^{\mbox{intra}}\right]^{\prime} \!\!, {\cal H}_Z  \right] =0
\end{equation}

\section{Operator Equivalents}

Finding suitable operator equivalents for irreducible tensor operators in terms of the familiar angular momentum operators 
$J_z, J_{\pm} = J_x \pm iJ_y$, and 
$({\bf J} \cdot {\bf J})  = J^2 \equiv J(J+1)$ has been a long-standing challenge in fields such as magnetic resonance. As a result,  there is an 
extensive literature over more than seven decades devoted to this problem  in NMR, EPR and ENDOR. A reasonably comprehensive list of the salient literature references on operator equivalents can be found in a recent publication by Ryabov \cite{ryabov:equiv}.  In their studies, Ryabov \cite{ryabov:equiv} himself, as well as Ambler et al. \cite{ambler:traces},  Ohlsen \cite{ohlsen},  and Biedenharn and Louck  \cite{biedenharn}, for example,  have tabulated a significant number of these operator equivalents.    Few however are those studies that focus on the operator equivalences between the spin polarization operators \cite{varshal1:ang} $\hat{T}_{\lambda \mu}^{(j)}$  and the Chebyshev polynomial operators 
$f_{\lambda}^{(j)} (J_z) $. Work by Meckler \cite{meckler:angular}, Corio  \cite{corio:ortho}, Marinelli et al. \cite{werb:tensor} and by Normand and Raynal \cite{NormRay} constitutes the essential core of the research that has elucidated the role of Chebyshev polynomials in developing operator equivalents, and in this section, we provide a brief summary of this work.

 In considering the history of  how Chebyshev polynomial operators $f_{\lambda}^{(j)} (J_z) $ were used to develop operator equivalents,  two developments stand out. First, 
 the operator equivalents for projection-zero tensor operators  $\hat{T}_{\lambda 0}^{(j)}$ were identified  as the Chebyshev polynomial 
 operators  $f_{\lambda}^{(j)} \!(J_z)$  \cite{meckler:angular,corio:ortho, werb:tensor,NormRay}. Second, an intriguing relationship between irreducible (spherical) tensor operators of arbitrary projection $T^{(j)}_{\lambda \mu}$  and  successive partial derivatives of the Chebyshev 
polynomial operators $f^{(j)}_{\lambda} \!(J_z)$ \cite{werb:tensor} was discovered \cite{werb:tensor}.  In this section, we begin in Section {\bf 7.1} by revisiting a  Chebyshev polynomial operator equivalent for projection-zero tensor operators $\hat{T}_{\lambda 0}^{(j)}$ that we had used in Section {\bf 3.1}. Then, in Section {\bf 7.2}, we use Table VIII to illustrate that Chebyshev polynomial operators $f^{(j)}_{\lambda} (J_z) $ have been used to develop operator equivalents for any irreducible tensor operator $\hat{T}_{\lambda \mu}^{(j)}$.

\subsection{Operator equivalent matrix elements}
As we illustrated in Section {\bf 3.1}, among the most useful operator equivalences involving the Chebyshev polynomial operators is 
\begin{equation}
\hat{T}_{L0}^{(j)} \equiv f_L^{(j)} \!(J_z)  \label{tensorcheb}
\end{equation}
Meckler's  \cite{meckler:angular} recognition of this equivalence between the Chebyshev polynomial operators $f_{L}^{(j)} (J_z) $ and the projection-zero spin polarization operators \cite{varshal1:ang} $\hat{T}_{L0}^{(j)}$ anticipated subsequent work by several 
investigators \cite{filippov2:thesis,corio:ortho,NormRay,werb:tensor,fillipov1:qubit}, who independently established this operator equivalence, and who provided specific and extensive tables illustrating this operator equivalence  \cite{corio:ortho,NormRay,werb:tensor}.

Ambler et al. \cite{ambler:traces} tabulated operator equivalents for the irreducible tensor operators $T_q^{(k)}\!(J)$ 
(equivalent to the spin polarization operators $ \hat{T}_{kq}^{(J)}$ in our notation), as did Corio 
\cite{corio:ortho} for his orthonormal operator basis $U^{(n)}_r \!(J)$, whose relation to the irreducible tensor operators $T_q^{(k)}\!(J)$ was determined by 
these relations \cite{corio:ortho}
\begin{eqnarray}
n+r & = & k \nonumber \\
r=q
\end{eqnarray}
In addition to showing how to construct operator equivalents for his orthonormal basis $U^{(n)}_r \!(J)$, Corio \cite{corio:ortho}, unaware of Meckler's earlier work \cite{meckler:angular}, also independently identified the 
$U^{(k)}_0 \!(J) \equiv \hat{T}_{k0}^{(J)}$ operators, the tensors of rank $k$ and projection $0$,   as the Chebyshev polynomial operator equivalents $f_k^{(J)} \! (J_z)$. Although a few others  \cite{werb:tensor,NormRay}  have also independently made this identification of the $\hat{T}_{k0}^{(J)}$ operators with  
the Chebyshev polynomial operators  $f_k^{(J)} \! (J_z)$, neither they nor others 
\cite{biedenharn,ryabov:equiv,ohlsen} who have developed operator equivalents for $\hat{T}_{k0}^{(J)}$ tensors  have cited the original work by 
Meckler \cite{meckler:angular} and Corio  \cite{corio:ortho} .

\subsection{A comparison of operator equivalents}
In Table VIII, we compare specific examples of tensor operator equivalents developed by Corio \cite{corio:ortho} (top row) with those developed by  Marinelli et al. \cite{werb:tensor} (bottom row). The examples have been chosen to demonstrate that all the equivalents have the same form (to within 
$j$-dependent normalization constants). The first column compares operator equivalents for the irreducible tensor operators $\hat{T}_{60}^{(j)}$, which both Corio \cite{corio:ortho} and 
Marinelli et al. \cite{werb:tensor} recognized as equivalent to the Chebyshev polynomial operator $f^{(j)}_6(J_z)$. The second column compares operator equivalents for the irreducible tensor operators $\hat{T}_{41}^{(j)}$. Although both equivalents have the same form,  Marinelli et al. \cite{werb:tensor} went one step further than 
 Corio \cite{corio:ortho}  did to show that any irreducible (spherical) tensor operator $T^{(k)}_L\! (S) $ of rank $L$ and 
projection $k>0$ could be represented by suitable linear combinations of successive partial derivatives (with respect to $S_z$) of the Chebyshev 
polynomial operators $f^{(S)}_L (S_z) $ as follows
\begin{eqnarray}
T^{(k)}_L \!(S)  & \sim & \left(S_{+}\right)^{|k|} \displaystyle\sum_{n=|k|}^L \; A_{Ln}^k\!  \left(\displaystyle\frac{\partial}{\partial S_z}\right)^{\!\!n}  
\!\left[ f^{(S)}_L (S_z) \right]
\end{eqnarray}
A table of the required coefficients $ A_{Ln}^k$ was provided by Marinelli et al. \cite{werb:tensor}. 
As an example of this linear combination of successive partial derivatives, the table entry in Table VIII for Marinelli's \cite{werb:tensor} operator equivalent of $ T^{(1)}_4 \!(S)$ we have  
calculated as follows
    \begin{eqnarray}
& & \displaystyle\sum_{n=1}^4 \; A_{4n}^1\!  \left(\displaystyle\frac{\partial}{\partial S_z}\right)^{\!\!n}  
\!\left[ f^{(S)}_4 (S_z) \right] \\
 & = &  \frac{1}{1!}  [14 \,S_z^3 - (6K-5)S_z] 
+ \frac{1}{2!} [42 \, S_z^2 -(6K-5)] 
+  \frac{1}{3!} [84 \, S_z] 
+ \frac{1}{4!}[84] \, \mathds{1}  \;\;\;\;\;\;\;\;  \label{werbsum} \\
& = & 14 S_{\!z}^3 + 21S_z^2 +(19-6K) S_z +3(2-K)\, \mathds{1}
\end{eqnarray}
$\mbox{where} \;\;\;\;  f^{(S)}_4 (S_z)  \sim  35 S^4_z -5 S^2_z (6K-5)+3K(K-2)$. Each summand in Eq.(\ref{werbsum}) is the product of a partial-derivative of $f^{(S)}_4 \! (S_z)$ (in square brackets), and coefficients 
$A_{4n}^1 $ tabulated in Marinelli et al. \cite{werb:tensor}. 

In conclusion, the examples discussed in Table VIII illustrate the fact that Chebyshev polynomial operators $f^{(S)}_L (S_z) $ can be used to develop an operator equivalent for any irreducible tensor operator.  

\section{Conclusion}

The Chebyshev polynomials $f_{\lambda}^{(j)} (m) $ first introduced in a physics application  by Meckler 
 \cite{meckler:majorana,meckler:angular}, are special 
functions \cite{askey},  a particular case of one member of the family of classical orthogonal polynomials of a discrete variable known as the Hahn polynomials \cite{bateman,nikiforov2,Nikiforov,vilenkin:specfuncbook,olver}.  Beginning with a close examination of the Meckler formula \cite{meckler:majorana,meckler:angular} for spin transition probabilities, we have described how Chebyshev polynomials of a discrete variable can be applied in physics. Applications of these very special special functions include spin physics, spin tomography, and the development of operator expansions and operator equivalents.  

Beyond their role as special functions, the Chebyshev polynomials $f_{\lambda}^{(j)} (m)$ double as the Clebsch-Gordan  coupling coefficients (3$j$-symbols \cite{brinksatch:ang}) $C_{jmj-m}^{\lambda \, 0}$ of angular momentum theory:
\begin{equation}
f_{\lambda}^{(j)} (m) = (-1)^{j-m} \;  C_{jmj-m}^{\lambda \, 0} \label{chebyclebschb}
\end{equation}  We have in this article often taken advantage of this duality of the Chebyshev polynomials to prove identities from their Clebsch-Gordan coupling coefficient homologs. The Chebyshev polynomial operators 
$f_{\lambda}^{(j)} (J_z) $ are identical to the projection-zero spin polarization operators $\hat{T}_{\lambda 0}^{(j)}$
\begin{equation}
f_{\lambda}^{(j)} \!(J_z)   \equiv \hat{T}_{\lambda 0}^{(j)} 
\end{equation}  
a relationship first recognized by Meckler \cite{meckler:angular}.
The similarity transform of this equivalence defines the Chebyshev polynomial operators $f_{\lambda}^{(j)} ( {\bf \hat{n}} \cdot {\bf J}) $ as the direct product 
 of two rank-$\lambda$ tensors, one the spin tensor ${\bf T}_{\lambda}({\bf J})$,  and the other, the spatial Racah tensor ${\bf C}_{\lambda}({\bf \hat{n}})$  \cite{brinksatch:ang}:
\begin{equation}
f_{\lambda}^{(j)} ( {\bf \hat{n}} \cdot {\bf J}) = {\bf C}_{\lambda}({\bf \hat{n}}) \cdot  {\bf T}_{\lambda}({\bf J})  
\end{equation}
Whether we consider the Chebyshev polynomial scalars  $f_{\lambda}^{(j)} (m)$, or the 
Chebyshev polynomial operators $f_{\lambda}^{(j)} (J_z) $  and  $f_{\lambda}^{(j)} ( {\bf \hat{n}} \cdot {\bf J}) $, the Chebyshev polynomials are distinguished as special functions by their very close connection to angular momentum theory and spin physics. As we have described in Section {\bf 4}, a vivid reminder of this connection is the Meckler formula 
 \cite{meckler:majorana,meckler:angular}  for the spin transition probability 
$\mbox{ P}^{(j)}_{mm^{\prime}} (t)$, whose calculation relies on angular momentum theory \cite{meckler:angular,schwinger:majorana}. As we have described in Section {\bf 5}, additional  reminders of this connection are the Chebyshev polynomial operator expansions of projection operators, the rotation operator, the Stratonovich-Weyl operator, and the tomographic reconstruction of the density operator.

\newpage

\section{Appendix A}
 
\subsection{Legendre polynomial operators}
In this section, two versions of Legendre polynomial operators  are discussed, one version due to Zemach \cite{zemach}, and the other due to 
Schwinger  \cite{schwinger:majorana}. The Legendre polynomial operators $ \overline{P}_{\lambda}({\bf \hat{n}} \cdot {\bf J})  $ defined by Zemach \cite{zemach} are closely related to the Chebyshev polynomial operators $  f_{\lambda} ^{(j)} ({\bf \hat{n}} \cdot {\bf J})$. On the other hand, the matrix elements of the  Legendre polynomial operators $P_{\lambda}({\bf J}) $ defined by Schwinger  \cite{schwinger:majorana} are closely related to the Chebyshev polynomials  $f_{\lambda}^{(j)} (m)  $. 
\subsubsection{Legendre polynomial operators $ \overline{P}_{\lambda}({\bf \hat{n}} \cdot {\bf J}) $}

From the addition theorem for spherical  harmonics, it is well-known that the direct product of rank-$\lambda$ spherical  harmonics tensors can be expressed in scalar form as the following  Legendre polynomial of order $\lambda$:
\begin{equation}
{\bf C}_{\lambda}({\bf \hat{n}}) \cdot {\bf C}_{\lambda}({\bf \hat{n}}^{\prime}) = P_{\lambda} ({\bf \hat{n}} \cdot {\bf \hat{n}}^{\prime}) \label{legscal}
\end{equation}
Not so well-known is how the direct product of rank-$\lambda$  spatial and spin tensors ${\bf C}_{\lambda}({\bf \hat{n}}) \cdot  {\bf T}_{\lambda}({\bf J})$ could  be expressed. 
Certainly the direct product of rank-$\lambda$  spatial and spin tensors ${\bf C}_{\lambda}({\bf \hat{n}}) \cdot  {\bf T}_{\lambda}({\bf J})$ could  not be a scalar function of $({\bf \hat{n}} \cdot {\bf J}) $, but  an operator function of $({\bf \hat{n}} \cdot {\bf J}) $. Could it  be expressed  as a Legendre polynomial operator? In developing tensor representations for application to angular-momentum problems in elementary-particle reactions, Zemach  \cite{zemach} was the first to develop such an expression. In the notation of this article, this direct product takes the following form 
\begin{eqnarray}
{\bf C}_{\lambda}({\bf \hat{n}}) \cdot  {\bf T}_{\lambda}({\bf J}) & = & \sqrt{\displaystyle\frac{2\lambda+1}{(2j+1) \left[ {\bf J}^2 \right]^{l} }} \; \overline{P}_{\lambda}({\bf \hat{n}} \cdot {\bf J}) \label{legop} \\
\mbox{where}\;\;\;\;  \left[ {\bf J}^2 \right]^{l} & = &  \prod_{n=0}^{l}
\left[  {\bf J}^2  -\tfrac{1}{4}(n^2-1) \right]
\end{eqnarray}
Just as the addition theorem of Eq.(\ref{legscal})  can be taken as the definition \cite{zemach} of the Legendre polynomials 
$P_{\lambda} ({\bf \hat{n}} \cdot {\bf \hat{n}}^{\prime}) $, the direct product expression of Eq.(\ref{legop}) can be used to define \cite{zemach}  the Legendre polynomial operators $\overline{P}_{\lambda}({\bf \hat{n}} \cdot {\bf J}) $. Comparing this expression with the corresponding direct product relations of 
Eq.(\ref{directcorio}) or (\ref{dotprod}) also defines the  Chebyshev polynomial operators $  f_{\lambda} ^{(j)} ({\bf \hat{n}} \cdot {\bf J})$ in terms of the Legendre polynomial operators $\overline{P}_{\lambda}({\bf \hat{n}} \cdot {\bf J}) $:
\begin{equation}
\overline{P}_{\lambda}({\bf \hat{n}} \cdot {\bf J}) =
\sqrt{\displaystyle\frac{(2j+1) \left[ {\bf J}^2 \right]^{l} }{2\lambda+1}} \,  f_{\lambda} ^{(j)} ({\bf \hat{n}} \cdot {\bf J})
\end{equation}
The recursion relation for the Legendre polynomials $P_{\lambda}(x)$ can be compared with its counterpart for the 
 Legendre polynomial operators $ \overline{P}_{\lambda}({\bf \hat{n}} \cdot {\bf J})  $ \cite{zemach}:
\begin{eqnarray}
(2\lambda+1)\, x \,  P_{\lambda}(x) & = & (\lambda+1) \, P_{\lambda+1}(x) + \lambda \, P_{\lambda-1}(x) \\
(2\lambda+1) ({\bf \hat{n}} \cdot {\bf J}) \,  \overline{P}_{\lambda}({\bf \hat{n}} \cdot {\bf J})   & = & (\lambda+1)\,  
\overline{P}_{\lambda+1}({\bf \hat{n}} \cdot {\bf J})  + \lambda  
\left[  {\bf J}^2  -\tfrac{1}{4} (\lambda^2-1)\right]   \overline{P}_{\lambda-1}({\bf \hat{n}} \cdot {\bf J})  \;\;\;  \;\;\; \;\;\;  \;\;\; \label{zemachrecur}
\end{eqnarray}
The recursion relation of Eq.(\ref{zemachrecur}) was used to generate the 
 Legendre polynomial operators $ \overline{P}_{\lambda}({\bf \hat{n}} \cdot {\bf J})  $ \cite{zemach} tabulated in Table IX. Inspection of this table confirms 
Zemach's observation  \cite{zemach}  that although the functional forms of the Legendre polynomial operators $ \overline{P}_{\lambda}({\bf \hat{n}} \cdot {\bf J})$ agree with those of the Legendre polynomials $P_{\lambda}(\cos \theta)$ for $\lambda=0,1,2$, the non-commutivity of components of ${\bf J}$ makes a difference for $\lambda \geq 3$. As we shall see in the next section, the same issue arises for the same reasons when comparing the 
Legendre polynomial operators $P_{\lambda}({\bf  J})$ introduced by Schwinger   \cite{schwinger:majorana}  with the Legendre polynomials $P_{\lambda}(\cos \theta)$.

\subsubsection{Legendre polynomial operators $P_{\lambda}({\bf  J})$}
In Section {\bf 4.1.2}, we discussed how Schwinger \cite{schwinger:majorana} calculated the spin transition probability $\mbox{ P}^{(j)}_{mm^{\prime}} (t)$. Using Schwinger's notation \cite{schwinger:majorana}, the final outcome of his calculation of the spin transition probability took the following form
\begin{equation}
\mbox{ P}^{(j)}_{mm^{\prime}} (t) = \left|{\cal D}_{m m^{\prime}}^{(j)}(\psi, {\bf \hat{n}})\right|^2 = \frac{1}{(2j+1)}
\sum_{l=0}^{2j} (2l+1)\,  P_{l}(j,m) \; P_{l}(j,m^{\prime})  \;  P_{l}(\cos \beta)  \label{probswing}
\end{equation}
in which the $P_{l}(j,m) $ functions were defined by Schwinger \cite{schwinger:majorana} as the matrix elements of the Legendre 
polynomial operators $P_{l}({\bf J})$:
\begin{equation}
P_{l}(j,m)  = \langle jm| \, P_{l}({\bf J})  \, | jm \rangle \label{ppdeff}
\end{equation}
Schwinger  \cite{schwinger:majorana} introduced the Legendre polynomial operators $P_{l}({\bf J})$ by expressing them in terms of the solid harmonic functions of the angular momentum vector ${\bf J}$:
\begin{eqnarray}
P_{l}({\bf J})  & = & \left[ \frac{2l+1}{4\pi} \left[ {\bf J}^2 \right]^{\!l}  \right]^{-1/2} \! \! \mathcal{Y}_{l 0}({\bf J}) \label{solidharm}  \\
\mbox{where \cite{ournote}}   \;\;\;\;\;  \left[ {\bf J}^2 \right]^{\!l}  & = & \prod_{n=0}^{l-1}
\left[  {\bf J}^2  - \bfrac{n}{2}\left( \bfrac{n}{2}\ +1 \right) \right]
\end{eqnarray}

The $ \mathcal{Y}_{lm}({\bf J})$ are operator analogues \cite{ligarg} of the solid harmonics $r^l \, Y_{lm} ({\bf \hat{n}})$ \cite{varshal1:ang}, which include an extra factor $r^l$ over the surface harmonics $Y_{lm} ({\bf \hat{n}})$. They are in fact irreducible tensor operators, proportional to the spin polarization operators $ \hat{T}^{(j)}_{lm}$ according to the following relation
\begin{equation}
 \mathcal{Y}_{lm}({\bf J}) =  \left[ \frac{2j+1}{4\pi} \left[ {\bf J}^2 \right]^{\!l}  \right]^{\!1/2}  \hat{T}^{(j)}_{lm} \label{solidpolar}
\end{equation}
It is evident from this proportionality that  $Y_{lm} ({\bf \hat{n}})$ and $ \mathcal{Y}_{lm}({\bf J})$ transform identically under rotations. 
They satisfy the following trace relation \cite{ligarg}
\begin{eqnarray}
 \mbox{Tr} \! \left[ \mathcal{Y}_{lm}({\bf J}) \; \mathcal{Y}^{\dagger}_{l^{\prime}m^{\prime}}({\bf J}) \right] &  = & \frac{2j+1}{4\pi} (a_{jl})^2 \;
\delta_{ll^{\prime}}\; \delta_{mm^{\prime}} \\
\mbox{where} \;\;\;\; (a_{jl})^2 & =  & \prod_{n=1}^{l} \left[ \left(j+\tfrac{1}{2} \right)^{\!2}-\tfrac{1}{4} n^2  \right] \equiv \left[ {\bf J}^2 \right]^{\!l} 
\end{eqnarray}

The solid harmonic operator functions $ \mathcal{Y}_{l 0}({\bf J})$ in Eq.(\ref{solidharm}) were defined \cite{schwinger:majorana} by the generating function 
 \cite{schwinger:majorana,schwingerosti,schwingerbiedenvandam,schwingerbook} 
\begin{eqnarray}
&  &\frac{1}{2^{l}l!} \left[ \frac{2l+1}{4\pi}  \right]^{\!1/2}  \left[-z_{+}^2(J_x +iJ_y) + z_{-}^2(J_x-iJ_y)+2z_{+}z_{-}J_z \right]^{l}
\nonumber \\
& = & \sum_{m=-l}^{l} \; \frac{ z_{+}^{l+m}\; z_{-}^{l-m} } { \left[(l+m)!(l-m)!\right]^{1/2}} \;  
\mathcal{Y}_{lm}({\bf J}) \label{swinggen}
\end{eqnarray} 
After evaluating $ \mathcal{Y}_{l 0}({\bf J})$ with this generating function, Eq.(\ref{solidharm}) was used to  tabulate examples of the 
Legendre polynomial operators $P_{l}({\bf J})$ in Table IX. Recognizing that the $ \mathcal{Y}_{lm}({\bf J})$ are operator analogues  of the solid harmonics $r^l \, Y_{lm} ({\bf \hat{n}})$ \cite{ligarg}, we should expect to see a difference in the functional forms of $P_{l}({\bf J})  \sim \mathcal{Y}_{l 0}({\bf J}) $ and 
$P_l(\cos \theta) \sim Y_{l0}({\bf \hat{n}})$ because the order of factors in the operator case is significant. Indeed, as this table demonstrates, for 
$\lambda \geq 3$, we can expect such a difference  taking into account the angular momentum commutation relations
\begin{equation}
\left[J_i,J_j \right] = J_k \;\;\;\;(i,j,k =x,y,z)
\end{equation}

Using the relation of Eq.(\ref{solidpolar}), and Eq.(\ref{solidharm}), the matrix element of Eq.(\ref{ppdeff}) which defines the $P_{l}(j,m) $ functions which appear in the transition probability formula  of Eq.(\ref{probswing}) can be related to the Chebyshev polynomials of a discrete variable 
$f^{(j)}_{l} (m)  $ as follows
\begin{eqnarray}
P_{l}(j,m)  = \langle jm| \, P_{l}({\bf J})  \, | jm \rangle & = & \sqrt{ \frac{2j+1}{2l+1}}  \langle jm| \,  \hat{T}^{(j)}_{l 0}  \, | jm \rangle \nonumber \\
& = & \sqrt{ \frac{2j+1}{2l+1}}  \langle jm| \,  f^{(j)}_{l} (J_z)  \, | jm \rangle \nonumber \\
& = &  \sqrt{ \frac{2j+1}{2l+1}}  \;  f^{(j)}_{l} (m)  \label{swingmatrixelement}
\end{eqnarray}
The Legendre polynomial operators  \cite{schwinger:majorana}  $P_{l}({\bf J}) $ and the Legendre polynomial operators \cite{zemach} 
$ \overline{P}_{\lambda}({\bf \hat{n}} \cdot {\bf J}) $ described in  the previous section are related to the Chebyshev polynomial operators $f_{\lambda} ^{(j)} (J_z) $ according to
\begin{equation}
P_{\lambda}({\bf J}) = 
  \left[  \left[ {\bf J}^2 \right]^{\!l}  \right]^{\!-1/2} \, \overline{P}_{\lambda}(J_z) =\sqrt{\displaystyle\frac{2j+1}{2\lambda+1}  }  \,  f_{\lambda} ^{(j)} (J_z) 
\end{equation}
This relation shows that the Chebyshev polynomials $  f_{\lambda} ^{(j)} (m) $ can be calculated from the diagonal matrix elements of the Legendre 
polynomial operators $ \overline{P}_{\lambda}(J_z) $ or  $P_{\lambda}({\bf J})$. 

Although Schwinger  \cite{schwinger:majorana}  did not cite a reference for the generating function of Eq.(\ref{swinggen}), Schwinger himself had derived this generating function. His derivation was included in a  1952 technical report \cite{schwingerosti}, later published in a book collection of reprints and original papers \cite{schwingerbiedenvandam}, and  as a book \cite{schwingerbook} with the same title as the technical report. Additional discussions of this generating function can be found in books by Schwinger et al. \cite{schwingerbook2} and by Garg \cite{garg}. 

 As we shall demonstrate, Schwinger's generating function \cite{schwinger:majorana,schwingerosti,schwingerbiedenvandam,schwingerbook}  is equivalent to the Herglotz generating function \cite{couranth,herglot,ligarg} for the solid harmonic operators $\mathcal{Y}_{lm}({\bf J}) $
\begin{eqnarray}
e^{\zeta {\bf \hat{a}} \cdot {\bf J}} & = & \sum_{lm} \sqrt{\frac{4\pi}{2l+1}} \;
\frac{\zeta^{l} \lambda^{m}}{\sqrt{(l+m)!(l-m)!}} \; \mathcal{Y}_{lm}({\bf J})   \\
\mbox{where }  \;\;\;\;\;  {\bf \hat{a}} & = & {\bf \hat{z}} -\frac{\lambda}{2} ({\bf \hat{x}} +i {\bf \hat{y}}) + \frac{1}{2\lambda}  ({\bf \hat{x}} -i {\bf \hat{y}})
\label{herglotz}
\end{eqnarray}
Introducing the definitions 
\begin{eqnarray}
{\hat A} & = &  J_z -\frac{\lambda}{2} (J_x+iJ_y) + \frac{1}{2\lambda}  (J_x - iJ_y) \equiv ({\bf \hat{a}} \cdot {\bf J})  \nonumber \\
\lambda & = & z_+/ z_{-}
\end{eqnarray}
and after a slight rearrangement, Schwinger's  generating function \cite{schwinger:majorana,schwingerosti,schwingerbook}  of Eq.(\ref{swinggen}) can be rewritten as 
\begin{equation}
\frac{{\hat A}^l}{l !} =  \sum_{m=-l}^{l} \; \sqrt{\frac{4\pi}{2l+1} }\; \frac{ \lambda^{m} } { \left[(l+m)!(l-m)!\right]^{1/2}} \; 
\mathcal{Y}_{lm}({\bf J}) \label{swingsimp}
\end{equation}
After multiplying both sides of Eq.(\ref{swingsimp}) by $\zeta^l$, and summing both sides over $l$, we obtain the Herglotz generating function \cite{couranth,herglot,ligarg} of Eq.(\ref{herglotz}):
\begin{equation}
\sum_{l=1}^{\infty} \frac{[\zeta{\hat A}]^l}{l !} = \sum_{l=1}^{\infty} \frac{(\zeta{\bf \hat{a}} \cdot {\bf J})^l}{l !} \equiv e^{\zeta {\bf \hat{a}} \cdot {\bf J}}   = 
 \sum_{l=1}^{\infty}  \sum_{m=-l}^{l} \; \sqrt{\frac{4\pi}{2l+1} }\; \frac{\zeta^l \lambda^{m} } { \left[(l+m)!(l-m)!\right]^{1/2}} \; 
\mathcal{Y}_{lm}({\bf J})
\end{equation}
Appendix A  of the first English edition of Courant and Hilbert \cite{couranth,herglot} published in 1953 cited Herglotz for his formula, but none of the pre-war German editions had this appendix. This meant that the generating function \cite{schwinger:majorana,schwingerosti,schwingerbook} documented by Schwinger in a technical report  \cite{schwingerosti} a year earlier in 1952 anticipated the Herglotz generating function \cite{couranth,herglot,ligarg}, and so there is some justification for renaming this generating function the Schwinger-Herglotz generating function. 

\subsection{Unit tensor (Wigner) operators}

In their comments on Schwinger's interpretation \cite{schwinger:majorana}  of the Majorana formula  \cite{emajorana} , Biedenharn and Louck 
\cite{biedenharn} observe that $ P_{l}(j,m) $ denotes the matrix element of the unit tensor operator (alias Wigner operators \cite{biedenharn})
\begin{equation}
 P_{l}(j,m)  \equiv \langle jm | 
\left\langle\begin{array}{ccc} & l & \\
2l & & 0 \\
& l & 
\end{array} \right\rangle | jm \rangle \label{unittensor}
\end{equation}
The Wigner operators can be defined by their action on the angular momentum basis $|jm \rangle$ \cite{biedenharn}, and in particular, the shift action is defined by  \cite{biedenharn}
\begin{equation}
\left\langle\begin{array}{ccc} & J+\Delta & \\
2J & & 0 \\
& J+M & 
\end{array} \right\rangle  | jm \rangle = C_{jmJM}^{j+\Delta\;  m+M} \;  |j+\Delta,m+M \rangle
\end{equation}
and so in particular
\begin{eqnarray}
\left\langle\begin{array}{ccc} & l & \\
2l & & 0 \\
& l & 
\end{array} \right\rangle | jm \rangle &  = & C_{jml0}^{jm} \;  | jm \rangle \\
& = & \sqrt{\frac{2j+1}{2l+1}}\;\boxed{ (-1)^{j-m} \;  C_{jmj-m}^{l0} } \; | jm \rangle \label{wigbox} \\
& = & \sqrt{\frac{2j+1}{2l+1}}\; f_l^{(j)}(m) \;  | jm \rangle \label{wigbox2} 
\end{eqnarray}
The ``boxed" term in Eq.(\ref{wigbox}) defines the Chebyshev polynomial $f_l^{(j)}(m)$ in Eq.(\ref{wigbox2}). Then, using the result of Eq.(\ref{wigbox2}), the unit tensor matrix element of Eq.(\ref{unittensor}) is evaluated as
\begin{equation}
P_{l}(j,m)  \equiv \langle jm | 
\left\langle\begin{array}{ccc} & l & \\
2l & & 0 \\
& l & 
\end{array} \right\rangle | jm \rangle = \sqrt{\frac{2j+1}{2l+1}}\; f_l^{(j)}(m)
\end{equation}
in agreement with the matrix element $P_{l}(j,m)  = \langle jm| \, P_{l}({\bf J})  \, | jm \rangle $ of Schwinger's \cite{schwinger:majorana} Legendre polynomial operator $P_{l}({\bf J}) $   in Eq.(\ref{swingmatrixelement}).

\newpage

\section{Appendix B}
As noted by Varshalovich et al. \cite{varshal1:ang}, an alternative explicit form of
  ${\cal D}^J_{MM^{\prime}}(\psi, {\bf \hat{n}})  \equiv{\cal D}^J_{MM^{\prime}}(\psi; \Theta, \Phi)$ can be obtained directly from 
${\cal D}^J_{MM^{\prime}}(\alpha, \beta, \gamma) $ by changing variables $(\psi; \Theta, \Phi) \rightarrow (\alpha, \beta, \gamma)$ with the aid 
of the following relations (and the corresponding inverse relations):
\begin{eqnarray}
\sin \bfrac{\beta}{2} & = & \sin \Theta\; \sin \bfrac{\psi}{2} \nonumber \\
\tan \frac{\alpha + \gamma}{2} & = & \cos \Theta \; \tan \bfrac{\psi}{2} \nonumber \\
\frac{\alpha - \gamma}{2} & = & \Phi -\bfrac{\pi}{2}
\label{ieuanglec}
\end{eqnarray}
The result of this variable change is the following \cite{varshal1:ang}
\begin{eqnarray}
{\cal D}^J_{MM^{\prime}}(\psi; \Theta, \Phi) &  = &  i^{M-M^{\prime}} \; e^{-i(M-M^{\prime})\Phi}
\left( \displaystyle\frac{1-i\tan \frac{\psi}{2} \cos \Theta}{\sqrt{ 1+\tan^2 \frac{\psi}{2} \cos^2 \Theta}} \right)^{\!\!M+M^{\prime}}
d^J_{MM^{\prime}}(\xi) \label{EulerEquiv}\;\;\;\;  \label{Dangleaxis} \\
\mbox{where}\;\;\;\; \sin \frac{\xi}{2} & =  & \sin \bfrac{\psi}{2} \; \sin \Theta \label{sinequiv}
\end{eqnarray}
The expression for ${\cal D}^J_{MM^{\prime}}(\psi; \Theta, \Phi) $ given in Eq.(\ref{Dangleaxis}) will now be used  to evaluate 
${\cal D}^J_{MM^{\prime}}(\psi, {\bf \hat{n}})$ in two cases described in the next sections. 

\subsection{Case I: ${\cal D}_{00}^{L}(\psi, {\bf \hat{n}})  = d_{00}^{\,L}(\xi) \equiv P_L(\cos \xi) = P_L(\cos \beta)$}

Since $M=M^{\prime}=0$, from Eq.(\ref{EulerEquiv}) we find
\begin{eqnarray}
{\cal D}_{00}^{L}(\psi, {\bf \hat{n}}) &  =  & d_{00}^{\,L}(\xi) \equiv P_L(\cos \xi)  \label{LLegend}\\
\mbox{where \cite{brinksatch:ang}}\;\;\;\; d_{00}^{\,L}(\xi) & = & P_L(\cos \xi)
\end{eqnarray}
Using Eq.(\ref{sinequiv}), the argument of the Legendre polynomial $P_L(\cos \xi)$ in Eq.(\ref{LLegend}) can therefore be rewritten as
\begin{equation}
\cos \xi \equiv 1-2  \sin^2 \displaystyle\frac{\xi}{2}  = \boxed{1-2  \sin^2\displaystyle\bfrac{\psi}{2} \sin^2 \theta =
\cos \beta} \label{convert}
\end{equation}
so that 
\begin{equation}
{\cal D}_{00}^{L}(\psi, {\bf \hat{n}}) = P_L(\cos \beta)
\end{equation}
The ``boxed" equivalence of Eq.(\ref{convert}) follows from this relation \cite{abragamtext}
\begin{equation}
\cos \beta = 1-2  \sin^2\displaystyle\bfrac{\psi}{2} \sin^2 \theta \label{abragam}
\end{equation}
where $\psi=|\gamma {\bf H}_e|t$, and $\theta$ is the angle between the direction of the effective field {\bf H}$_e$ and the static applied field 
{\bf H}$_0$.

\subsection{Case II: ${\cal D}^{\lambda}_{\mu 0}(\theta, {\bf \hat{n}}_{\bot}) \equiv {\cal D}^{\lambda}_{\mu 0}(\theta; \frac{\pi}{2}, \phi + \frac{\pi}{2} ) 
\equiv C_{\lambda \mu}^{\star}(\theta, \phi)$}

Since $\Theta =\pi/2$, $\sin \Theta  =  1$, $\cos \Theta  =  0$, and $ \xi = \theta$,  and so using the expression of Eq.(\ref{EulerEquiv}), we find
\begin{eqnarray}
{\cal D}^{\lambda}_{\mu 0}(\theta, {\bf \hat{n}}_{\bot}) &  \equiv & {\cal D}^{\lambda}_{\mu 0}(\theta; \bfrac{\pi}{2}, \phi + \bfrac{\pi}{2} )\\
& = & i^{\mu} \, e^{-i\mu \phi} \, e^{-i \mu \frac{\pi}{2}} \; \boxed{d^{\lambda}_{\mu 0} (\theta)} \label{reducedd}\\
& = & e^{-i\mu \phi} \, \boxed{(-1)^{\mu} \sqrt{\displaystyle\frac{(\lambda-\mu)!}{(\lambda-\mu)!}} \; P_{\lambda}^{\mu} (\theta)} \label{replacerdd}\\
& = & C_{\lambda \mu}^{\star}(\theta, \phi)
\end{eqnarray}
The reduced matrix element in the ``boxed" term of Eq.(\ref{reducedd}) has been replaced by its equivalent \cite{brinksatch:ang} in the ``boxed" term 
of Eq.(\ref{replacerdd}). 

\newpage

\section{Appendix C}
\subsection{A closed-form expression for the  Fourier-Legendre series expansion of the spin transition probability $\mbox{P}^{(j)}_{j,-j} (t) $}
Consider the following Fourier-Legendre series expansion  (identity {\bf 5.10.1(17)} in Prudnikov et al. \cite{prudnikov})
\begin{equation}
(a-1) \left[\frac{1-x}{2} \right]^{\!(a-2)} = \;  \sum_{k=0}^{\infty} (2k+1) \frac{(2-a)_k}{(a)_k} \, P_k(x)   \;\;\;\;\;\;\;\;(|x|<1; \;\;a>5/4) \label{identprud}
\end{equation}
where the Pochhammer symbol $(a)_n$ is defined in terms of the Gamma function  $\Gamma(z)$ as follows \cite{prudnikov}
\begin{equation}
(a)_n = \frac{\Gamma(a+n)}{\Gamma(a)}
\end{equation}
If we let $a=n+2$ in the identity of Eq.(\ref{identprud}), where $n \geq 0$ is an integer, then we obtain
\begin{equation}
 \left[\frac{1-x}{2} \right]^{\!n} = \frac{1}{n+1} \sum_{k=0}^{\infty} (2k+1) \frac{(-n)_k}{(n+2)_k} \, P_k(x) \label{identprud2}
\end{equation}
Then using the following Pochhammer symbol expressions \cite{prudnikov}  
\begin{eqnarray}
(-n)_k & = &  (-1)^k \frac{n!}{(n-k)!} \\
(n+2)_k & = &\frac{ \Gamma(n+2+k)}{\Gamma(n+2)} =\frac{(n+1+k)!}{(n+1)!}
\end{eqnarray}
we see that the Fourier-Legendre series expansion of Eq.(\ref{identprud2}) terminates when $k=n$, and that  this expansion now simplifies to read
\begin{equation}
\left[\frac{1-x}{2} \right]^{\!n} =  \left[n!\right]^2\sum_{k=0}^{n} \frac{(-1)^k (2k+1)}{(n-k)!(n+1+k)!} \, P_k(x) \label{identprud3}
\end{equation}
Mathematica$^{\circledR}$ \cite{wolfram} can be also used to verify this expansion. For example, by using the following input command to implement the summation of 
Eq.(\ref{identprud3}) with $n=101$
\vspace{8mm}
\begin{lstlisting}
In[1] = Simplify[2^(101) Sum[(-1)^L (2 L + 1)(101!)^2 
   LegendreP[L, x]/((101 - L)!(101 + L + 1)!),{L, 0, 101}]]
\end{lstlisting}
the expected output is obtained as
\vspace{8mm}
\begin{lstlisting}
Out[2] = -(-1 + x)^101
\end{lstlisting}
An amusing consequence of Eq.(\ref{identprud3}) is the following Fourier-Legendre series expansion of 
$[a+b]^n$:
\begin{equation}
 [a+b]^n= [2a]^n \; \!\! \left[n!\right]^2\sum_{k=0}^{n} \frac{(2k+1)}{(n-k)!(n+1+k)!} \, P_k(b/a)
\end{equation}

\newpage

\section{Appendix D}
\subsection{Using $f^{(j)}_{\lambda}({\bf \hat{n}} \cdot {\bf J})$  to evaluate the generalized characters $\mbox{{\large $\chi$}}_{\lambda}^{(j)}(\psi) $ }
Using the ``boxed" trace relation of Eq.(\ref{charcheby}), the generalized character functions can be written as
\begin{equation}
\mbox{{\large $\chi$}}_{\lambda}^{(j)}(\psi) =  i^{3 \lambda}  \sqrt{\displaystyle\frac{2j+1}{2\lambda+1}} \; 
\mbox{Tr} \! \left[ f^{(j)}_{\lambda}({\bf \hat{n}} \cdot {\bf J})  \; \hat{{\cal D}}^{(j)} \! (\psi, {\bf \hat{n}}) \right]
\end{equation}
Then, putting $\lambda=1$ and $J=1/2$, the generalized character function {\large $\chi$}$_1^{1/2}(\psi) $ can be evaluated as 
\begin{eqnarray}
 \mbox{{\large $\chi$}}_1^{(1/2)}(\psi) & = & -i \sqrt{\frac{2}{3}} \; \mbox{Tr} \! \left[ f^{(1/2)}_{1}({\bf \hat{n}} \cdot {\bf J}) \; \hat{{\cal D}}^{(1/2)}  (\psi, {\bf \hat{n}}) \right] \\
& = & -i \sqrt{\frac{2}{3}}  \; \mbox{Tr} \! \left[   \sqrt{2} \; J_z \left(\mathds{1} \cos \bfrac{\psi}{2} + 2i \, J_z \, \sin \bfrac{\psi}{2} \right)    \right] \label{tracesimp} \\
& = & \frac{4}{\sqrt{3}} \, \sin \bfrac{\psi}{2}  \; \mbox{Tr} \! \left[  J_z^2 \right] \label{charex} \\
& = & \frac{2}{\sqrt{3}} \, \sin \bfrac{\psi}{2} \label{charex2}
\end{eqnarray}
a result which agrees with $ \mbox{{\large $\chi$}}_1^{(1/2)}(\psi) $ tabulated in Varshalovich et al. \cite{varshal1:ang}. Because the trace is invariant with respect to a change in basis, the trace of Eq.(\ref{tracesimp}) was evaluated in a representation in which $({\bf \hat{n}} \cdot {\bf J})$ is diagonal.  In addition, the following identities have been used to arrive at the final result in Eq.(\ref{charex2}):
\begin{eqnarray}
f^{(1/2)}_{1}({\bf \hat{n}} \cdot {\bf J}) & = &  \sqrt{2} \; J_z \;\;\; \;\;\; \;\;\;\;\;\;\;\ \;\;\; \;\;\;\;\;\;\;\ \;\;\; \;\;\;\;\mbox{(see Table II)} \\
\hat{{\cal D}}^{(1/2)}  (\psi, {\bf \hat{n}})  & =  & \mathds{1} \cos \bfrac{\psi}{2} + 2i \, J_z \, \sin \bfrac{\psi}{2}  
  \;\;\; \;\;\;\;\;\;\;\mbox{(Reference \cite{varshal1:ang})} \\
 \mbox{Tr} \! \left[  J_z^2 \right]  & = & \sum_{m=-1/2}^{1/2} \!m^2 = \frac{1}{2} \\
 \mbox{Tr} \! \left[  J_z \right]  & = & 0
\end{eqnarray}

\newpage

\section{Appendix E}

The composite irreducible product tensor  ${\bf X}^K_Q $ of rank $K$ is defined by the following combination of irreducible tensors 
${\bf T}^k_q $ and ${\bf U}^{k^{\prime}}_{q^{\prime}} $ of rank $k$ and $k^{\prime}$, respectively:
\begin{equation}
{\bf X}^K_Q =  \left\{ {\bf T}^{k}_{q} \circledast {\bf U}^{k^{\prime}}_{q^{\prime}} \right \}^{K}_{Q} = \sum_{\substack{q,q^{\prime} \\q+q^{\prime}=Q }}
 {\bf T}^{k}_{q} \; {\bf U}^{k^{\prime}} _{q^{\prime}}  \;
C^{KQ}_{kqk^{\prime}q^{\prime}}
\end{equation}
where $C^{KQ}_{kqk^{\prime}q^{\prime}}$ is a Clebsch-Gordan coefficient which vanishes unless $q + q^{\prime} =Q$. 
In general, 
  \begin{equation}
 \left\{{\bf R}^{(k)}   \circledast  {\bf S}^{(k)} \right\}^0_0 =\sum_{\substack{q,q^{\prime} \\ q+q^{\prime}=0}} {\bf R}^{(k)}_q  \;  
 {\bf S}^{(k)}_{q^{\prime}} \; C^{00}_{kqkq^{\prime}}
\label{dot}
 \end{equation}
 But since the Clebsch-Gordan coefficient is evaluated as~\cite{brinksatch:ang}
  \begin{equation}
   C^{00}_{kqkq^{\prime}} = \frac{(-1)^{k-q}}{\sqrt{2k+1}} \; \delta_{q,-q^{\prime}}
  \end{equation}
 the double sum of Eq.(\ref{dot}) reduces to a single sum
 \begin{eqnarray}
  \left\{{\bf R}^{(k)}   \circledast {\bf S}^{(k)} \right\}^0_0 & = &  \frac{(-1)^{k}}{\sqrt{2k+1}} \sum_{q} (-1)^{-q} \;  {\bf R}^{(k)}_q \; {\bf S}^{(k)}_{-q}  \nonumber \\
 & = &  \frac{(-1)^{k}}{\sqrt{2k+1}} \;  {\bf R}^{(k)}  \cdot  {\bf S}^{(k)} 
 \end{eqnarray}
where ${\bf R}^{(k)}  \cdot  {\bf S}^{(k)}  = \displaystyle \sum_{q} (-1)^{-q} \;  {\bf R}^{(k)}_q \; 
 {\bf S}^{(k)}_{-q}$ defines  the scalar product of two tensors.

A more general result for the recoupling of four arbitrary 
commuting tensors can be defined in terms of a 9$j$-symbol \cite{brinksatch:ang} as follows
\begin{eqnarray}
&  & \left\{ \left\{ {\bf S}^{k_1}_{q_1} \circledast {\bf T}^{k_2}_{q_2} \right \}^{k_{12}}_{q_{12}} \circledast \left\{ {\bf U}^{k_3}_{q_3} \circledast {\bf V}^{k_4}_{q_4} \right \}^{k_{34}}_{q_{34}} \right \}^K_Q  \nonumber \\
& = &
\sum_{k_{13},k_{24}} \Pi_{k_{13} k_{24} k_{12}k_{34}}  
 \left\{\begin{array}{ccc} k_1 & k_2 & k_{12} \\
k_3 & k_4  & k_{34} \\
k_{13} & k_{24} & K\end{array}\right\} \left\{ \left\{ {\bf S}^{k_1}_{q_1} \circledast {\bf U}^{k_3}_{q_3} \right \}^{k_{13}}_{q_{13}} \circledast
\left\{ {\bf T}^{k_2}_{q_2} \circledast {\bf V}^{k_4}_{q_4} \right \}^{k_{24}}_{q_{24}} \right \}^K_Q \;\;\;\; \;\;\;\; \;\;\;\;
\label{rec4ang}
\end{eqnarray}
where as a matter of notation, in this equation, and all that follow, we find it convenient to define 
\cite{varshal1:ang}:
\begin{equation}
\Pi_ {abc \ldots d} = \sqrt{(2a+1)(2b+1)(2c+1) \ldots (2d+1)}
\end{equation}

In Eq.(\ref{rec4ang}), set $K=Q=0$, from which it follows that $k_{12} = k_{34}$ and $k_{13} = k_{24}$. In addition, set $k_{12} = k_{34}=0$, and consider the particular case where we recouple four irreducible tensors of rank 1, so that $k_1=k_2=k_3=k_4=1$. In particular, we suppose 
\begin{eqnarray}
 {\bf S}^{k_1} & \equiv  &  {\bf I}_1 \nonumber \\
 {\bf U}^{k_3} & \equiv &  {\bf I}_2 \nonumber \\
 {\bf T}^{k_2} & = &  {\bf V}^{k_4} \equiv  {\bf r} 
 \end{eqnarray}
Then, from Eq.(\ref{rec4ang}), we have 
\begin{equation}
\Big\{  \left\{ {\bf I}_1^{}  \circledast {\bf r} \right\}^0 \circledast  \left\{ {\bf I}_2 \circledast {\bf r} \right\}^0 \Big\}^0_0 = \sum_{k_{13},k_{24}}\Pi_{k_{13} k_{24}} 
 \left\{\begin{array}{ccc} 1 & 1 & 0 \\
1 & 1  & 0 \\
k_{13} & k_{24} & 0\end{array}\right\}
\Big\{ \left\{ {\bf I}_1 \circledast {\bf I}_2 \right\}^{k_{13}} \circledast \left\{ {\bf r} \circledast {\bf r} \right\}^{k_{24}} \Big\}^0_0
\label{appl}
\end{equation}
Now since $k_{13}$ and $k_{24}$ satisfy the triangle inequalities
\begin{eqnarray}
|k_1-k_3| & \leq & k_{13} \leq |k_1+k_3| \nonumber \\
|k_2-k_4| & \leq & k_{24} \leq |k_2+k_4|
\end{eqnarray}
and $k_1=k_2=k_3=k_4=1$, $k_{13}$ and $k_{24}$ are summed from 0 to 2. Note however that when $k_{24} =1$, $ \left\{ {\bf r} \circledast {\bf r} \right\}^1$ vanishes since 
$ \left\{ {\bf r} \circledast {\bf r} \right\}^1  \propto  ({\bf r} \times {\bf r}) =0$. Therefore, in the sum of Eq.(\ref{appl}), only two terms contribute: 
\begin{eqnarray}
& & \Big\{ \left\{ {\bf I}_1 \circledast {\bf r} \right\}^0 \circledast  \left\{ {\bf I}_2 \circledast {\bf r} \right\}^0 \Big\}^0_0 \nonumber \\
&  =  & 
\left\{\begin{array}{ccc} 1 & 1 & 0 \\
1 & 1  & 0 \\
0 & 0 & 0\end{array}\right\}
\Big\{ \left\{ {\bf I}_1 \circledast {\bf I}_2 \right\}^0 \circledast \left\{ {\bf r} \circledast {\bf r} \right\}^0 \Big\}^0_0 
 + 5 \left\{\begin{array}{ccc} 1 & 1 & 0 \\
1 & 1  & 0 \\
2 & 2 & 0\end{array}\right\}
\Big\{ \left\{ {\bf I}_1 \circledast {\bf I}_2 \right\}^2 \circledast  \left\{ {\bf r} \circledast {\bf r} \right\}^2 \Big\}^0_0  \nonumber \\
& & 
\label{sum9j}
\end{eqnarray}
The 9$j$-symbols in Eq.(\ref{sum9j}) can be evaluated by expressing them in terms of 6$j$-symbols as follows \cite{varshal1:ang}
\begin{equation}
 \left\{\begin{array}{ccc} a & b & e \\
c & d  & e \\
f & f & 0\end{array}\right\} = \frac{(-1)^{b+c+e+f}}{\sqrt{(2e+1)(2f+1)}} \left\{\begin{array}{ccc} a & b & e \\
d & c  & f \end{array}\right\}
\end{equation}
and in turn, all the 6$j$-symbols required can then be evaluated using tables  \cite{roten:3j6j}. Replacing all the recoupling coefficients 
in Eq.(\ref{sum9j}) then yields the following relation
\begin{equation}
\Big\{ \left\{ {\bf I}_1  \circledast   {\bf r} \right\}^0  \circledast   \left\{ {\bf I}_2  \circledast   {\bf r} \right\}^0 \Big\}^0_0 = 
\frac{1}{3} 
\Big\{ \left\{ {\bf I}_1  \circledast   {\bf I}_2 \right\}^0  \circledast   \left\{ {\bf r}  \circledast   {\bf r} \right\}^0 \Big\}^0_0
+ \frac{\sqrt{5}}{3} 
\Big\{ \left\{ {\bf I}_1  \circledast   {\bf I}_2 \right\}^2  \circledast    \left\{ {\bf r}  \circledast  {\bf r} \right\}^2 \Big\}^0_0
\label{ssum9j}
\end{equation}
Then, after expressing the rank zero recoupled tensors in Eq.(\ref{sum9j}) as scalar products using the general relation
\begin{equation}
\left\{ {\bf a}  \circledast   {\bf b} \right\}^0_0  =  -\frac{{\bf a} \cdot {\bf b}}{\sqrt{3}} 
 \label{rex1}
 \end{equation}
 and using the following relations \cite{brinksatch:ang}
  \begin{eqnarray}
  \left\{ {\bf r}  \circledast  {\bf r} \right\}^2_q & = &  \sqrt{\frac{2}{3}} \; {\bf r}^2 \;  {\bf C}^2_q \nonumber \\
  \mbox{where}\;\; {\bf C}^2_q & = & \sqrt{\frac{4 \pi}{5}} \; {\bf Y}^2_q
  \label{Racah}
  \end{eqnarray}
  the following recoupling expression for the classical dipolar Hamiltonian is obtained
\begin{eqnarray}
W_{12} & = & a \left[ {\bf I}_1 \cdot  {\bf I}_2 -3 \frac{({\bf I}_1 \cdot {\bf r}) ({\bf I}_2 \cdot {\bf r}) }{r^2 }\right] \nonumber \\
 & = & -\sqrt{6} a \left\{ {\bf I}_1  \circledast   {\bf I}_2 \right\}^2 \cdot  {\bf C}^2_q \nonumber \\
& = & -\sqrt{6} a \left\{ {\bf I}_1  \circledast  {\bf I}_2 \right\}^2 \cdot  \sqrt{\frac{4 \pi}{5}}  {\bf Y}^2_q  \nonumber \\
& = & -\sqrt{\frac{24 \pi}{5}} \sum_q (-1)^q \left\{ {\bf I}_1  \circledast   {\bf I}_2 \right\}^2_q  {\bf Y}^2_{-q}
\end{eqnarray}

 \newpage
 {\bf Table I}\\

 Title: Chebyshev polynomial basis operators $f^{(j)}_{\lambda}({\bf \hat{n}} \cdot {\bf J}) \;\;(\lambda \leq 2j; \;\; \kappa \equiv j(j+1))$ \\
\vspace{5mm}
 Caption: As described in Section  {\bf 2.2}, all elements were generated following the procedure described by Corio \cite{corio:siam}.

 \ \\ 
\begin{sidewaystable}
\begin{tabular}{||c||c|c|c||} \hline & & & \\  \hspace{1mm} $\lambda$  \hspace{1mm}&  {\large $j=1/2$} & {\large $j=1$}
& {\large $j=3/2$}  \\
& & &  \\   \hline \hline
 & & &  \\
 \hspace{1mm} 0  \hspace{1mm} &  {\large $f^{(\frac{1}{2})}_0({\bf \hat{n}} \cdot {\bf J})= \displaystyle\frac{1}{\sqrt{2}} \,\mathds{1} $} &  {\large $f^{(1)}_0({\bf \hat{n}} \cdot {\bf J}) = \displaystyle\frac{1}{\sqrt{3}} \,\mathds{1} $} & \hspace{5mm}  {\large $f^{(\frac{3}{2})}_0({\bf \hat{n}} \cdot {\bf J})= \displaystyle\frac{1}{\sqrt{4}} \,\mathds{1} $}
\\
 & & &  \\
\hline

 & &  & \\ 

 \hspace{1mm} 1  \hspace{1mm}& \hspace{1mm} {\large $f^{(\frac{1}{2})}_1({\bf \hat{n}} \cdot {\bf J})= \sqrt{2} \, \left[({\bf \hat{n}} \cdot {\bf J})\right] $} \hspace{1mm}
 &   {\large $f^{(1)}_1({\bf \hat{n}} \cdot {\bf J}) = \displaystyle\frac{1}{\sqrt{2}} \, \left[({\bf \hat{n}} \cdot {\bf J})\right] $} &  {\large $f^{(\frac{3}{2})}_1({\bf \hat{n}} \cdot {\bf J})= \displaystyle\frac{1}{\sqrt{5}} \, \left[({\bf \hat{n}} \cdot {\bf J})\right] $}
\\
& &  & \\
 \hline
 & & &  \\ 
 \hspace{1mm} 2  \hspace{1mm}& {\large   --- }  &   \hspace{1mm} {\large $f^{(1)}_2({\bf \hat{n}} \cdot {\bf J}) = \displaystyle\frac{1}{\sqrt{6}} \left[3 ({\bf \hat{n}} \cdot {\bf J})^2-\kappa \,\mathds{1} \right] $}  \hspace{1mm}
&   {\large $f^{(\frac{3}{2})}_2({\bf \hat{n}} \cdot {\bf J}) = \displaystyle\frac{1}{6} \left[3 ({\bf \hat{n}} \cdot {\bf J})^2-\kappa  \,\mathds{1} \right] $}   \\ 
& & &  \\ 
\hline
& &  &  \\ 
 \hspace{1mm} 3  \hspace{1mm} &  {\large   --- }  &  {\large   ---}  
&   \hspace{1mm}    {\large $f^{(\frac{3}{2})}_3({\bf \hat{n}} \cdot {\bf J}) = \displaystyle\frac{1}{3\sqrt{5}} \left[5 ({\bf \hat{n}} \cdot {\bf J})^3-(3\kappa -1)({\bf \hat{n}} \cdot {\bf J})  \right] $}  \hspace{1mm}    \\
& &  & \\ 
\hline

\end{tabular}
\end{sidewaystable}

\newpage
 {\bf Table II}\\
 
 Title: A comparison of Chebyshev polynomial scalars $f^{(j)}_{\lambda}(m)  $ and operators $f^{(j)}_{\lambda}({\bf \hat{n}} \cdot {\bf J})$. \\
\vspace{5mm} 
 Caption: First column compares the definitions of the Chebyshev polynomial scalars adopted by various authors, while the second column compares 
the definitions of the Chebyshev polynomial operators  adopted by the same authors. Note that Corio  \cite{corio:ortho, corio:siam} suppressed  the explicit dependence on the spin angular momentum $j$ in his definitions of both the Chebyshev polynomial scalars $  Z_n(x)    $ and operators $U_{n}({\bf \hat{n}} \cdot {\bf J})$, while Meckler  \cite{meckler:angular} suppressed the explicit dependence on  the spin angular momentum $S$ in his definition of the Chebyshev polynomial operators $A^{(n})/g_n $, where 
$g_n = [n!]^2 \sqrt{2S+1+n}\left[2^n (2n-1)!! \sqrt{(2n+1)(2S-n)!}\right]^{-1}$.  Using Filippov's notation  \cite{filippov2:thesis} 
for the Chebyshev polynomial operators, Meckler's operators \cite{meckler:angular} can be written as $A^{(n})/g_n \equiv f^{(S)}_n({\bf \hat{a}} \cdot {\bf S}) $.  Just as the $(2j+1)$ Chebyshev polynomial  scalars $ p_n(S,-j), \ldots,  p_n(S,j)$ defined by Meckler  \cite{meckler:angular}   are the diagonal matrix elements of $A^{(n)}/g_n  $ in a representation where 
$({\bf \hat{a}} \cdot {\bf S})  $ is diagonal,  the $(2j+1)$ Chebyshev polynomial  scalars $ Z_n(0), \ldots,  Z_n(2j)$ defined by Corio 
\cite{corio:ortho, corio:siam}  are the diagonal matrix elements of $ U_{n}({\bf \hat{n}} \cdot {\bf J})  $ in a representation where 
$({\bf \hat{n}} \cdot {\bf J})  $ is diagonal.

 \ \\ 

\begin{tabular}  {||c|c|c||}  \hline  &  & \\  
{\large Scalars}  & {\large Operators} & {\large Authors} \\
 & & \\ 
\hline  \hline
 & & \\
 {\large $f^{(j)}_{\lambda}(m)  $} &
 {\large $f^{(j)}_{\lambda}({\bf \hat{n}} \cdot {\bf J})$} & {\large Filippov} \cite{filippov2:thesis} \\
 \hspace{8mm} & & \\  & & \\   \hline
  & & \\

{\large$  p_n(S,m)  $} & 
{\large $\displaystyle\frac{A^{(n)}}{g_n}        $} & {\large Meckler} \cite{meckler:majorana, meckler:angular}
\\
 
 &  &  \\
 {\large $(n \equiv \lambda; S \equiv j) $} &  {\large $(n \equiv \lambda) $} & \\
 
&  &  \\
 \hline
 & & \\ 

{\large $  Z_n(x)    $} & {\large $U_{n}({\bf \hat{n}} \cdot {\bf J})   $} & {\large Corio} \cite{corio:ortho, corio:siam}
\\
&  &  \\
 {\large $(n \equiv \lambda; x \equiv j+m) $} & {\large $(n \equiv \lambda) $}   & \\
&  &  \\
\hline

\end{tabular}

\newpage
 {\bf Table III}\\
 
 Title: Chebyshev polynomial definitions and recursion relations. \\
\vspace{5mm}

 Caption: The top box in the first column gives  Filippov's definition  \cite{filippov2:thesis} of the Chebyshev polynomials $ f_L^{(j)} (m) $ in terms of 
 Bateman's \cite{bateman} Chebyshev polynomials $ t_{L}(j+m,2j+1) $, which are defined in the bottom box of the first column using 
 finite differences \cite{footnote2}. Two versions of the recursion relations are given in the second column. The form of the first version is that given by 
Filippov \cite{filippov2:thesis} for the Chebyshev polynomials $ f_L^{(j)} (m) $, although equivalent forms were derived previously by Meckler \cite{meckler:majorana, meckler:angular} and by Corio \cite{corio:ortho, corio:siam}. The second version of the recursion relation is that given by 
Varshalovich et al. \cite{varshal1:ang} for the Clebsch-Gordan coefficients $C^{L \, 0}_{jmj-m}$, which are identical to the Chebyshev polynomials 
$ f_L^{(j)} (m) $ to within a phase-factor. The bottom boxes  also define functions $F(L,j)$ and $ G(a,b) $  used for the matrix elements (top box, first column)  and recursion relations (top box, second column), respectively. 

 \ \\ 
\begin{sidewaystable}
\begin{tabular}  {||l|l|c||} \hline  &   \\   {\large Matrix Element Definition via Bateman's \cite{bateman} }
&  \hspace{42mm} {\large Recursion Relations}  \\
 \hspace{8mm}{\large Chebyshev Polynomials  \;  {\large $\boxed{t_{L}(j+m,2j+1) }$}} & \\  & \\   \hline \hline
  & \\

    \hspace{22mm}  {\large $   f_L^{(j)} (m) =  \langle jm| \; f_{L}^{(j)} ( J_z)  \;  |jm \rangle         $}  
&    {\large $   G(L+1,j) \,  f_{L+1}^{(j)} (m) -2m  \, f_L^{(j)} (m)  + G(L,j)   \, f_{L-1}^{(j)} (m) =  0               $}   \\ 

&           \\

  \hspace{39mm}   {\large $ = F(L,j) \;t_L(j+m,2j+1)  $}   &    {\large $   G(L+1,j) \,  C^{L+1 \, 0}_{jmj-m} -2m  \, C^{L \, 0}_{jmj-m} + G(L,j)   \, C^{L-1 \, 0}_{jmj-m} =  0               $}         \\
&          \\
 \hspace{39mm}   {\large $ = (-1)^{j-m} \; C^{L \, 0}_{jmj-m}  $}  &          \\
&          \\
\hline \hline

&          \\
&          \\
 {\large $(1) \;\;\; F(L,j)  = \left[\displaystyle\frac{(2L+1)(2j-L)!}{(2j+L+1)!} \right]^{1/2}$} & 
 {\large $ G(a,b)  =   \left[\displaystyle\frac{a^2 ((2b+1)^2-a^2)}{4a^2-1} \right]^{1/2}  $}     \\
&          \\

&          \\
{\large $ (2) \;\;\;  t_{L}(j+m,2j+1) = L!\, \Delta^{L} \! \left[ H_{L}^j(m) \right]  $} &          \\
&          \\
{\large \mbox{where}} \;\; {\large $ H_{L}^j(m) = \left[\displaystyle {j+m \choose L} {m-j-1 \choose L}\right] $}  &          \\
&          \\

\hspace{17mm}  {\large $ \Delta h(m) = h(m+1) - h(m) $} & \\
&          \\
\hspace{17mm}  {\large $ \Delta^{k+1} h(m) = \Delta [\Delta^{k} h(m)] $} & \\
&          \\
\hline

\end{tabular}
\end{sidewaystable}

\newpage
 {\bf Table IV}\\
 
 Title: A comparison of expansions, traces, matrix elements, Hermitian conjugates and density operators $\hat \rho$ for the spin polarization operators $\hat{T}^{(j)}_{\lambda \mu}$ and the Chebyshev polynomial 
operators $ f_{\lambda} ^{(j)}( {\bf \hat{n}} \cdot {\bf J}) $.

\begin{sidewaystable}
\begin{tabular}  {||c|c|c||} \hline  &  & \\  
{\large Operator }& {\large Spin Polarization} & {\large Chebyshev Polynomial}
\\
& & \\
\hline

  {\large $\hat{O}$} &   {\large $ \hat{O} = \hat{T}_{\lambda \mu}^{(j)} $}   &
 {\large $\hat{O} = f^{(j)}_{\lambda}({\bf \hat{n}} \cdot {\bf J}) = \displaystyle\sum_{\mu=-\lambda}^{\lambda} C_{\lambda \mu}^{\star}({\bf \hat{n}}) \; \,
\hat{T}^{(j)}_{\lambda \mu}  $}  \\
   \hline \hline

{\large Expansions} & 
{\large$  \hat{A}  = \displaystyle\sum_{\lambda=0}^{2j}\sum_{\mu=-\lambda}^{\lambda} A_{\lambda \mu}^{(j)} \;  \hat{T}_{\lambda \mu}^{(j)} $ } &
 {\large $ \hat{B}  = \displaystyle\sum_{\lambda=0}^{2j} B_{\lambda}^{(j)}  \,  f_{\lambda} ^{(j)}( {\bf \hat{n}} \cdot {\bf J}) $}
\\
& & \\
 &{\large $ A_{\lambda \mu}^{(j)} = \mbox{Tr} \! \left[ \left[ \hat{T}_{\lambda \mu }^{(j)} \right]^{\! \dagger}  \! \hat{A}    \right] $}&
{\large $B_{\lambda}^{(j)} = \mbox{Tr} \! \left[  f_{\lambda} ^{(j)}( {\bf \hat{n}} \cdot {\bf J})   \,  \hat{B}   \right]$} \\

 \hline
 
 {\large Traces} & 
{\large$ \mbox{Tr} \! \left[\hat{T}_{\lambda \mu}^{(j)} \left[ \hat{T}_{\lambda^{\prime} \mu^{\prime} }^{(j)} \right]^{\! \dagger}  \right] 
= \delta_{\lambda \lambda^{\prime} } \; \delta_{\mu \mu^{\prime} }     $}  &  
{\large $  \mbox{Tr} \! \left[ f^{(j)} _{\lambda}({\bf \hat{n}} \cdot {\bf J}) \;   f^{(j)} _{{\lambda}^{\prime}}({\bf \hat{n}} \cdot {\bf J}) \right] = 
\delta_{\lambda \lambda^{\prime} }$ }
 \hspace{3mm}

\\

\hline

 & & \\
 {\large $\langle m| \; \hat{O} \;  |m^{\prime} \rangle$}  &   {\large $  C_{jmj-m^{\prime}}^{\lambda \,(m-m^{\prime})} \; (-1)^{j-m^{\prime}} $} & 
{\large $  C_{\lambda \, (m-m^{\prime})}^{\star}({\bf \hat{n}}) \; C_{jmj-m^{\prime}}^{\lambda \, (m-m^{\prime})} \; (-1)^{j-m^{\prime}} 
$} \\
   &  & \\

\hline

{\large Hermitian} & {\large $ \left[ \hat{T}_{\lambda \mu }^{(j)} \right]^{\! \dagger}    = (-1)^{\mu} \;\, \hat{T}_{\lambda -\mu}^{(j)}$}&
 {\large $ \left[ f_{\lambda}^{(j)} ( {\bf \hat{n}} \cdot {\bf J})  \right]^{\! \dagger} =  f_{\lambda}^{(j)} ( {\bf \hat{n}} \cdot {\bf J})  $}  \\
{\large Conjugate}& & \\

\hline

{\large $\hat \rho$}   &   {\large $ \displaystyle\sum_{\lambda=0}^{2j}\sum_{\mu=-\lambda}^{\lambda}  \mbox{Tr} \! \left[ \hat \rho \,   
\left[ \hat{T}_{\lambda \mu }^{(j)} \right]^{\! \dagger}  \right]  \hat{T}_{\lambda \mu }^{(j)}   $}   & {\large $ \;\;\;   \displaystyle\sum_{\lambda=0}^{2j} \frac{(2\lambda+1)}{4\pi} 
\! \! \int _{{\bf S}^2}  f_{\lambda}^{(j)}( {\bf \hat{n}} \cdot {\bf J}) \;  
\mbox{Tr} \!\left[ \hat \rho \,  f_{\lambda}^{(j)}( {\bf \hat{n}} \cdot {\bf J}) \right]   d{\bf \hat{n}}  \;\;\;$}   \\
&     &      \\

\hline 

\end{tabular}
\end{sidewaystable}

\newpage
{\bf Table V}\\
 
 Title: Similarity Transforms \\
\vspace{5mm}

 Caption: Similarity transforms of $J_z$   and $f^{(j)}_{\lambda}(J_z)$. 
\newpage
\begin{tabular}  {||c|c||}  \hline  &   \\  
 &   \\  
 {\large Operator} & {\large Similarity Transform}  \\  
 &   \\  
 \hline  \hline &   \\  
 & \\
{\large $\hat O$}  & {\large $ \hat{{\cal D}}^{(j)}(\theta, {\bf \hat{n}}_{\bot}) \, \hat O   \left [\hat{{\cal D}}^{(j)} (\theta, {\bf \hat{n}}_{\bot}) \right ]^{\!\dagger}$ } \\
 &  \\ 
\hline 
 &  \\
 {\large $J_z $} & {\large $({\bf \hat{n}} \cdot {\bf J})$} 
 \\
 \hspace{8mm} & \\   \hline
  &  \\

{\large $f^{(j)}_{\lambda}(J_z)$}  & 
 {\large $f^{(j)}_{\lambda}({\bf \hat{n}} \cdot {\bf J})$} 
\\

&    \\
 \hline

\end{tabular}

\newpage
 {\bf Table VI}\\
 
 Title: A comparison of the relative orientations of Meckler's instantaneous axis ${\bf \hat{b}}(t)$ \cite{meckler:majorana} and Abragram's magnetic moment ${\bf \hat{m}}(t)$ \cite{abragamtext}.

\begin{sidewaystable}
\begin{tabular}  {||c|c|c||} \hline  &  & \\  
&   {\large Meckler \cite{meckler:majorana}}  & {\large Abragam \cite{abragamtext}} \\
  & & \\
  & & \\  \hline \hline
  & & \\

{\large$  t=0 $} & 
{\large$  {\bf \hat{b}}(0)  \parallel   {\bf \hat{a}}$}  & {\large $  {\bf \hat{m}}(0)  \parallel  \mbox{H}_0 \, {\bf \hat{z}}$}
\\
 & & \\
& & \\

 \hline
 & & \\ 
 {\large$  t > 0$} & 
{\large$  {\bf \hat{b}}(t)  \cdot{\bf \hat{a}} =1-\displaystyle\frac{\lambda^2}{u^2}(1- \cos ut)  $}  &  
{\large $  {\bf \hat{m}}(t)  \cdot{\bf \hat{m}}(0) =1-\displaystyle\frac{\omega^2_1}{a^2}(1- \cos at)  $ }
 \hspace{3mm}

\\
 & & \\
& {\large $ \!\!\!\! \!\!\!\! \!\!\!\!   \!\!\!\! \!\!\! \!\!\equiv Z $}  & {\large $ \!\!\!\! \!\!\! \equiv \cos \alpha $ } \\

& & \\

\hline
  & & \\ 
   {\large Nutation } & 
{\large$  u=\sqrt{\lambda^2+(\omega- \omega_0)^2} $}  & {\large $  a=\sqrt{\omega_1^2+(\omega-\omega_0)^2} $ }
\\
  {\large frequency}&  &  \\
&  &  \\
\hline
& & \\

{\large Excitation} & {\large $ \lambda$}& {\large $ \omega_1$}  \\
{\large radiofrequency}& & \\

&     &      \\

\hline

\end{tabular}
\end{sidewaystable}

\newpage
 {\bf Table VII}\\
 
 Title: Probability distributions and tomographic reconstruction relations.  \\
\vspace{5mm}
 Caption:  Probability distributions defined in the left-hand column are used to define the corresponding tomographic reconstructions of the density matrix $\hat \rho$ 
in the right-hand column. The function $F_{\lambda}({\bf \hat{n}} ) $ used in the definitions of the probability distributions is defined at the bottom of the left-hand column, and the function ${\cal F}_{\lambda}({\bf \hat{n}} ) $ used in the definitions of the tomographic reconstruction relations is defined at the bottom of the right-hand column.

\begin{sidewaystable}
\begin{tabular}  {||l|l||}  \hline  &   \\  
  
 \hspace{14mm} {\large Probability Distributions}& \hspace{14mm} {\large Tomographic Reconstruction Relations}  \\

 &  \\ \hline \hline
  &  \\

 \hspace{6mm} {\large$  w^{(j)}(m,{\bf \hat{n}})  =  \mbox{Tr} \! \left[ \hat \rho \;  \mbox{{\boldmath $\Pi$}}^{(j)}(m,{\bf \hat{n}})  \right]$}  & 
\hspace{6mm} {\large$ \hat \rho=  \displaystyle\sum_{\lambda=0}^{2j} \displaystyle\sum_{m=-j}^j  f^{(j)}_{\lambda}(m)
 \int_{{\bf S}^2}  w^{(j)}(m,{\bf \hat{n}}) \; {\cal F}_{\lambda}({\bf \hat{n}} )\; d{\bf \hat{n}} $}  \hspace{6mm}
\\
  \hspace{29mm} {\large$ =  \displaystyle\sum_{\lambda=0}^{2j} f^{(j)}_{\lambda}(m) \; F_{\lambda}({\bf \hat{n}} ) $}  \hspace{6mm} & \\
&    \\
\hline
&    \\ 
\hspace{6mm} {\large $  Q({\bf \hat{n}}) =  \mbox{Tr} \! \left[ \hat \rho \;  \mbox{{\boldmath $\Pi$}}^{(j)}(j,{\bf \hat{n}})  \right]$} &
\hspace{6mm} {\large $  \hat \rho=   \displaystyle\sum_{\lambda=0}^{2j} \left[  f^{(j)}_{\lambda}(j) \right]^{-1} 
 \int_{{\bf S}^2}   Q({\bf \hat{n}}) \; {\cal F}_{\lambda}({\bf \hat{n}} ) \; d{\bf \hat{n}} $}
 \\
\hspace{18.5mm} {\large$ =  \displaystyle\sum_{\lambda=0}^{2j} f^{(j)}_{\lambda}(j) \; F_{\lambda}({\bf \hat{n}} ) $}  \hspace{6mm} & \\
&    \\
&    \\
\hline
&    \\ 
 \hspace{6mm}{\large $  W({\bf \hat{n}}) =   \mbox{Tr} \! \left[ \hat \rho \;  \Delta^{\!(j)}({\bf \hat{n}})  \right]$}  &  
 \hspace{6mm}   {\large $ \hat \rho=    \displaystyle\sum_{\lambda=0}^{2j} \sqrt{\frac{2j+1}{2\lambda+1}} \;  \displaystyle\int_{{\bf S}^2}  W({\bf \hat{n}}) \; {\cal F}_{\lambda}({\bf \hat{n}} ) \; d{\bf \hat{n}} $} \\
  \hspace{18.5mm}   {\large $ = \displaystyle\sum_{\lambda=0}^{2j} \sqrt{ \frac{2\lambda+1}{2j+1}} \;  F_{\lambda}({\bf \hat{n}} )   $} 
 &    \\
&  \\
\hline \hline

&  \\  

 \hspace{10mm} {\large   $\;\;F_{\lambda}({\bf \hat{n}} ) =  \mbox{Tr} \! \left[ \hat \rho \; f^{(j)}_{\lambda}({\bf \hat{n}} \cdot {\bf J}) \right] $ } 
& \hspace{0mm} {\large $\;\; {\cal F}_{\lambda}({\bf \hat{n}} ) = \left(\displaystyle\frac{2\lambda+1}{4\pi} \right)   f^{(j)}_{\lambda}({\bf \hat{n}} \cdot {\bf J}) 
$ }  \\

&  \\ 
\hline

\end{tabular}
\end{sidewaystable}

\newpage
 {\bf Table VIII}\\
 
 Title: A comparison of tensor operator equivalents expressed in terms of Chebyshev polynomials.\\
\vspace{5mm}
 Caption: First row is work done by Corio \cite{corio:ortho}, and second row is work done by Marinelli et al. \cite{werb:tensor}.

\begin{sidewaystable}
\begin{tabular}  {||l|l||}  \hline  &   \\  
  &  \\
 \hspace{26mm} {\large $\hat{T}_{60}^{(j)}  \equiv f^{(j)}_6(J_z)$}& \hspace{46mm} {\large $\hat{T}_{41}^{(j)} $}  \\
  &  \\  \hline \hline
  &  \\

 \hspace{6mm} {\large$  U^{(6)}_0 \!(J) \sim  \Big\{231  J_{\!z}^6-105(3\kappa-7) J_{\!z}^4  $}  \hspace{6mm} & 
\hspace{6mm} {\large$  U^{(3)}_1 \!(J) \sim J_{+} \Big\{14 \, J_{\!z}^3 + 21J_z^2 +(19-6\kappa) \,J_z  $}  \hspace{6mm}
\\
&    \\
\hspace{6mm} {\large $\;\;\;\;\;\;\;\;\;\;\;+21(5\kappa^2 -25 \kappa +14) J_{\!z}^2 $}&\hspace{6mm} {\large $ \;\;\;\;\;\;\;\;\;\;\;\;\;\;\;+3(2-\kappa) \mathds{1}$}
 \Big \}    \\
&    \\
 \hspace{6mm}{\large $\;\;\;\;\;\;\;\;\;\;\;\;\;\;\; -5\kappa(\kappa^2 -8 \kappa +12) \mathds{1} \Big\} $} &    \\
 &    \\
\hspace{6mm} {\large $\;\;\;\;\;\;\;\;\;\;\;\;\;\;\;\boxed{\kappa \equiv J(J+1)}$} & {\large $\;\;\;\;\;\;\;\;\;\;\;\;\;\;\;\boxed{\kappa \equiv J(J+1)}$}   \\
&  \\
&  \\

 \hline
 &  \\ 
 \hspace{6mm} {\large$  T^{(0)}_6 \!(S) \sim \Big\{231 S_{\!z}^6-105(3K-7)  S_{\!z}^4  $} & 
\hspace{6mm} {\large$  T^{(1)}_4 \!(S)\sim S_{+} \Big \{14 S_{\!z}^3 + 21S_z^2 +(19-6K) S_z  $}  
 \hspace{3mm}

\\
&    \\
\hspace{6mm}  {\large $\;\;\;\;\;\;\;\;\;\;\;+21(5K^2 -25 K +14) S_{\!z}^2 $}&\hspace{6mm}  {\large $ \;\;\;\;\;\;\;\;\;\;\;\;\;\;\;+3(2-K)\, \mathds{1}$} \Big \}    \\
\hspace{6mm} {\large $\;\;\;\;\;\;\;\;\;\;\;\;\;-5K(K^2 -8 K +12)  \mathds{1} \Big\} $} & 
\hspace{6mm} {\large $ \;\;\;\;\;\;\;\;\;\; \sim  S_{+} \! \displaystyle\sum_{n=1}^4 \; A_{4n}^1 \, \left(\displaystyle\frac{\partial}{\partial S_z}\right)^{\!\!n}  
\left[ f^{(S)}_4 (S_z) \right]$}   \\
&   \\
\hspace{6mm} {\large $\;\;\;\;\;\;\;\;\;\;\;\;\;\;\;\boxed{K \equiv S(S+1)}$} & {\large $\;\;\;\;\;\;\;\;\;\;\;\;\;\;\;\boxed{K \equiv S(S+1)}$} \\
&  \\

&  \\

\hline

\end{tabular}
\end{sidewaystable}

 \newpage
 {\bf Table IX}\\

 Title: Legendre polynomial  operators.   \\
\vspace{5mm}
 Caption: A comparison of the Legendre polynomial operators defined by Schwinger \cite{schwinger:majorana} ($P_{\lambda}({\bf J})  $ in the second column) and by Zemach \cite{zemach} ($ \overline{P}_{\lambda}({\bf \hat{n}} \cdot {\bf J}) $ in the third column) with the Legendre polynomials
 ($ P_{\lambda}(\cos \theta) $ in the fourth column). ($ \kappa \equiv {\bf J} \cdot {\bf J} =  j(j+1))$

 \ \\ 
\begin{sidewaystable}
\begin{tabular}{||c||c|c|c||} 

 \hline & & & \\ {\large  $\lambda$ } & {\large $P_{\lambda}({\bf J})$} &
 \hspace{2mm}   {\large $ \overline{P}_{\lambda}({\bf \hat{n}} \cdot {\bf J}) $} \hspace{2mm}  
& {\large $ P_{\lambda}( \cos \theta ) $}  \\
& & &  \\   \hline \hline
 & & &  \\
{\large 0} & {\large $ \mathds{1}  $} & {\large $ \mathds{1}      $} & {\large $  1  $}
\\
 & & &  \\
\hline

 & &  & \\
{\large  1} & {\large $ \left[ \kappa^2  \right]^{-1/2} J_z   $} & {\large $  ({\bf \hat{n}} \cdot {\bf J})  $} & {\large $  \cos \theta   $}
\\
& &  & \\
 \hline
 & & &  \\ 
{\large  2} & {\large $  \left[ \kappa^2 \!  \left(\kappa^2-\tfrac{3}{4}\right) \right]^{-1/2} \, \tfrac{1}{2} \! \left[3J_z^2- \kappa \mathds{1} \right]  $}  &   
{\large $  \tfrac{1}{2} \! \left[3 ({\bf \hat{n}} \cdot {\bf J})^2-\kappa  \,\mathds{1} \right]    $}  
&    {\large $ \tfrac{1}{2} (3 \cos^2 \theta-1) $}   \\ 
& & &  \\ 
\hline
& &  &  \\ 
{\large  3} & \hspace{2mm}    {\large $ \left[ \kappa^2 \!  \left(\kappa^2-\tfrac{3}{4}\right) \! \left(\kappa^2-2\right) \right]^{-1/2}\,  \tfrac{1}{2} \!
 \left[5J_z^3 -\left(3 \kappa  -1 \right) J_z \right]    $ } \hspace{2mm}   
&    \hspace{5mm}   {\large $  \tfrac{1}{2} \! \left[5 ({\bf \hat{n}} \cdot {\bf J})^3-(3\kappa -1)({\bf \hat{n}} \cdot {\bf J}) \right]  $} \hspace{5mm}    & 
 {\large $ \tfrac{1}{2}(5\cos^3\theta -\cos \theta)$}    \\
& &  & \\ 
\hline

\end{tabular}
\end{sidewaystable}


\begin{thebibliography}{99}
\bibitem{corio:correction}
In Corio \cite{corio:siam},  the term $(2i \sin \Phi)^n$ in Eq.(3.2) should be corrected to read  $(2i \sin (\tfrac{1}{2}\Phi))^n$
\bibitem{footnote2}
Given the equivalence between Chebyshev polynomials and Clebsch-Gordan coefficients, the finite difference expression for the Chebyshev polynomials  in this table implies  that Clebsch-Gordan coefficients are also related to finite differences. \cite{varshal1:ang}
\end{thebibliography}
 \end{document}